\documentclass[twocolumn, twocolappendix]{aastex6}
\usepackage{graphicx}
\usepackage{amssymb,amsfonts,amsmath,amstext,amsgen,amsopn,amsxtra,indentfirst,times}

\usepackage{multirow}
\usepackage{setspace}
\usepackage{enumitem}
\usepackage{longtable} 
\usepackage{epstopdf}
\newcommand{\kms}{km\,s$^{-1}$}

\newcommand{\fil}{$f_\mathrm{IL}$}
\newcommand{\fnl}{$f_\mathrm{NL}$}
\newcommand{\wil}{$W_\mathrm{IL}$}
\newcommand{\wnl}{$W_\mathrm{NL}$}

\newcommand{\dnl}{$\Delta \mu_\mathrm{NL}$}

\newcommand{\Ha}{H$\alpha$}
\newcommand{\ghafas}{\texttt{GH$\alpha$FaS}}
\newcommand{\osiris}{\texttt{OSIRIS}}

\newcommand{\sex}{\textsc{Sextractor}}

\begin{document}

%===================================================================== % FRONT MATTER %=====================================================================

\title{Balmer filaments in Tycho's supernova remnant: an interplay between cosmic-ray and broad-neutral precursors}

%\correspondingauthor{Sladjana Kne\v{z}evi\'{c}}
%\email{sladja.ast@gmail.com}

\author{Sladjana Kne\v{z}evi\'{c}\altaffilmark{1,}\altaffilmark{2}}
\altaffiltext{1}{Department of Particle Physics and Astrophysics, Faculty of Physics, The Weizmann Institute of Science, P.O. Box 26, Rehovot 76100, Israel}
\altaffiltext{2}{Benoziyo Fellow, Email: sladja.ast@gmail.com, ORCID: 0000-0003-1416-8069}

\author{Ronald L\"asker\altaffilmark{3}}
\altaffiltext{3}{Finnish Centre for Astronomy with ESO (FINCA), University of Turku, V\"ais\"al\"antie 20, FI-21500 Kaarina, Finland}

\author{Glenn van de Ven\altaffilmark{4}}
\altaffiltext{4}{Max Planck Institute for Astronomy, K\"{o}nigstuhl 17, D-69117, Heidelberg, Germany}

\author{Joan Font\altaffilmark{5}}
\altaffiltext{5}{Instituto de Astrof\'{i}sica de Canarias, V\'{i}a L\'{a}ctea, La Laguna, Tenerife, Spain}

\author{John C. Raymond\altaffilmark{6}}
\altaffiltext{6}{Harvard-Smithsonian Center for Astrophysics, 60 Garden Street, Cambridge, MA 02138, U.S.A.}

\author{Coryn A.~L. Bailer-Jones\altaffilmark{4}}
%\affiliation{Max Planck Institute for Astronomy, K\"{o}nigstuhl 17, D-69117, Heidelberg, Germany}

\author{John Beckman\altaffilmark{5}}
%\affiliation{Instituto de Astrof\'{i}sica de Canarias, V\'{i}a L\'{a}ctea, La Laguna, Tenerife, Spain}

\author{Giovanni Morlino\altaffilmark{7}}
\altaffiltext{7}{INFN – Gran Sasso Science Institute, viale F. Crispi 7, 67100 L'Aquila, Italy}

\author{Parviz Ghavamian\altaffilmark{8}}
\altaffiltext{8}{Department of Physics, Astronomy and Geosciences Towson University, Towson, MD 21252, U.S.A.}

\author{John P. Hughes\altaffilmark{9}} 
\altaffiltext{9}{Department of Physics and Astronomy, Rutgers University, 136 Frelinghuysen Road, Piscataway, NJ 08854, U.S.A.}

\author{Kevin Heng\altaffilmark{10}}
\altaffiltext{10}{University of Bern, Center for Space and Habitability, Sidlerstrasse 5, CH-3012, Bern, Switzerland}

\begin{abstract} 

We present \Ha\ spectroscopic observations and detailed modelling of the Balmer filaments in the supernova remnant Tycho (SN\,1572). We used \ghafas\ (Galaxy \Ha\ Fabry-P\'erot Spectrometer) on the William Herschel Telescope with a 3.4\arcmin$\times$3.4\arcmin\ field-of-view, 0.2\arcsec\ pixel scale and $\sigma_\mathrm{instr}=8.1$\,kms$^{-1}$ resolution at 1\arcsec\ seeing for $\sim10$ hours, resulting in 82 spatial-spectral bins that resolve the narrow \Ha\ line in the entire SN\,1572 northeastern rim. For the first time, we can therefore mitigate artificial line broadening from unresolved differential motion, and probe \Ha\ emission parameters in varying shock and ambient medium conditions. Broad \Ha\ line remains unresolved within spectral coverage of 392\,kms$^{-1}$. 
We employed Bayesian inference to obtain reliable parameter confidence intervals, and quantify the evidence for models with multiple line components. The median \Ha\ narrow-line full-width at half-maximum of all bins and models is $W_\mathrm{NL}=(54.8\pm1.8)$\,kms$^{-1}$ at the 95\% confidence level, varying within [35,\,72]\,kms$^{-1}$ between bins and clearly broadened compared to the intrinsic (thermal) $\approx$\,20\,kms$^{-1}$. Possible line splits are accounted for, significant in $\approx18\%$ of the filament, and presumably due to remaining projection effects. We also find wide-spread evidence for intermediate-line emission of a broad-neutral precursor, with median $W_\mathrm{IL}=(180\pm14)$\,kms$^{-1}$ (95\% confidence). Finally, we present a measurement of the remnant's systemic velocity, $V_\mathrm{LSR}=-34$\,kms$^{-1}$, and map differential line-of-sight motions. Our results confirm the existence and interplay of shock precursors in Tycho's remnant. In particular, we show that suprathermal narrow-line emission is near-universal in SN\,1572 and that, in absence of an alternative explanation, collisionless supernova remnant shocks constitute a viable acceleration source for Galactic TeV Cosmic-Ray protons.

\end{abstract}


%=====================================================================
% BEGIN MAIN BODY %
%=====================================================================

 %============================= section 1 ================================
 \section{Introduction}
 \label{sec:intro}
 %=====================================================================

Supernova remnant (SNR) shocks are suspected to be the long-sought Galactic cosmic-ray (CR) sources. Observational evidence for a particle acceleration at work in SNRs has been seen throughout synchrotron emission from electrons, gamma radiation, signatures of an amplified magnetic field, higher compression ratio than predicted by jump conditions and through optical line profiles from the shocks propagating in a partially ionized plasma (for a review see \cite{helder12}). Since most of the CRs that we detect on Earth consist of protons, finding evidence for accelerated CR protons in SNRs with the energies up to the 'knee' ($\sim1\,\mathrm{PeV}$) is thus of a particular importance. Insight in CR protons is possible either through gamma ray spectrum as a result of neutral pion decay \citep{ackermann13}, or throughout optical wavelength window by carefully analyzing the \Ha-line profiles \citep{helder09,helder10,nikolic13}. 
The latter shock emission, usually also referred to optical emission from Balmer-dominated shocks (BDSs), is the subject of study of this paper.

The spectra of BDSs, typically observed around SNRs that originate from Type Ia supernova explosions, show the presence of strong two-component hydrogen lines \cite{heng10}. When a shock wave encounters partly ionized interstellar medium (ISM), the cold pre-shock hydrogen atoms overrun by the shock can either be excited by hot post-shock gas resulting in the narrow \Ha-component emission, or enter a charge exchange (CE) process with the hot post-shock plasma producing hot neutrals whose collisional excitation then give rise to the broad \Ha\ component. The two-component \Ha-line parameters provide valuable information on the CR precursor existence in the shocks. A narrow line broadened beyond 10--20\,\kms\ gives direct evidence of the non-thermal particle presence in the shock precursor \citep{mor3i13}. This is due to the fact that hydrogen atoms are ionized at temperatures larger than $\approx$\,10\,000\,K, but also because the lifetime of the neutral hydrogen in the post-shock region is too short for the collisional interaction to broaden the line profiles \citep{smith94}.
The CRs will heat the cold neutrals in the interstellar medium, resulting in broadening of the narrow \Ha-line, but the CRs will also reduce
the broad \Ha-line width by removing energy from the protons in the post-shock region. 

Several authors proposed that narrow line broadening can also arise from a broad-neutral (BN) precursor: hot neutrals created in CE processes between hot protons and slow neutrals streaming to the pre-shock region \citep{hes94,smith94}. Recent theoretical studies \citep{blasi12,mor2i12} show that a BN precursor does not broaden the narrow component, but rather introduces a third intermediate component with the FWHM (full width at half maximum) of around $\sim$\,150\,\kms\ and depends on the shock speed. The reason is that only a small number of incoming neutrals does interact with ions in the BN precursor because its extent, which corresponds to the interaction length of the returning neutrals, is much smaller than the CE interaction length of the incoming neutrals. 
Therefore, Balmer lines can be used to study the microphysics of collisionless shocks and are currently the only mean that give an insight into the collisionless shocks.

We have observed Tycho's SNR, which has already been well studied across all wavelength ranges \citep{reynolds92,stroman09,bamba05,katsuda10,lkt04,tian11,acciari11,giordano12} and the evidence for particle acceleration in shocks of Tycho's SNR including acceleration up to the knee in the CR spectrum was found \citep{warren05, cc08, eriksen11, slane14}. In 1572, the star exploded as a Type\,Ia supernova,
leaving a remnant at an estimated heliocentric distance of 2.3$\pm$0.5\,kpc \citep{ckr80} \footnote{Several optical studies based on modelling the observed \Ha-line spectra give a distance 2--3\,kpc. Radio, X-ray and gamma-ray observations prefer larger distances 3--5\,kpc (for a review see \citealt{hayato10}).}. At that distance, the remnant's diameter of 8\,\arcmin\ corresponds to $\approx$\,5\,pc. Density gradients in the medium around the remnant \citep{williams13} modified the evolution of the shock, which in turn resulted in the asymmetric remnant. The lower shock velocity inferred in the northeastern (NE) part suggests the shock interaction with a dense ambient medium, namely a diffuse cloud \citep{reynolds99,lkt04}.

\begin{figure*}[!t]
	\centering
		\includegraphics[width=0.4\textwidth,height=0.9\textwidth,angle=-90]{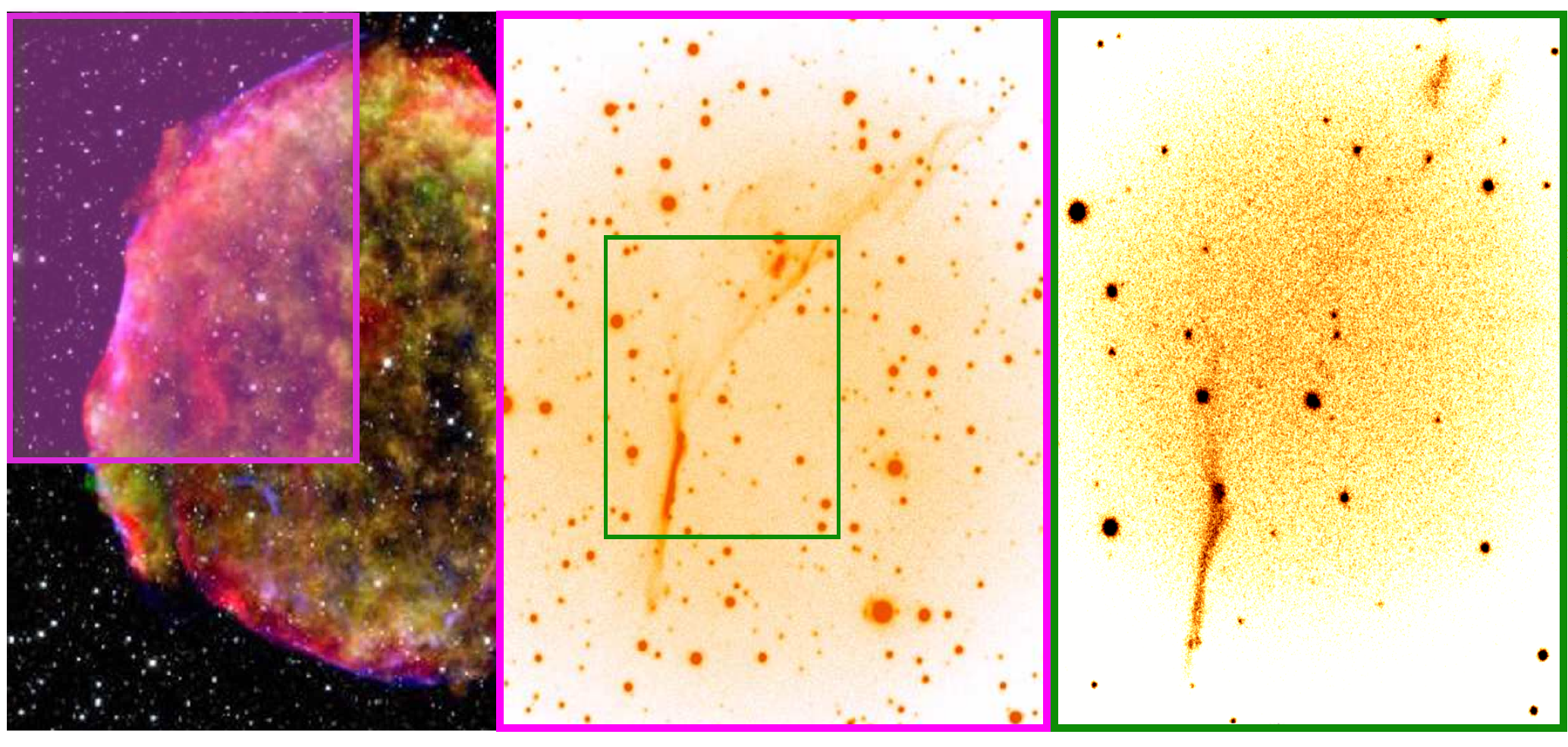}
	\caption{\small The left panel shows a composite image of the remnant ($\sim$8\,\arcmin\ in diameter) of Tycho Brahe's 1572 
	supernova, combining data from the Chandra X-ray Observatory (yellow, green, blue; NASA/CXC/SAO), Spitzer Space Telescope 
	(red; NASA/JPL-Caltech), and the Calar Alto Observatory (white stars; Krause et al.). The transparent magenta box indicates 
	the pointing of the ACAM (Auxiliary-port CAMera) on the Cassegrain focus of the WHT with a FOV of 
	4\,\arcmin$\times$4\,\arcmin. The center panel shows a zoom-in on the ACAM FOV. Using the same pointing as with ACAM, we 
	have covered the same region with the \ghafas\ Fabry-P\'erot interferometer with a FOV of 3\arcmin.4$\times$3\arcmin.4. 
	The green box marks the region which is zoomed-in in the right panel to show our reduced and integrated \ghafas\ \Ha\ image.}
	\label{fig:fig1}
\end{figure*}

Previous optical studies of Tycho's SNR have shown indications for CRs (e.g. \citealp{ghava00,lee07,lee10}). However, these studies focused on the \Ha-bright, but very complex 'knot\,g', where multiple or distorted shock fronts can contribute to the measured narrow-line broadening and thus partially mimic the effect that CR acceleration would have. Using the Fabry-P\'{e}rot instrument \ghafas\ (Galaxy \Ha\ Fabry-P\'erot System) on the William Herschel Telescope (WHT), we observed a great portion of the shock front in the NE region of the remnant. The high spatial and spectral resolution together with the large field-of-view (FOV) of the instrument, allow us to measure the narrow \Ha-line width across individual parts of the shocks simultaneously, and thereby study the indicators of CR presence in a large variety of shock front conditions. In particular, the spatial resolution allows us to distinguish intrinsic line broadening from line broadening originating in geometric distortions and differential kinematics. Moreover, our observational setup provides us with a unique insight into the intermediate component existence along the entire filament previously only reported for the bright 'knot\,g' \citep{ghava00,lee07}.
Apart from vastly enhancing the amount and quality of the spectroscopic data available for the NE filament, our study also improves the analysis: 
instead of fitting line models, we employ Bayesian inference to obtain full information and realistic uncertainties on the line parameters, 
as well as quantitative, reliable evidence for the presence of an intermediate line (IL) originating in a BN precursor and multiple shock fronts.
The interpretation of the results is based on predictions of the state-of-the-art shock models that include the effects of BN and CR precursor on the observed \Ha\ profiles \citep{mor2i12,mor3i13}. With the \ghafas\ spectral coverage of approximately 400\,\kms\ we are not able to resolve the broad \Ha-component that was found to be about $\approx$\,2000\,\kms\ in previous studies \citep{ckr80,ghava01}.

%============================= section 2 ================================
\section{Observations \& Data reduction}
\label{sec:observations}
%========================================================================

In order to resolve the narrow \Ha\ lines along the rim of Tycho's SNR we have used the instrument \ghafas\ mounted on the Nasmyth focus of the 4.2\,m WHT \citep{hernandez08}, which operates at the Observatory del Roque de Los Muchachos in La Palma, Canary Islands. \ghafas\ is a Fabry-P\'{e}rot interferometer-spectrometer with a FOV of 3\arcmin.4$\times$3\arcmin.4. Its detector is an Image Photon Counting System (IPCS) for which the absence of readout noise is an advantage for observations of diffuse emission from extended objects. IPCS cameras are almost insensitive to cosmic rays and thus do not require cosmic ray rejection. We used a high resolution mode, acquiring data on 1024$\times$1024 pixels$^2$ with R$\sim$21\,000 resolving power and a pixel scale of nearly 0.2\arcsec/pixel. The free spectral range (FSR) of the etalon was 8.56\,\AA\ or 392\,\kms\ centered at 6561\,\AA\ and split into 48 {\it channels} that differ in their central wavelength, leading to a sampling velocity resolution of 8.16\,\kms. The instrument response function is well approximated by a Gaussian with FWHM of 19\,\kms\ \citep{blasco10}. 

The observations were conducted on 15-19 November 2012 under FWHM$\simeq$1\arcsec\ seeing conditions. Successive exposures differ by one channel, and 
thus 48 successive exposures complete one {\it cycle} that covers the full spectral range. Total integration time was $\approx$ 9.6\,h, comprising 72 cycles and 3456 $10\,\mathrm{s}$-exposures. Observations conducted in several cycles provide homogeneous airmass and atmospheric conditions for all channels.
We reduced the data (see Figure~\ref{fig:fig1}) by first applying the phase correction to all exposures individually, where we follow the standard 
procedure for \ghafas\ data described in \citep{hernandez08}. The phase correction is a process of designating the photons' positions for the interference rings and assigning the corresponding wavelength to each position $(x,y)$ on the image, $\lambda_i=\lambda_i(x,y)$. As a result, from each exposure $D_i(x,y)$, we build a data-{\it subcube} $D_i(x,y,\lambda)$ with 48 monochromatic images. In order to use the largest possible \ghafas\ FOV, we did not use the optical derotator. Therefore, we have to align and derotate the observed data-subcubes before co-adding them. To determine the exposures' relative pointing and 
orientation, we measure the centroid positions of bright point sources (stars) on a stack that combines each exposure with the $2 \times 4$ exposures that 
precede and suceed it. Thus, we assure that at least three bright point sources are detected with enough flux for a $<1$ pixel centroid precision. 
This requirement means that we have to discard the first four and the last four $D_i$ of each "run" of consecutive observations, but this concerns only 57 exposures (2\%). Along with removing 13 cycles that suffer from reflected light or other defects that we noticed in visual inspection of the data, and also excluding 7 cycles that lack reference sources for derotation, we retain 2439 exposures in 52 cycles. The final datacube $D(x,y,\lambda)$ results from 
summing all data-subcubes and consists of 48 calibrated constant-wavelength slices. 

Besides the datacube, that is, the stack of all aligned data-subcubes, we produce a background-cube and a flatfield-cube. To this end, we model the background flux in individual exposures as well as the flatfield image (the position-dependent throughput of the optical system), and subsequently process the individual background and flatfield frames in exactly the same manner as the corresponding data frames. By constructing the co-added background- and flatfield-cube and including them in our parametric models of the observed shock emission, as opposed to subtracting the background from the individual exposures and dividing them by the flatfield image before calibration and co-addition, we preserve the photon (Poisson) statistics in the data-cube and simultaneously account for the variable effective exposure time in the stack of aligned subcubes. Details on modelling of the flatfield and background frames are given in Appendix \ref{DR}.

\newpage
%============================= section 3 ================================
\section{Analysis}
\label{sec:analysis}
%========================================================================

Our analysis is guided by the following goals and principles:

\begin{enumerate}
 \item to determine the \Ha\ narrow-line (NL) width across a maximal area of the observed shock
 \item to achieve maximum spatial resolution, implying minimal binning (bin size) and signal per bin
 \item to still extract line parameters reliably and, in particular, characterize their uncertainties accurately
 \item to do so even when including up to two additional lines (10 model parameters)
 \item to compare single-NL and multi-line models and to quantify their relative evidence
\end{enumerate}

In order to achieve these aims we decide to perform parameter estimation and model comparison using Bayesian inference instead of traditional (maximum-likelihood, minimum-$\chi^2$) fitting routines. We also account for the Poisson statistics of the data, as opposed to the often tacitly applied Gaussian approximation. Details on our method can be found in subsection~\ref{bayes}, Appendices~\ref{mcmc} and \ref{cv}.

\subsection{Motivation for a multi-line analysis}
\label{motivation}

Narrow \Ha\ lines in non-radiative shocks around SNRs are conventionally modeled by a single Gaussian, which theoretically has a width (FWHM) of $W_\mathrm{NL} \sim 15$\kms\ in accordance with the pre-shock temperature of $\sim$\,10$^4$\,K expected for the warm interstellar medium. However, visual inspection of our data indicates that this basic model may not be descriptive; theoretical considerations also justify investigation of more complex models for the shock emission spectrum. Since the spatial elements (bins) cover a small but finite part of the shock, and moreover the observed filament is the projection of an extended shock section along the line-of-sight (LOS), a single-NL model is only suitable if one assumes that within each projected resolution element (bin), the following conditions are realized:
\begin{itemize}
\item the pre-shock ambient medium is homogeneous with constant temperature;
\item the velocity distribution is uniform, without differential bulk velocity components along the LOS; and
\item there are no precursors (classical BDS).
\end{itemize}

If the three conditions above are all satisfied, the projection effects cannot modify the NL width, because both the width and the centroid are the same everywhere, corresponding to the upstream plasma temperature and bulk speed, respectively. Unresolved or projected inhomogeneous pre-shock temperature causes the superposition of Gaussians of different width. The presence of a CR precursor can alter the NL in two different ways: the cold neutrals in the interstellar medium will be heated, resulting in the NL being broadened beyond the normal 10--20\,\kms gas dispersion \citep{mor3i13} and they acquire a bulk speed up to few \% of the shock speed. Therefore, inhomogeneous CR emission can be one reason for a non-Gaussian NL. CRs also transfer momentum to the pre-shock neutrals and potentially introduce a Doppler shift between the gas in the precursor and the pre-shock gas not affected by the precursor if the shock is not moving strictly perpendicular to the LOS \citep{lee07}, in addition to shifts from any pre-existing differential bulk motion in the ambient medium. These projected or spatially unresolved shifts will also alter the line shape, and if the effect is pronounced enough, even lead to a split in the NL. One way to account for those distortions of the Gaussian line shape is by allowing generalized Gaussians with non-zero third- and fourth-order moments, for example using a Gauss-Hermite-polynomial. However, we visually identify splits in the narrow line, and choose to represent a non-Gaussian NL by a sum of two Gaussian components. 

Apart from CRs there is another possible precursor -- a BN precursor, which introduces a new line component: the so-called intermediate line (IL) that can be described by a Gaussian\footnote{Strictly speaking the IL is not a perfect Gaussian, because it results from the population of neutrals undergoing CE in the BN precursor and they have not enough time to thermalize to a single temperature \citep{mor2i12}.} with FWHM in the range 100--350\,\kms\ for a typical shock velocity in the range of [1500, 3500]\,\kms \citep{mor2i12}. We therefore also investigate models that include this IL.
 
Previous observations focused on the brightest \Ha\ knot, where the spectrum was measured by combining all pixels across the so-called `knot~g' \citep{ghava00,lee07} which, as HST imaging shows, has a complex structure \citep{lee10}. It was then used to estimate the NL and IL line widths of 'knot~g', with best-fit 44$\pm$4\,\kms\ and 150\,\kms \citep{ghava00}, or 45.3$\pm$9.0\,\kms\ and 108$\pm$4\,\kms\ \citep{lee07}.

We set out to check if these results hold when `knot~g'  is spatially resolved and its parts analyzed individually, or if spatially averaging the spectrum introduced an artificially large $W_\mathrm{NL}$ and an IL that may have mimicked the effects of a CR or BN precursor. We also vastly extend the areal coverage of those earlier studies to include the lower-surface brightness parts of the filament, and there, too, exploit the high spatial resolution provided by the \ghafas\ instrument. In this way we can investigate whether signature of the existence of CR and BN precursor are present also in regions less complex than the 'knot\,g' and looking for possible difference in the physical properties of the shock.

We find that the contribution of the emission arising in the photo-ionization precursor (PIP), previously suggested and measured by \cite{ghava00} and \cite{lee07}, is negligible (see Appendix~\ref{pip}), so that we do not need to account for it in our filament flux models.

\subsection{Definition of models}
\label{models}

For each location (bin), we consider several parametrized models ($S$) to characterize the shock \Ha\ emission. Regardless of the parametrization (type of model) or specific parameter values ($\theta$), we factor in the local flatfield spectrum ($F$) and add to it the observed background spectrum ($B$) before comparing the model with the data. This has the advantage of preserving the correct photon statistics, and contrasts with the common approach of subtracting the background from the data and dividing by the flatfield before modelling. Hence, for a given location (bin), the full model is represented by $M(\theta) = S(\theta) \times F +B$, while by model we mostly refer just to $S$, the intrinsic, or "source" component.  Note that in this expression, $F$ and $B$ are the result of binning the flatfield cube and the background cube in the same way as the data, and therefore they only depend on wavelength ($\lambda$) for the bin that $M$ describes. $F$ and $B$ are constructed separately from the data and inserted into the model without free parameters. More information on how we established $F$ and $B$ can be found in Section~\ref{sec:observations} and Appendix~\ref{DR}.

Every $S$ consists of a NL with Gaussian profile as well as a constant component ($c$) that accounts for the sum of the continuum level and the broad \Ha\ line (BL), which is several times wider than our spectral range. Depending on its type, $S$ may further include one or two Gaussian components that represent an IL or an additional NL. Overall, in each bin we therefore have four different models $S$ to compare with the data spectrum:

\begin{enumerate}
\renewcommand{\theenumi}{\Roman{enumi}}
 \item \textbf{NL}     - constant plus single narrow line 
 \item \textbf{NLNL}   - constant plus two narrow lines 
 \item \textbf{NLIL}   - constant plus one narrow and one intermediate line 
 \item \textbf{NLNLIL} - constant plus two narrow lines and one intermediate line 
\end{enumerate}

In connection with calculating the models' relative evidence, we also consider a "no-line model" (0L, i.e. with the constant-spectrum as the only component), which gives an auxilary baseline for the relevance of at least the NL being present in the data. A possible choice of model parameters are the component fluxes ($f_c=f_\mathrm{constant}$ and $f_i=f_\mathrm{\{ NL,\,NL1,\,NL2,\,IL\}}$), line centroids ($\mu_i$), and lines' FWHM ($W_i$). The general form of this model is
\begin{equation}
 S(\lambda)=f_c+(2\pi)^{-1/2}\sum_i f_i \exp\left(-(\lambda-\mu_i)^2/(2\sigma_i^2)\right) /\sigma_i~, 
\end{equation} 
where $\sigma_i=\sqrt{W_i^2+W_\mathrm{instr}^2}/\sqrt{4\ln 4}$ is the observed (instrumentally broadened) Gaussian dispersion, and $W_\mathrm{instr}$ the FWHM of the instrumental response.
 
In practice, our models employ a transformed version of those parameters, which has the advantage of their more direct interpretation, and a simpler functional form of the desired parameter priors (see next paragraph). For example, instead of the two NL centroids, $\mu_\mathrm{NL1}$ and $\mu_\mathrm{NL2}$, the NLNL and NLNLIL models use the NL centroid mean ($ < \mu_\mathrm{NL} > $) and the separation between the NLs ($\Delta\mu_\mathrm{NL}$). Similarly, the IL centroid is specified by its offset from the NL centroid or NLNL centroid mean, $\Delta\mu_\mathrm{IL}=\mu_\mathrm{IL}-\mu_\mathrm{NL}$. We replace the component fluxes by the total flux and the components' flux fractions, and use the logarithm of the total flux and line widths as they are strictly positive quantities. A more detailed account of the definition of model parameters can be found in Appendix~\ref{param}.

\begin{figure*}[t!]
	\centering
		\includegraphics[width=0.9\textwidth,height=\textwidth,angle=-90]{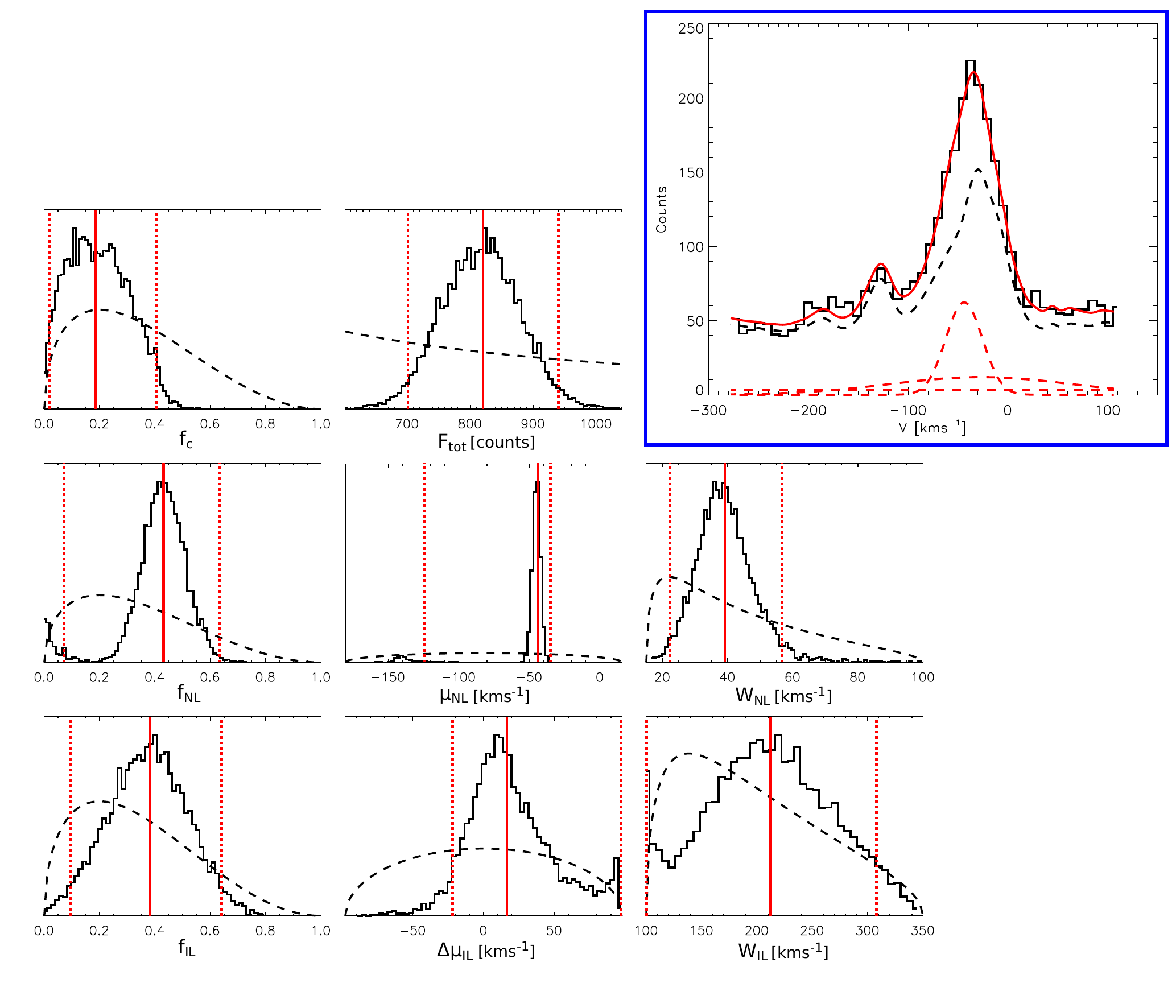}
\vspace{-0.5cm}
	\caption{\small Parameter estimation via Bayesian inference for a bin in the NE filament of Tycho's SNR. 
	The top-right blue panel shows the observed spectrum (solid-black line), the background
model (dashed-black line) and components of the intrinsic median NLIL model (dashed-red lines). The median
model is overplotted with the solid red line. The remaining 8 panels are 1D-marginalized posteriors over model
parameters (solid-black lines): total flux ($F_\mathrm{tot}$ in counts), flux fractions in the continuum ($f_\mathrm{c}$) 
and lines ($f_\mathrm{NL}$ and $f_\mathrm{IL}$), NL centroid ($\mu_\mathrm{NL}$), IL 
offset from the NL centroid ($\Delta \mu_\mathrm{IL}$), and intrinsic line widths ($W_\mathrm{NL}$ and $W_\mathrm{IL}$) with the latter three quantities given in \kms. 
Dashed-black lines are prior distributions and vertical red lines are the estimated parameters of the median model, 
i.e. the median values (solid red), and the boundaries enclosing the highest density 95\% confidence intervals (dashed red).
 }
	\label{fig:fig_post}
\end{figure*}

\begin{figure*}[t!]
	\centering
		\includegraphics[width=0.7\textwidth,height=\textwidth,angle=-90]{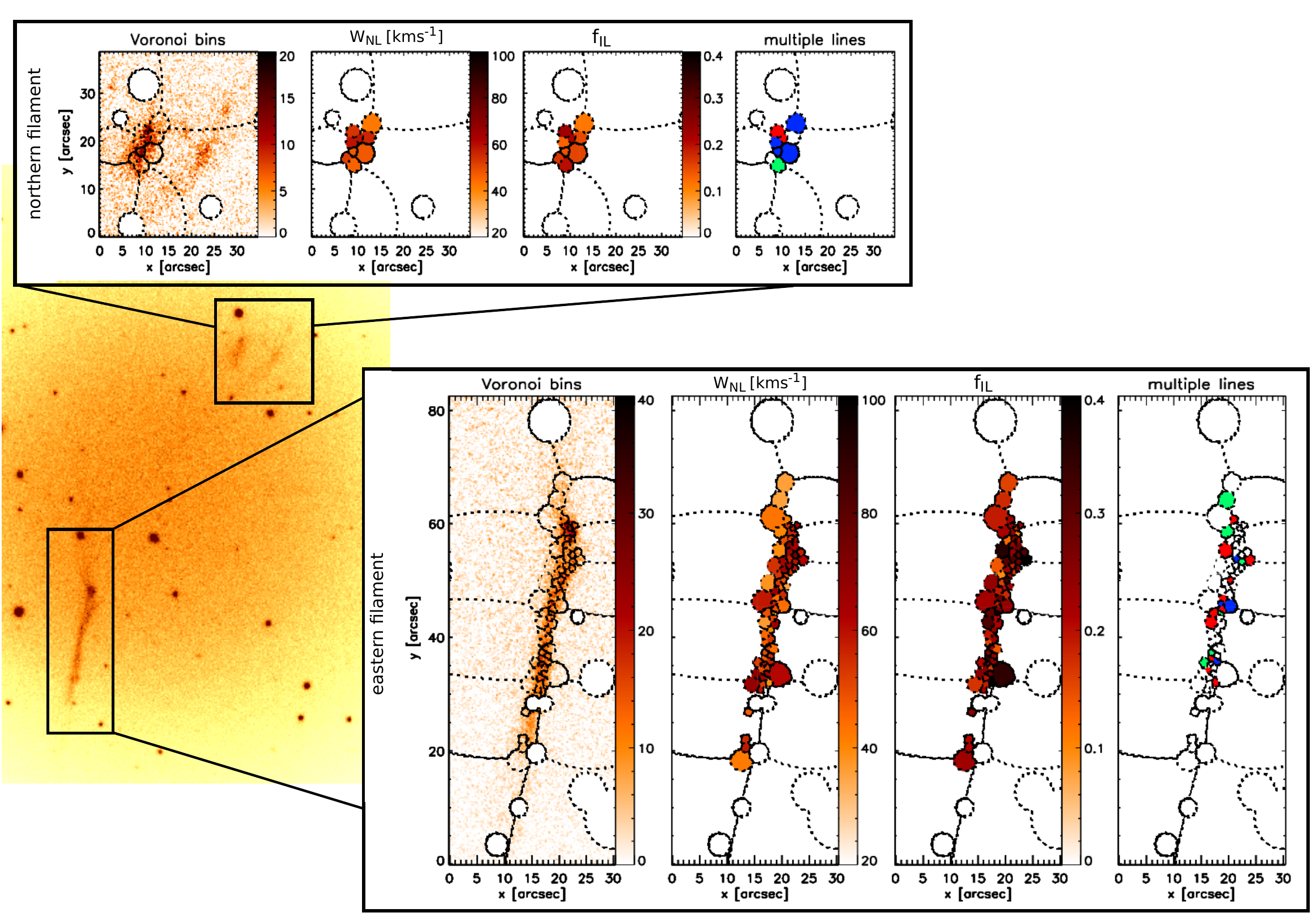}
	\caption{\small NE filament of Tycho's SNR. The two boxes on the \ghafas\ \Ha\ image show northern and eastern shock filaments. 
	The four panels in each box represent: the bin contours overplotted on the background subtracted cube, spatial variation of the 
	median values of evidence-weighted NL width (in \kms) and IL flux fraction posteriors, and bins that show necessity of 
	IL (red), second NL (blue), and $\textit{both}$ the second NL and IL (green). 
	In this last (rightmost) panel, white indicates either bins with too little flux, or those where no line is required in addition to the single NL to describe the data.}
	\label{fig:fig_spatial}
\end{figure*}

The parameter priors, $P(\theta)$, are an integral part of the model definition: they encapsulate what we know (or assume) about the relative probabilities of parameter values {\it a priori}, before considering the data. In particular, they can impose parameter boundaries by way of being zero outside of those. The line centroid parameters are effectively restricted to our spectral window. In our models the FWHM parameters $W_\mathrm{NL}$ and $W_\mathrm{IL}$ are limited to [15, 100]\,\kms\ and [100, 350]\,\kms\, respectively. The lower boundary of $W_\mathrm{NL}$ reflects the lower limit of the pre-shock temperature ($\approx$5000\,K), while the upper boundary is based on theoretical models that include the effects of CR precursor \citep{mor3i13}. The $W_\mathrm{IL}$ range is the theoretical expectation for shock velocities around 2000\,\kms\ and a range of shock parameters (see Figure~10 in \citealt{mor2i12}). 

Inside of the parameter boundaries, we desire to not strongly favor any particular parameter values a priori. However, on physical grounds we prefer a smooth transition of the prior to zero for parameter values approaching the boundaries (see Figure~\ref{fig:fig_post}). We therefore employ shifted and scaled Beta distributions with $\alpha=\beta=1.5$ for the centroid parameters and logarithmic line widths, and a Dirichlet distribution with $\alpha=1.5$ for the component flux fractions which have the constraint of being summed up to unity. For comparison, the special case of "flat" Beta and Dirichlet distributions would be realized by setting $\alpha(=\beta)=1.0$. Thus, our choice slightly favors the centre of the allowed parameter range. The logarithm of the total flux has a flat unbounded prior. Model parameters and their priors are summarized in the Table~\ref{tab:param}, and detailed definitions presented in Appendix~\ref{param}.

\newpage
\subsection{Binning}
\label{binning}

We analyze two shock filaments, one in the more eastern part of the NE rim which contains 'knot~g',  and the other in the more northern part (Figure~\ref{fig:fig_spatial}). We use the Weighted Voronoi Tessellation \citep{ds06} with the adaptive bin size to spatially bin the pixels and obtain a signal-to-noise ratio ($S/N$) of $\gtrsim 10$ (Appendix \ref{voronoi}) in the wavelength-integrated signal that remains after subtraction of the background. This implies an average minimum $S/N$ of 1.4 per spectral element. Due to the seeing of 1\arcsec\ we require at least 5 pixels across, so that for a round bin this implies a minimum of 19 pixels. We exclude bins that would require $>400$ pixels for our target $S/N$, so that unaccounted-for residual background variations, which we estimate to be at most $\sim 2\%$ of the background level, do not significantly effect our measurements. Following these criteria, we study 73 Voronoi bins in the eastern and 9 Voronoi bins in the northern filament.

\subsection{Parameter estimation and model comparison}
\label{bayes}

For each Voronoi bin, we want to find which of the four models, $M$ (see Sec.~\ref{models}), and which vector of model parameters $\theta$, best explain its data (spectrum). Since data are noisy, model comparison and parameter estimation are inherently probabilistic. 
For both tasks and the reasons discussed below, we use Bayesian inference.

The "standard" approach to estimate parameters are the maximum-likelihood (or minimum-$\chi^2$) method.
It relies on a well-defined likelihood maximum (mode), and convergence of the optimizer deteriorates when high noise or multiple modes are present. Both conditions are met in our study: we desire high spatial resolution and therefore small bin size, implying low $S/N$.
Our models are non-linear and comprise up to 10 parameters, implying generally multiple modes. We wish to characterize all those modes, not just the "main" (global) maximum, towards reliable, high-confidence level parameter uncertainties ("errors"), instead of the minimal but common 68\%-confidence ("$1-\sigma"$) error. Often, and in our models with their non-trivial likelihood function, error propagation over a large parameter range is cumbersome or impossible. Finally, the maximum of the likelihood does not provide a quantitative, well-defined measure for relative probability of different models with different parametrization (model comparison). These circumstances make maximum-likelihood or other fitting methods insufficient for our purposes. 

Bayesian inference, by contrast, provides the full, multivariate parameter probability distribution function (PDF), the so-called {\it posterior} $P(\theta|D,M)$, as well as its integral (marginalization) over all parameters, $P(D|M)$ -- the {\it evidence}. Evidences of models are the relative model probabilities, without reference to any specific (e.g., best-fit) parameter values. Bayes' theorem states that the posterior as a function of $\mathrm{\theta}$ is
\begin{equation}
 P(\theta|D,M) =  \frac{P(D|\theta,M) P(\theta|M)}{P(D|M)}  ~. \label{eqn:Bayes}
\end{equation}
It is proportional to the product of likelihood, $L=P(D|\theta,M)$, and {\it prior}, $P(\theta|M)$. $L$ reflects the model and is the probability of the data for given model parameters and measurement errors. Our IPCS instrument counts photons, which are described by a Poisson distribution that therefore represents the measurement error, with expectation value and variance equal to the flux predicted by the spectral model. The prior is the parameter PDF that we know or assume before taking into account the data at hand. Apart from making these assumptions or knowledge explicit (fitting methods {\it implicitly} assume a flat prior in the chosen parameters), it has the advantage of naturally facilitating self-consistent parameter changes.

We represent the $N$-dimensional posterior PDF of the $N$ model parameters as a sample, which we obtain using a Markov chain Monte Carlo (MCMC) method. For details on the sampling algorithm, see Appendix~\ref{mcmc}. The posterior can be summarized in many ways. One is the posterior maximum; we do not emphasize as it is a relatively noisy estimator and its computation is not unique. Instead, we provide the median and the highest-density (shortest) 95\%-confidence intervals of the one-dimensional marginalized distributions that result from integrating $P(\theta | D)$ over all but one parameter. We deem 95\%  to be the minimal confidence level worth quoting, and more reliable than the frequently employed 68\% ("$1-\sigma$") level, which carries a high probability (32\%) of not including the optimal parameter values.

For the parameter estimation, we are interested only in the posterior's shape (relative parameter probabilities)which does not necessitate normalization by the evidence. However, the Bayes factor, i.e. the evidence ratio of two models, is a probabilistically well-defined, quantitative measure for comparing models. For a given $M$, the evidence is defined as $P(D|M) = \int_{\theta} P(D|\theta,M)P(\theta|M)\mathrm{d}\theta$. It is the probability of observing the data when assuming that the model is "true" but parameters are not specified. In practice, the evidence integral is often high-dimensional and therefore computationally intensive. In order to approximate it numerically, we use the \textit{cross validation} (CV) likelihood \citep{cbj12}, particularly the \textit{leave-one-out} CV likelihood: $P(D|M) = L_\mathrm{LOO-CV}(D|M)$, see also Appendix~\ref{cv}. To compute it, samples are drawn from the data partition posteriors instead of the prior as in "standard" evidence integrals, and it has the advantage that it depends on the prior only to second order.
Each bin spectrum has 48 elements and the LOO-CV is applied to these elements by predicting each of them from the remaining 47 elements under the model $M$  (marginalized over the parameters).

Since $P(D|M)$ can be a very large (or small) number, its absolute value is meaningless, and only relative values are needed for different models, we express it as the base-10 logarithm of the ratio with some reference model. As a matter of choice, we consider 0.5\,dex log-evidence differences as "significant" to clearly prefer one model over another. This choice is somewhat conservative; testing on simulated data reveals that we start to distinguish the correct models at 0.2\,dex, while our numerical precision is around 0.05\,dex.

With 0.5\,dex criterion per bin we do not rule out the respective other models, but rather indicate that significant evidence exists that an IL or double-NL (or both) is present in the data as a "population" in the filament overall, or conversely, that such an additional line emission is most likely not present if the evidence for the NL model relative to other models is larger than 0.5\,dex. 

The Bayes factors, and in turn the fraction of bins that show significant evidence for a double-NL or an additional IL, depend on the line width used to distinguish an NL from an IL. Our choice of FWHM=100\,\kms\ as lower IL width limit corresponds to shock speeds as low as 1500\,\kms\, while previous results show the shock speed of Tycho's SNR to be at least 2000\,\kms\ \citep{ghava01}. \wil\ $<$ 100\,\kms\ requires unrealistic full electron-proton temperature equilibrium \citep{caprioli15}, or shock speeds lower than 1500\,\kms\ and zero equilibration, i.e. $T_e/T_p = m_e/m_p$. The latter case was already debated by \citealt{mor2i12}, and furthermore $T_e/T_p \ll  0.01$ was never measured in any of the remnants \citep{ghava13}. 
On the other hand, NL widths larger than 60\,\kms\ have not been observed before, and $>100$\,\kms\ would require cosmic ray acceleration efficiency $>$ 40\% \citep{mor3i13}, compared to the more realistic 10--20\% efficiency in SNR shocks (as was also found in Tycho's SNR by \citealt{mc12}). Therefore, our NL-IL separating line width limit robustly distinguishes NL and IL that arise from different processes (cold neutral excitation and broad-neutral precursor).

%==================================================================
\section{Results}
\label{res}
%==================================================================

Following analysis described above, we calculated posterior parameter distributions for every model and Voronoi bin, which we then summarized by the median of the 1D-marginalized posteriors as a central parameter estimator, and the boundaries of the shortest (highest-density) 95\% confidence interval. We also compared models using the Bayes factors (evidence ratios) calculated by the CV likelihood method. Our focus is on the measured NL width, evidence for and magnitude of a split in the NL, evidence for an IL, its strength and width, the variation of LOS velocities across the filament, and possible correlations among the line parameters.  

In Figure~\ref{fig:fig_post}, we illustrate the application of our analysis to one of the bins from our data set. The observed spectrum (solid-black line) is shown in the top blue panel, and the median NLIL model in solid red, while the model components are shown with dashed lines: the nonparametric background model (dashed black), and the parametrized "source"  (SNR emission) model components -- a constant component, one NL and one IL -- in dashed red. The background spectrum, which has been derived and fixed independently (see Appendix \ref{DR}) shows geocoronal and Galactic \Ha\ emission, including the Galactic [NII] line at around -130\,\kms\ LOS velocity. The 1D-marginalized posterior parameter distributions (solid black) and prior parameter distributions (dashed black) are shown in the remaining 8 panels. The prior parameter distributions are overplotted for comparison with the posterior. The priors are chosen to not strongly prefer any parameter values over the allowed parameter range, but taper off smoothly towards zero at the range limits. Most posteriors are significantly different from the priors, thus all the parameters are well constrained by the data. The parameters of the median model are denoted with solid-red vertical lines. In this example, the median NL FWHM width is $W_{NL}\approx 40$\,\kms. The IL is $W_{IL}\approx$\,210\,\kms\ wide and comprises $\approx$\,40\% of the total flux. With dashed-red vertical lines we marked the boundaries of the 95\%-confidence interval. 
Compared to all other model parameters, the posterior shape of \wil\ is more sensitive to the choice of prior (Appendix F). Still, the median and 95\%-confidence intervals of \wil\ of a flat and Beta prior agree within $\approx$\,10\%. This relatively benign difference occurs because by coincidence the peak of the likelihood is close to the centre of the prior range, and because the 95\%-confidence interval boundaries are similar to the minimum/maximum limits of the prior range.

\begin{figure*}[t!]
	\centering
		\includegraphics[width=0.9\textwidth,height=\textwidth,angle=-90]{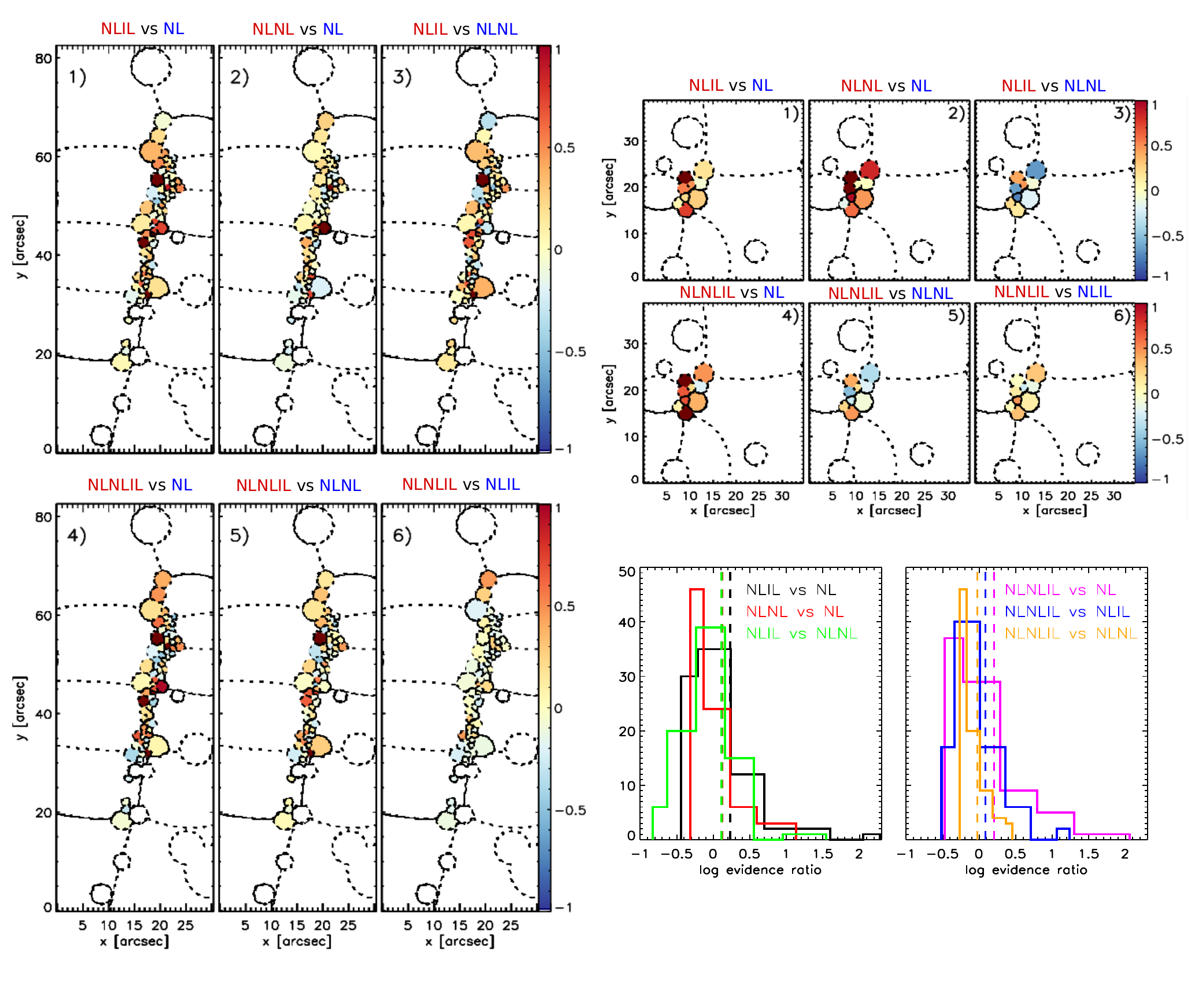}
\vspace{-0.5cm}
	\caption{\small Spatial variation and histogram representation of logarithmic (base 10) evidence ratios across the NE rim in Tycho's remnant.
	                The logarithmic evidence ratios are shown in the range $\pm$\,1\,dex in the spatial maps, while the histograms show the entire range of log evidence ratio values. Bins with log evidence ratio of around +1\,dex ratio would appear as dark red, while bins with log evidence ratios of around -1\,dex appear as dark blue. For example, the dark red bins in the panel 2) of the eastern filament favor NLNL model when compared to NL model with the log evidence ratio between the two models of $>$\,1\,dex. Model comparison is also quantitatively presented in the histogram visualization of evidence ratios where we also plot the mean values of the distributions (vertical dashed lines).}
	\label{fig:fig_ev}
\end{figure*}

To summarize the parameter estimation of one bin, we consider the {\it evidence-weighted} 1D-posterior (see Figure~\ref{fig:globalpicture}, middle row), to which any of the four models that feature the parameter of interest contribute their marginalized posterior in proportion to their evidence. We then give the median of the evidence-weighted posterior, as well as the boundaries of its 95\%-confidence interval.

To summarize the results of all bins, we take three different routes.
\begin{itemize}
\item We consider the distribution of all 82 evidence-weighted posteriors' medians, as well as their $2 \times 82$ lower and upper 95\%-confidence interval boundaries (Figure~\ref{fig:globalpicture}, top panels). This yields information on the variability of line parameters across the filament.
\item Second, we combine (average) the evidence-weighted posteriors of all bins, providing a representation of the information that we {\it typically} find in {\it one} individual bin (Figure~\ref{fig:globalpicture}, middle row of panels, black curves). We also combine model-specific posteriors separately, to illustrate the relative contribution of different models and how parameters estimates depend on the model choice (coloured curves).
\item Finally, we evaluate the parameter constraints as imposed by all data (bins) combined. For that purpose, we sample from all bins' evidence-weighted posteriors, each time computing the median value over all bins. That is, we define one new parameter for each model parameter: the {\it cross-bin median}. Its posterior is plotted in the bottom row of Figure~\ref{fig:globalpicture}, and again summarized by its median and 95\% confidence interval boundaries. We emphasize that this measure is decidedly distinct from a modelling the spectrum of all bins combined, because it is still based on models constrained by all bins individually and independently, and, in particular, allows for local shifts in the line centroids to avoid artificial line broadening.
\end{itemize}

Apart from estimating model parameters, we compare the probability of models (different number of emission lines) against each other. In each bin, we define a model as favored if it has the highest evidence {\it and} a $>0.5$\,dex logarithmic evidence ratio ($>3:1$ probability) over the NL model (see Figure~\ref{fig:fig_spatial}, right panel). This threshold is a matter of choice; it is twice the 0.2\,dex by which the correct model is typically favored in our tests on simulated data, and reflects our approximate notion of the minimum for a "significant" probability. If no model satisfies the 0.5\,dex requirement, we consider the fiducial single-NL model as favored. In this scheme, the data in one bin may favor an IL (NLIL or NLNLIL model), a double-NL (NLNL or NLNLIL model), both (NLNLIL), or none of them (NL). Using this criterion for each bin separately, we ascertain the fraction of bins in which the evidence indicates an IL, as well as the fraction of double-NL occurrence.

\subsection{Narrow-Line width}
\label{sec:res_NL}

The example in Figure~\ref{fig:fig_post} is by no means the only one where the Narrow Line (NL) width, $W_\mathrm{NL}$, is much larger than the maximally allowed thermal NL broadening (20\,\kms). On the contrary: our central estimator, the median of the evidence-weighted $W_\mathrm{NL}$ posterior, is never lower than 35\,\kms, in \textit{any} bin. 
The spatial distribution of estimated $W_\mathrm{NL}$ is shown in panel~2 of Figure~\ref{fig:fig_spatial}. We do not recognize a strong spatial pattern of $W_\mathrm{NL}$; most of the bin-to-bin variations appear to be randomly distributed. However, in the northern part of the eastern filament there appears to be a trend of lower (higher) $W_\mathrm{NL}$ on the pre-shock (post-shock) side, and a generally higher $W_\mathrm{NL}$ in the southern part. The spatial variation of $W_\mathrm{NL}$ (Figure~\ref{fig:fig_spatial}) probably indicates variations in the amount of neutrals in the ambient medium: more neutrals imply a more efficient ion-neutral damping of magnetic waves excited by CRs, thus resulting in a pre-shock gas heated to larger temperatures.
The histogram of median and 95\%-confidence interval boundaries can be found in the top-left panel of Figure~\ref{fig:globalpicture}. As measured by the cross-bin median, the global $W_\mathrm{NL}$ is $(54.8\pm1.8)\,\mathrm{km\,s^{-1}}$ with 95\% confidence. Suprathermal NL widths are required even when only models with double-NL or an additional IL are considered; their average $W_\mathrm{NL}$ is only a few \kms\ lower (see next two subsections).

\begin{figure*}[ht!]
	\centering
		\includegraphics[width=0.9\textwidth,height=\textwidth,angle=-90]{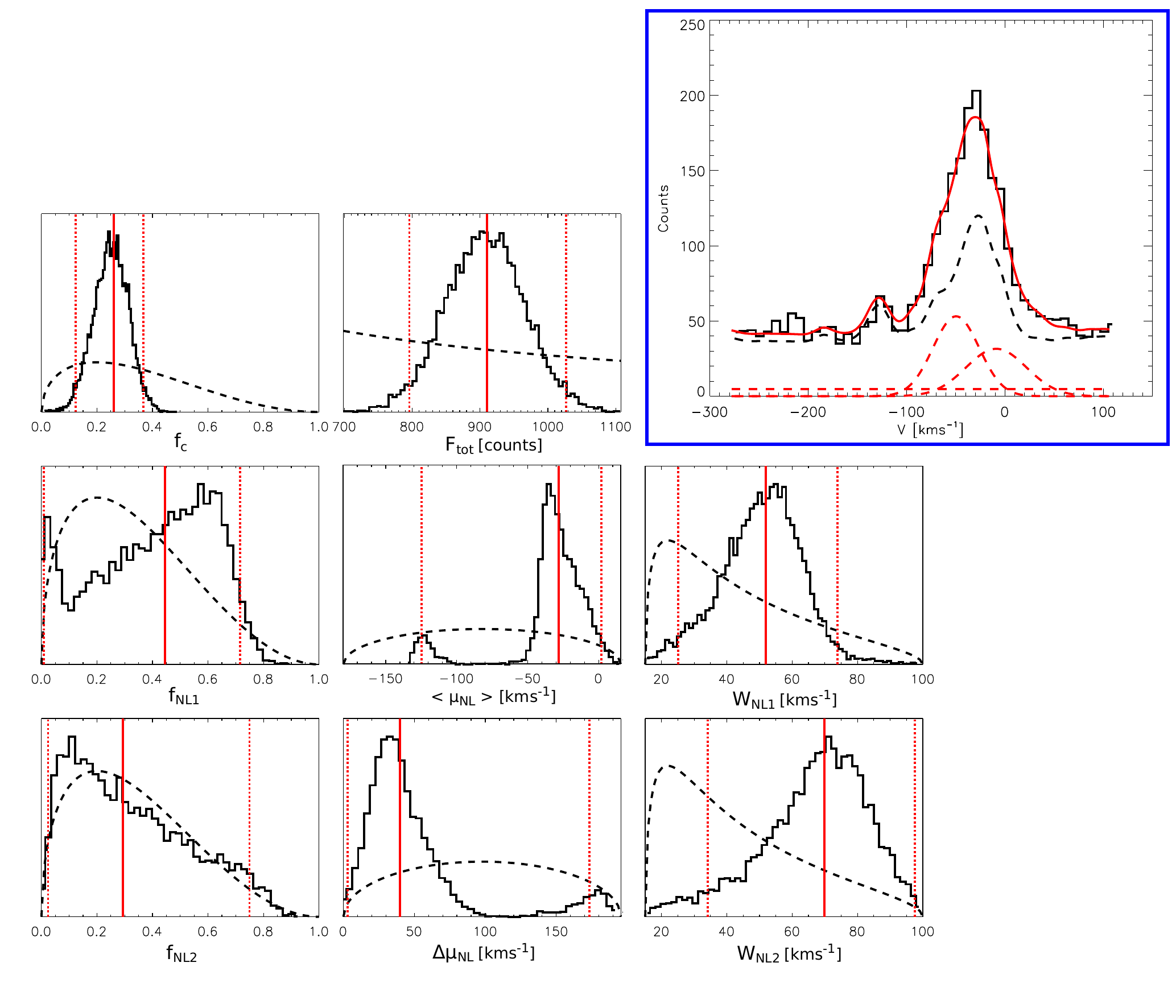}
\vspace{-0.5cm}
	\caption{\small Parameter estimation via Bayesian inference for a bin in the NE filament of Tycho's SNR that requires a second NL. 
	The top-right blue panel shows the observed spectrum (solid-black line), the background
model (dashed-black line) and components of the intrinsic median NLNL model (dashed-red lines). The median
model is overplotted with the solid red line. The remaining 8 panels are 1D-marginalized posteriors over model
parameters (solid-black lines): total flux ($F_\mathrm{tot}$ in counts), flux fractions in the continuum ($f_\mathrm{c}$) 
and two narrow lines ($f_\mathrm{NL1}$, $f_\mathrm{NL2}$), NL centroid mean $\langle \mu_\mathrm{NL} \rangle$, the separation between the two NLs $\Delta \mu_\mathrm{NL}$, and intrinsic NL widths ($W_\mathrm{NL1}$, $W_\mathrm{NL2}$), the latter four all given in \kms. 
Dashed-black lines are prior distributions and vertical red lines are the estimated parameters of the median model, 
i.e. the median values (solid red), and the boundaries enclosing the highest density 95\% confidence intervals (dashed red).
 }
	\label{fig:fig2_add}
\end{figure*} 

\begin{figure*}[t!]
	\centering
	\hspace{-1cm}
		\includegraphics[width=\textwidth, angle=-90]{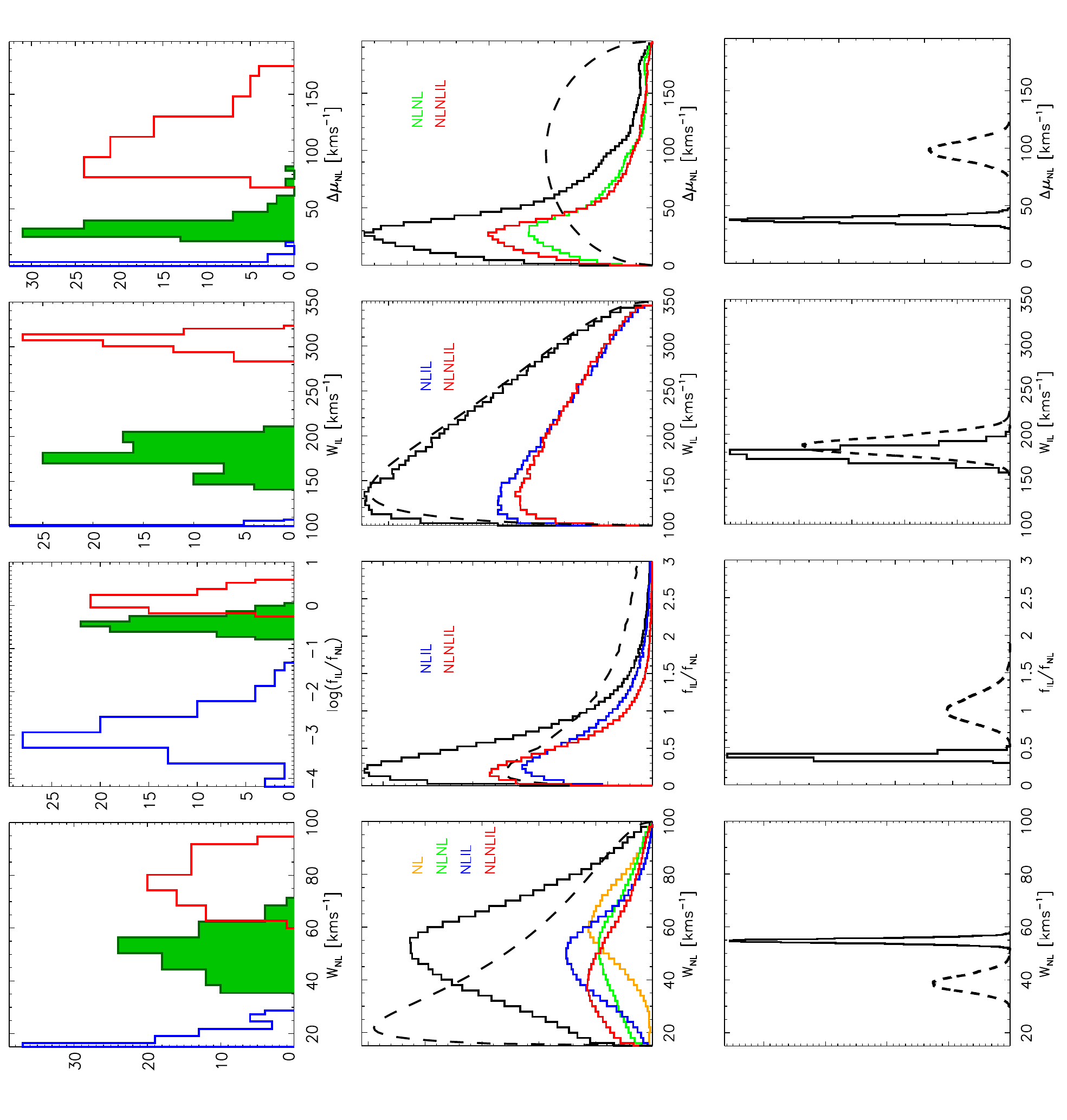}
	\caption{\small Summary of results for narrow-line width ($W_\mathrm{NL}$, left panels), intermediate-to-narrow line flux fraction (\fil/\fnl, centre-left), IL width ($W_\mathrm{IL}$, centre-right) and NL-centroid separation ($\Delta \mu_\mathrm{NL}$, right-most panels). All figures are based on the marginalized posteriors of all 82 bins and $82 \times 4$ models, weighted within individual bins by relative model evidence. {\it Top row:} Distribution of median (green) and highest-density 95\%-confidence interval boundaries (blue, red), quantifying the variation across the filament. For the $f_\mathrm{IL}/f_\mathrm{NL}$, we adopted a log-scale in order to make the low-end confidence interval boundary histogram more visible. {\it Middle:} Sum of all posteriors (solid black), illustrating the average posterior an individual bin and the typical relative contributions of single-NL (orange), NLNL (green), NLIL (blue) and NLNLIL (red) models. {\it Bottom:} Posterior of the median across all bins (solid), in effect using all data to constrain the respective parameter values. These are significantly narrower than the individual bin posterior. The prior is shown by the dashed curve in the middle and bottom panels, whereas the prior of the cross-bin median is given in the bottom panels.} \label{fig:globalpicture}
\end{figure*}
 
\subsection{Evidence for a split in the NL}
\label{sec:res_sNL}

In 18\% of the Voronoi bins (15 of 82) we find significant evidence for an NLNL(IL) model, i.e. for a split in the NL. For one of these bins (bin 5 in Table~\ref{tab:tycho_cv_N}), we show posteriors in Figure~\ref{fig:fig2_add}. In this example, NL centroids and widths are well determined, while the flux of the second (right, high-velocity) NL closely follows the prior distribution. The two NLs are separated by $\approx$40\,\kms\ and have median widths of $\approx$\,52\,\kms\ and 70\,\kms. That is, despite using two Gaussians, both are much wider than the thermal 20\,\kms, indicating that the broadening is mostly not an artifact of unresolved differential line-of-sight motion. Not only is the log evidence ratio of NLNL to NL model larger than 1\,dex in favor of NLNL model, but it is also more than 3 times ($\gtrsim$ 0.5\,dex) more probable than the NLIL or NLNLIL model. This implies highly significant evidence for the presence of a second (split) NL, and in this particular case, also significant evidence {\it against} an IL.

The example in Figure~\ref{fig:fig2_add} is not untypical: {\it even in locations where double-NL models are favored}, the average measured $W_\mathrm{NL}$ is 49\,\kms, again much larger than the upper limit of the intrinsic thermal NL width.
The median parameters and 95\%-confidence intervals for the favored model of each bin are listed in Tables~\ref{tab:tycho_param_E} and \ref{tab:tycho_param_N}. The cross-bin median NL-centroid separation constrained from all the bins in the NE rim is $\Delta \mu_\mathrm{NL} = 38.5 \pm 5.1$\,\kms\ (95\% confidence).

In case of a perfectly spherical shock and a homogeneous ambient ISM, the parts of the filament closest to the upstream are seen edge-on and we would expect to see a single NL, while the parts of the filament closer to the downstream are inclined to the LOS and should exhibit a split in the NL.  
Evidence for a double NL in the eastern filament is found in a few inner bins (blue bins in Figure~\ref{fig:fig_spatial}), but also in some outer (green) bins. Given the small number of double-NL occurrence and its scattered locations, any determination of the shell geometry would be vague. However, detection of double NLs with $W_\mathrm{NL}$ $\gg$ 20\,\kms\ clearly points toward heating and momentum transfer in the CR precursor.\\
\\
\\
\subsection{Intermediate-Line evidence and parameters}
\label{sec:res_IL}

As quantified by the Bayes factors, we find that 34\% of the Voronoi bins (28 bins out of 82) are significantly better explained when a line is added to the fiducial NL-only model. In 74\% of those, the NLIL or NLNLIL model is also preferred over an NLNL model, which means 24\% of the bins overall.  
This is illustrated in Figure~\ref{fig:fig_ev}, where logarithmic (base 10) ratios of all models versus one another are shown in separate panels and each panel shows the spatial variation of the corresponding ratio.
These values are visualized by the histograms, but also tabulated in the Tables~\ref{tab:tycho_cv_E} and \ref{tab:tycho_cv_N}. Apart from just a few bins, there is no clear favor of NLNLIL model over NLIL model (panel 6). The situation changes when it comes to comparison between NLNLIL and NLNL (panel 5), further confirming the necessity of an IL component. One example (bin 17 in Table~\ref{tab:tycho_cv_E}) is illustrated in Figure~\ref{fig:fig_post}, where the logarithmic evidence ratio of the NLIL relative to the NL and NLNL models is larger than 1\,dex. 
This means that an IL in addition to a single NL explains our data more than 10 times better than a simple NL model or NLNL model, irrespective of any particular parameter values.   
Both NLIL and NLNLIL models are equally likely for this bin which implies no need for the second NL component. For this particular bin, we present posteriors for all other models in Figures~\ref{fig:fig_post_nl}, \ref{fig:fig_post_nlnl}, \ref{fig:fig_post_nlnlil}. 

Prominent ILs are seen across the entire filament (panel 3 in Figure~\ref{fig:fig_spatial}), with IL flux fraction of up to 42\%. If we consider only bins that favor NLIL or NLNLIL model, we get an IL flux fraction of 28\% on average, and the intermediate-to-narrow flux fraction $f_\mathrm{IL}/f_\mathrm{NL}$ is estimated at 0.61 with a 95\% confidence interval of (0.01--1.87).

In contrast to the IL flux, $W_\mathrm{IL}$ is not well constrained in individual bins, as indicated by the similarity of prior and average bin-specific posterior (Figure~\ref{fig:globalpicture}, middle panels). 
However, several bins with high $S/N$ and/or strong IL emission (as the bin in Figure~\ref{fig:fig_post}) have good constraints on IL width. 
Moreover, the combination of the data in all bins provides even better information: the cross-bin median is $W_\mathrm{IL} \in [166.77, 194.15]$\,\kms\ (95\% confidence, see bottom of Figure~\ref{fig:globalpicture}). The global 95\%-confidence interval of $f_\mathrm{IL}/f_\mathrm{NL}$ is [0.34, 0.47]; however, we caution that it varies considerably between bins, with median $\log (f_\mathrm{IL}/f_\mathrm{NL}) \in [-0.79, 0.05]$.

\subsection{Line-of-sight velocity}
\label{sec:los}

\begin{figure}[h!]
	\centering
	\hspace{-1cm}
    \includegraphics[width=0.5\textwidth]{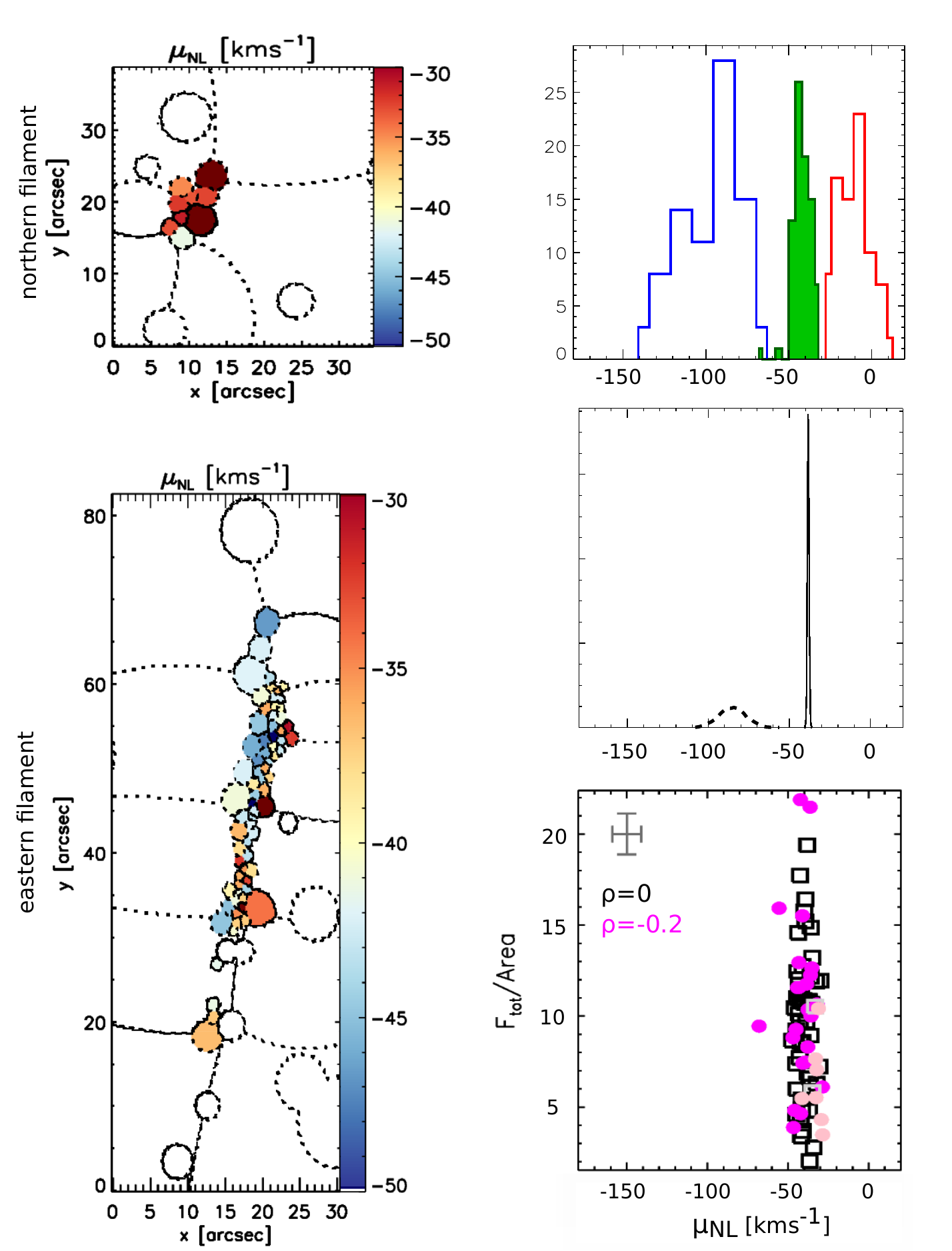}
	\caption{\small NL centroid (line-of-sight velocity) variation in the NE rim (left panels), distribution of median $\mu_\mathrm{NL}$ (green) and highest-density 95\%-confidence interval boundaries (blue, red) in the top-right panel, and posterior of the median across all bins (solid) using all data to constrain $\mu_\mathrm{NL}$ (middle-right panel). The prior is overplotted with the dashed line. The northern (top-left) filament moves systematically with respect to the eastern filament, and the latter exhibits significant internal differential motion. 
	The bottom-right panel shows that NL centroid does not correlate with the surface brightness. Symbols and notation are the same as for Figure~\ref{fig:fig_corr}.} \label{fig:fig_muNL}
\end{figure}

We present the observed line-of-sight (LOS) velocity, i.e. the NL centroid $\mu_\mathrm{NL}$, in Figure~\ref{fig:fig_muNL}. The map of the bin-specific $\mu_\mathrm{NL}$ is shown in the left panels. In the top-right panel, we show the corresponding histogram (green), as well as the distribution of 95\%-confidence interval boundaries. Across the entire NE rim, we find median LOS velocities in the range $[-67.9, -28.6]$\,\kms. Notably, the northern filament moves with respect to the eastern filament; $\mu_\mathrm{NL}=-32.9$\,\kms\ and $-40.5$\,\kms\ on average. The bin-to-bin variations in $\mu_\mathrm{NL}$ are most noticeable in the eastern filament, indicating inhomogeneities in the shock and, in turn, in the ISM density. The bulk (median) LOS velocity of the filament as a whole is $\mu_\mathrm{NL} = -38.3 \pm 1.5$)\,\kms\ at the 95\% confidence level (see middle-right panel). Converting to the local standard of rest (LSR), we obtain $V_\mathrm{LSR} \approx -34$\,\kms. We also check for the correlation between LOS velocities and surface brightness (bottom-right panel), but find them to be uncorrelated.

\subsection{'knot\,g'}
\label{sec:knotg}

This location in the eastern filament has been the target of previous observational \Ha\ studies, but it was not spatially resolved with spectroscopic data. We spatially resolve this 'knot\,g', and the entire NE filament, with our spectroscopic data for the first time. More precisely, the centroids of 6 of our bins drop within the \citealt{lee07} slit. Averaging over these bins, we find \wnl\ = 49 $\pm$ 15\,\kms\ (68\% confidence), which is in agreement with Lee et al.'s reported values of 44$\pm$4\,\kms. The mean LOS velocity is $-39.7$\,\kms\ enclosed by 68\% confidence interval of [$-$50.2, $-$34.7]\,\kms. The mean LSR velocity in 'knot~g' is found to be $\approx -36$\,\kms\ in agreement with previous estimate by Lee et al., but with a blue-shifted offset of nearly 6\,\kms. Typical measured parameters and their 68\% confidence intervals in 'knot~g' are \wil\ = 188 [105, 229]\,\kms, \fil/\fnl = 0.35 [0.05, 0.54], and \dnl\ = 32.4 [8.1, 46.7]\,\kms.

\subsection{Correlation in parameters}
\label{sec:global}

%\clearpage
\begin{figure*}[t!]
	\centering
	\hspace{-1cm}
		\includegraphics[width=0.73\textwidth,angle=-90]{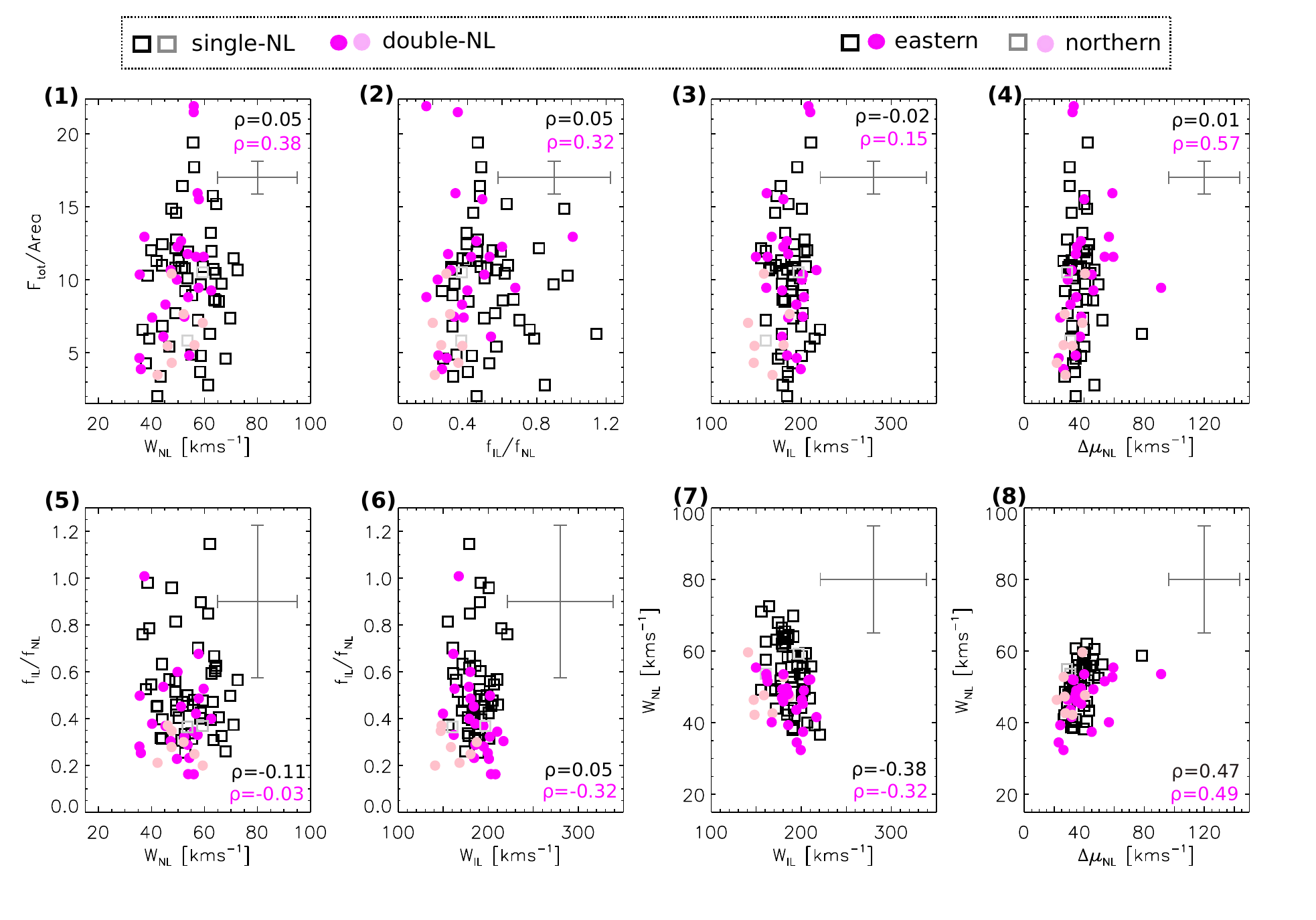}
\vspace{-0.2in}
	\caption{\small Parameter correlation: top row shows median of evidence-weighted $W_\mathrm{NL}$, $f_\mathrm{IL}/f_\mathrm{NL}$, $W_\mathrm{IL}$ and $\Delta \mu_\mathrm{NL}$ posteriors and their dependence on the surface brightness, i.e. total model $S$ flux over bin area (panels 1-4).
	                In the bottom-row panels 5-8 correlation among model parameters is tested. Squares (circles) refer to single (double) NL models in the eastern filament (black/magenta points) and northern filament (gray/pink points) based on the highest CV. Wherever we have NLNL(IL) model as the one with the highest CV, $W_\mathrm{NL}$ is set to the average of the two NLs widths. In panel 8 for bins favoring single-NL models we use the posterior results for their NLNL(IL) models. Labeled are Spearman's rank correlation coefficient $\rho$ that correspond to the color-relevant points, where black label refers to black and gray points together, magenta label to magenta and pink points. We indicate typical error bars on the surface brightness and parameters (average shortest 68\% confidence intervals) in the right corner of every panel.}
	\label{fig:fig_corr}
\end{figure*}

We investigate the correlation between various line parameters and surface brightness, as well as correlation among model parameters, using Spearman's rank correlation coefficient ($\rho$). It is sensitive to any monotonic relationship, even if it is not linear as for Pearson's correlation coefficient, and does not require normally distributed data. The result indicates strong correlation (or anti-correlation) for values close to $\pm$\,1. 

In the top row in Figure~\ref{fig:fig_corr} we plotted median of evidence-weighted $W_\mathrm{NL}$, $f_\mathrm{IL}/f_\mathrm{NL}$, $W_\mathrm{IL}$ and $\Delta \mu_\mathrm{NL}$ posteriors against the surface brightness (SB, total flux of the intrinsic model divided by the bin area). We separately test correlation for bins for which a single-NL model (NL or NLIL, marked with squares) has the highest evidence, and those bins for which a double-NL model (NLNL or NLNLIL, marked with circles) has the highest evidence of all models. We do not find any strong correlation: parameters of single-NL models are uncorrelated with surface brightness, while double-NL models show weak correlations with SB. The highest correlation is with $\Delta \mu_\mathrm{NL}$ ($\rho$=0.57), meaning that with increasing surface brightness, the two NL centroids tend to be more separated. 

In the bottom-row panels of the same Figure we estimated correlation among the parameters. Although we see a hint at positive correlation between NL width and NL separation ($\rho$=0.47 and $\rho$=0.49, see panel 8), we do not find a clear correlation among any of the parameters. 
We get very similar results when we apply Pearson's correlation: its coefficient is either very close to the Spearman's rank coefficient or closer to zero.
In addition, the typical uncertainty of all these points (plotted in the top-right corner of each panel and given as an average shortest 68\% confidence interval) propagates into correlation rank coefficient uncertainty. 

\section{Discussion}
\label{sec:discuss}

Previous \Ha\ observations of Tycho's `knot~g' \citep{lee07} were modeled by \cite{wagner09} computing a series of time-dependent numerical simulations of CR-modified shocks. Assuming a distance of 2.1\,kpc to the remnant, they found the CR diffusion coefficient of $\kappa$ = 2$\times$10$^{24}$\,cm$^2$s$^{-1}$ and the lower limit of the injection parameter $\xi_\mathrm{inj}$ = 4.2$\times$10$^{-3}$ to be in good agreement with the observations, suggesting that CR acceleration in the shock is efficient. Diffuse emission 1\arcsec\ ($\sim$10$^{16}$\,cm) ahead of the eastern filament was also detected by \cite{lee10} and interpreted as emission from the CR precursor with T$\sim$80 000--100 000\,K.

In what follows we will summarize the main results of our paper and look for theoretical explanations for our findings.  

\textit{Presence of a CR precursor}:
The NL width is much broader than 20\,\kms ($\approx$ 55\,\kms\ on average) in the entire NE rim regardless of describing the shock emission with a single or double NL. In other words, even when differential velocities (double NL) are present and accounted for, the NL is still significantly broadened.
This clearly points toward gas heating in a CR precursor \citep{mor3i13}. Furthermore, momentum transfer in a CR precursor might result in a split in the NL if we have two inclined shocks projected on our LOS. This is something that we observe, more precisely we find significant Bayesian evidence for it
in 18\% of the data. Apart from the separation between the two NLs being 38\,\kms\ on average, at the same time we find their intrinsic widths are around 49\,\kms. If we assume that we have contribution of two shocks inclined with the same angle to the LOS, and that neutrals in the CR precursor acquired 10\% of the previously estimated shock velocity of $2500$\,\kms\ \citep{ghava01}, we find that the shock normal inclination of 85$^{\circ}$--86$^{\circ}$ explains the centroid separation of 38\,\kms. Obviously, for the same shock speed and the smaller acquired bulk velocity, we would need the shocks slightly more inclined, i.e. with angles $<$\,85$^{\circ}$. However, since only few bins show evidence for a split in the NL, we do not have enough information to construct the geometry of the shock.

\textit{Presence of a BN precursor}:
The main signature of the BN precursor is an IL \citep{mor2i12}; 24\% of the bins demand an additional line being specifically IL. We find a median value for the intrinsic IL width of 180\,\kms, comprising on average a 41\% intermediate-to-narrow flux ratio. Moreover, the observed high pre-shock neutral fraction of 0.9 \citep{ghava01} in combination with the shock velocity of $\sim2500$\,\kms\ supports the BN presence since a large neutral fraction and the specified shock velocity contribute to efficient charge-exchange and larger number of created broad-neutrals that take essential part in the formation of and heating in the BN precursor. Broad-neutrals are ionized inside the BN precursor almost immediately when they cross the shock. The newly created protons move with a bulk speed larger than the Alfv\'en speed, hence they can trigger the streaming instability (and possibly other kind of instabilities) resulting in an increase of the ion
temperature in the BN precursor. In turn, the CE between pre-shock neutrals and warm ions creates warm neutrals that produce the IL.

\begin{figure}
	%\centering
	\hspace{-0.5cm}
		\includegraphics[width=0.35\textwidth,angle=-90]{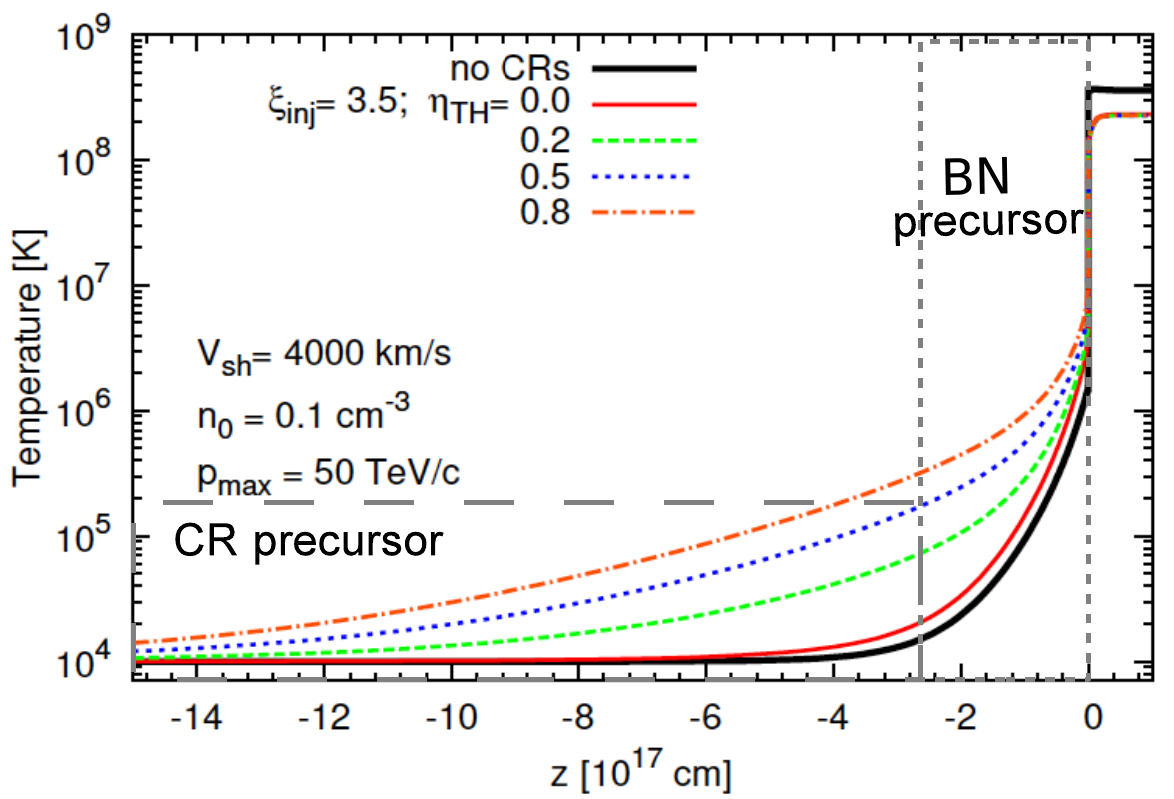}
	\caption{\small Heating and length scales of CR and BN precursor for a specific shock and CR properties \citep{mor3i13}. Zero marks the location of the shock front, while the negative distance from the shock front represents distance in the pre-shock region. As labeled, different lines present the extent and the level of heating in the precursor for different amount of turbulent heating $\eta_\mathrm{TH}$. The black solid line shows the case without CRs and clearly shows the extend of the BN precursor alone. Immediate pre-shock region is affected by both BN and CR precursors where the temperature reaches several 10$^{6}$\,K. CR precursor extends much further from the shock front ($>$\,10$^{17}$\,cm) where the gas is heated to several 10$^{5}$\,K.}
	\label{fig:prec}
\end{figure}

For illustrative purpose in Figure~\ref{fig:prec} we report the temperature profile of the pre-shock gas for specific values of shock parameters as calculated in \cite{mor3i13} where the CR and the BN precursor can be clearly distinguished. BN precursor acts on scales 10$^{16}$--10$^{17}$\,cm in the immediate pre-shock region and can heat the gas to temperature of $\sim$\,10$^{6}$\,K. CR precursor length is much larger, depends of the maximum energy of the accelerated particles and extends over several 10$^{17}$\,cm. The expected gas temperature in the CR precursor is $\sim$\,10$^{5}$\,K.

\textit{Collesionless shock model prediction in partially ionized medium}: 
Following results of \cite{mor3i13} and their Figure 9, 40--50\,\kms\ NL widths require efficient turbulent heating $\eta_\mathrm{TH}$ in the CR precursor, but also accelerated particles with the maximum momentum of $p_\mathrm{max}$ = 40 TeV/c or higher. However, this result is dependent on the shock velocity assumption which is 4000\,\kms\ in the cited paper. Furthermore, Figure 10 in the same paper shows the IL width as a function of the CR acceleration efficiency $\epsilon$ for a fixed shock speed $V_\mathrm{sh} = 4000$\,\kms, $\eta_\mathrm{TH}$ = 0.5 and $p_\mathrm{max}$ = 50 TeV/c and two values of downstream electron-to-proton temperature ratios $\beta_\mathrm{down}$ being either 0.01 or 1. Interestingly, our measured IL width of 180\,\kms\ on average can be explained with the mentioned shock parameters and the acceleration efficiency of 15-25\%, although any further constraints are difficult since we do not know  $\beta_\mathrm{down}$. The latter value, as well as the shock velocity, is something that can be constrained from the observations of the broad \Ha\ component.  

In order to obtain the measured $W_\mathrm{NL} \approx 40$\,\kms, one requires a combination of shock parameters (and CR acceleration included): $p_\mathrm{max}$ = 10 TeV/c, $V_\mathrm{sh}$=2500\,\kms\ and $\beta_\mathrm{down}$ of up to 0.1 \citep{mor3i13}. At the same time this configuration predicts IL width of around 300\,\kms\ and intermediate-to-narrow flux ratio of 1.6 on average for various values of $\eta_\mathrm{TH}$ and $\epsilon$. 
Both these IL parameters are much higher than the \textit{median} values we infer from our data and analysis. We speculate that somewhat higher $\beta_\mathrm{down}$ than 0.1 (see Figure 13 in \citealt{mor2i12}) and a shock velocity of around 3000\,\kms\ would possibly be able to explain the observed $W_\mathrm{IL} \approx 180$\,\kms, $f_\mathrm{IL}/f_\mathrm{NL} \approx 0.41$ and $W_\mathrm{NL} \approx 55$\,\kms. Finally, we also notice that keeping $\eta_\mathrm{TH}$ constant and increasing $\epsilon$ the model of \cite{mor3i13} predicts a simultaneous increase in $W_\mathrm{NL}$ and decrease in $W_\mathrm{IL}$ and $f_\mathrm{IL}/f_\mathrm{NL}$. We find a hint for such an anti-correlation between $W_\mathrm{NL}$ and $W_\mathrm{IL}$ (panel 7 in our Figure~\ref{fig:fig_corr}).

{\it Average and variation of velocities:}
Our result for the median LSR-corrected line-of-sight velocity, $V_\mathrm{LSR} = (-34 \pm 2)$\,\kms\ is in agreement with earlier study by \cite{lee07}, who reported narrow-line \Ha\ LSR velocity $-30.3$\,\kms\ of 'knot g'. Furthermore, \cite{ghava17err} reported a value of $-45.6$\,\kms, that falls within our range [$-$64,\,$-$25]\,\kms\ of observed LSR velocities in the NE rim. In HI 21-cm observations toward Tycho's NE rim \cite{reynoso99} found $V_\mathrm{LSR}$ = $-51.5$\,\kms\ and associated the location of Tycho's SNR and the HI cloud to Perseus arm. Similarly, $^{12}$CO emission was found at $V_\mathrm{LSR} = -62.5$\,\kms\ (\citealt{lkt04}, see also \citealt{zhou16}), and might be associated with SN1572's pre-shock gas. However, as pointed out by \cite{tian11}, there is no clear evidence that either HI cloud or CO cloud are physically associated with Tycho's NE rim. We leave further discussion of the $V_\mathrm{LSR}$ result and its possible interpretation to future work, where we will also investigate shock and CR properties in more detail (see Section~\ref{sec:summary}).

\section{Summary and Conclusions}
\label{sec:summary}

We present \Ha\ spectroscopic observations of Tycho's NE Balmer filaments. This study provides spectroscopic data that for the first time are spatially resolved (spectro-imagery), with large coverage that comprises and resolves the entire NE filament. Our analysis is based on Bayesian inference that enables a quantitative, probabilistic and well-defined model comparison, and a comprehensive, complete characterization of the parameter probabilities. 

We find that the broadening of the NL beyond 20\,\kms\ that was noted in previous studies was not an artifact of the spatial integration, and that it extends across the whole filament, not only the previously covered 'knot\,g'. NL width in the NE rim is typically found around 55\,\kms. Such a large width cannot be due to superposition of multiple lines. In fact, our data analysis allows us to take projection effects into account when interpreting the data. 
We are able to distinguish between single-NL and double-NL models where we find significant evidence for split in NL in 18\% of the Voronoi bins. The widths of the two NLs are around 49\,\kms\ and their centroid separations 38\,\kms\ on average. 

A NL width of 55\,\kms\ implies a temperature of the upstream gas of $\approx$\,68\,000\,K. If this were the temperature of the unperturbed ISM where the SNR is expanding, no neutral hydrogen would exist in the first place, contradicting the presence of \Ha\ emission. Hence our finding is the signature of the existence of a mechanism able to heat the upstream plasma in a region ahead of the shock much smaller than the collisional ionization length-scale. 
As shown by the previous study of \cite{mor3i13}, a CR precursor is the best candidate to explain the widening of the NL, opening the possibility to study particle acceleration at shock using \Ha\ emission. The fact that the NL width ranges from 35\,\kms\ to 72\,\kms\ across the NE rim suggests that the amount of neutrals in the ambient medium varies which imposes different degree of damping of magnetic waves excited by CR streaming.

Likewise, we confirm the suspected presence of an IL, and show it to be widespread (24\% of the bins). Typical IL widths and intermediate-to-narrow flux ratios are 180\,\kms\ and 0.41, respectively.

Our model parameters also comprise the line-of-sight velocity centroids. After correction to the local standard of rest, their median is $V_\mathrm{LSR} = (-34 \pm 2)$\,\kms, in agreement with the \cite{lee07} investigation of 'knot g'. 

Overall, our results reveal an interplay between two precursors in the Tycho's NE rim: broaden NL widths point toward the evidence for the presence of a CR precursor, while detected IL reveals the presence of a BN precursor. 

From the knowledge of the NL width only it is not possible to determine the CR acceleration efficiency, because such width depends on many parameters (shock speed, maximum energy of accelerated particles, electron-ion equilibration and turbulent heating). 
Nevertheless, we can conclude that, assuming a shock speed between 2500 and 3000\,\kms, our result is compatible with having a maximum CR energy $>$\,10\,TeV, a turbulent heating $>$\,10\% and an acceleration efficiency $>$\,few \%. The degeneracy between these parameters could be broken using other information coming from the broad-line width and intensity (giving more precise information on shock speed and electron/ion equilibration) and X-ray/gamma-ray observations (determining the maximum energy of accelerated particles). The difficulties in performing such calculations rely in the fact that the broad line is not known with the same accuracy as the NL and IL and that the gamma-ray emission do not have enough spatial and spectral resolution to fix unambiguously the maximum energy \citep{park15,mor16}. Improvements in this regards will surely come from the Cherenkov Telescope Array. Furthermore, parallel to the study of the \ghafas\ narrow \Ha-line profiles, we have conducted an investigation on the same part of the Tycho's remnant using \osiris\ (Optical System for Imaging and low-intermediate Resolution Spectroscopy) on the GTC (Gran Telescopio Canarias) to observe the broad \Ha-line profiles. In a forthcoming paper we will present results of broad-\Ha\ components of the same spatial locations (bins) along the filaments as presented in this paper, which combined with narrow-\Ha\ components and applied shock models will give a better handle on the overall conditions in the shock, and will enable us to quantify CR properties. 

\section*{Acknowledgements}
\label{sec:acknowledgements}
We would like to thank the anonymous referee for the constructive report and a valuable contribution towards the completeness of this paper.
We would also like to thank Ren\'e Andrae (MPIA, Heidelberg) abd Joonas N\"attil\"a (Tuorla Observatory, University of Turku) for fruitful discussions and helpful suggestions on Bayesian inference.

\clearpage
\newpage
%\twocolumn
%======================================================
\appendix
%\section*{Appendix}
\renewcommand{\thesubsection}{\Alph{subsection}}
\renewcommand{\thefigure}{A.\arabic{figure}}
\setcounter{figure}{0}
%======================================================

\subsection{Data, Flatfield and Background}
\label{DR}

In order to compare our models of Tycho's SNR spectra to the data, we need to account for the variable sensitivity ("flatfield") and the background flux. We eschew the standard method of applying background subtraction and flatfield correction to compensate for both effects in the individual exposures. Instead, we reconstruct the flatfield and background in the final product of the data reduction pipeline, which is a cube (position, wavelength) of co-added, "stacked" individual observations. Notably, the flatfield will also have a wavelength-dependence in addition to begin position-dependent. We eventually include flatfield and background in our models. The advantage of our approach is the preservation of photon statistics and therefore accurate uncertainties of model parameters and evidence.

Before describing the construction of individual as well as the co-added flatfield and background, we formalize the process by which the data are generated and how it is propagated by the data reduction pipeline.

\subsubsection*{Measurement process}

Each datum (measurement) $D_{xy,i}$ with pixel indices $(x,y)$ and exposure index $i$ is the response of the telescope and instrument to the incoming, seeing-convolved flux $d_0=d_0(\alpha,\delta,\lambda)$, which varies with sky coordinates $(\alpha,\delta)$ and wavelength $\lambda$:
\begin{equation}
D_{xy,i} = \int_{-\infty}^\infty d_{0,i}(x,y,\lambda) \, F_{xy} \, R_{xy,i}(\lambda) \, \mathrm{d}\lambda~, \label{eqn:D_xyi}
\end{equation}
where 
\begin{equation}
d_{0,i}(x,y,\lambda) \equiv d_0\left(\alpha=\alpha_i(x,y),\,\delta=\delta_i(x,y),\lambda\right) \notag
\end{equation}
has the astrometric solution, $\alpha=\alpha_i(x,y),\,\delta=\delta_i(x,y)$, and the integral over the area of pixel $(x,y)$ implicitly applied.

$F_{xy}$ is the spatially varying sensitivity of the detector and optical system -- the {\it flatfield image}. As expected, and as verified by us (see below), it is the same for all exposures.

$R_{xy,i}(\lambda)$ is the pixel- and exposure-specific wavelength filter imposed by the Fabry-P\'{e}rot interferometer (etalon). In our case, it is accurately described by a universal {\it line-spread function} (LSF), $r(\lambda)$:
\begin{equation}
R_{xy,i}(\lambda) = r(\Lambda_{xy,i}-\lambda)~, \label{eqn:LSF}
\end{equation}
where the {\it wavelength calibration} $\Lambda_{xy,i}$ returns the central, maximum-throughput wavelength for each pixel and exposure (tuning of the etalon). By definition, the LSF is centered on (peaks at) the origin. In our case it is a Gaussian with dispersion measured from calibration spectra.

Inserting (\ref{eqn:LSF}) in (\ref{eqn:D_xyi}), one sees that the datum $D_{xy,i}$ is the LSF-convolved spectral flux $d_i(x,y,\lambda$), evaluated at $\Lambda_{xy,i}\,$:
\begin{eqnarray}
D_{xy,i} &=& \int_{-\infty}^\infty d_{0,i}(x,y,\lambda) \, F_{xy} \, r(\Lambda_{xy,i}-\lambda) \, \mathrm{d}\lambda \notag \\
&=& F_{xy} \cdot \Big( d_{0,i} \ast_\lambda r \big) (x,y,\lambda=\Lambda_{xy,i} ) \label{eqn:d_LSF} \\
& \equiv & F_{xy} \cdot d_i(x,y,\Lambda_{xy,i})~.\notag
\end{eqnarray}

\subsubsection*{Data processing}

By way of converting the observed images into {\it data sub-cubes}, $D_{xy,i} \rightarrow D_{xyl,i}$, we make the wavelength information contained in them explicit and obtain a format that allows direct co-addition. Each of the 48 "slices" (third index) of a sub-cube corresponds to a wavelength, $\lambda_l$. Each observed $\Lambda_{xy,i}$ is bracketed by two slices, and the corresponding flux $D_{xy,i}$ is assigned to them via linear interpolation:
\begin{eqnarray}
D_{xyl,i} &=& D_{xy,i} \cdot T_{xyl,i} \notag \\
T_{xyl,i} &=& t_l(\Lambda_{xy,i}) \label{eqn:T_xyli} \\ 
& \equiv & \left\lbrace \begin{array}{rcl} 
1-\left| \Lambda_{xy,i} - \lambda_l \right| / \Delta\lambda &:& | \Lambda_{xy,i} - \lambda_l | \leq \Delta\lambda \\
0 &:& | \Lambda_{xy,i} - \lambda_l | > \Delta\lambda \end{array} \right.~, \notag
\end{eqnarray}
where $\Delta\lambda$ is the "size" of each slice, i.e. the distance between the slices' central wavelengths, $\lambda_l$. Another way to describe this assignment is that each slice imposes a triangle filter $t_l(\lambda)=\max\left(0,1-|\lambda-\lambda_l|/\Delta\lambda \right)$. We shall use it again to construct the co-added flatfield and background cubes.

As $D_{xy,i}$ itself is a filtered version of the flux, the sub-cube $D_{xyl,i}$ at a given pixel $(x,y)$ is nearly "empty", except for two slices $l$. Adequate coverage of the spectrum therefore necessitates multiple tunings of the etalon, or imaging the source in varying locations on the detector, since even for unchanged tuning, $\Lambda$ depends on $x$ and $y$. The resulting multiple sub-cubes are then coadded, albeit after projecting and spatially resampling them onto a common astrometrically calibrated frame $(x',y')$:
\begin{eqnarray}
D_{xyl,i} & \stackrel{\alpha,\,\delta}{\longrightarrow} & D_{x'y'l,i} \label{eqn:D_x'y'li} \\
D_{x'y'l} &=& \sum_i D_{x'y'l,i}~. \label{eqn:D_cube}
\end{eqnarray}
$D_{x'y'l}$ is the final, co-added data cube which, apart from spatial binning, directly constrains the SNR shock models. We account for the spatial sensitivity variations and any residual non-constant spectral sampling rate by including them in the models, in form of the co-added flatfield cube, $F_{x'y'l}$, which we derive below along with the background cube, $B_{x'y'l}$.

\subsubsection*{Propagation of flatfield and background}

We represent the incoming flux $d_{0,i}$ as the sum of "source" flux ($s_{0,i}$) that comprises celestial objects, in particular the SNR filament that we are interested in, and background flux ($b_{0,i}$). Because of the linearity of the integral in (\ref{eqn:d_LSF}), the same applies to the LSF-convolved flux $d_i$:
\begin{eqnarray*}
d_{0,i}(x,y,\lambda) &=& s_{0,i}(x,y,\lambda) + b_{0,i}(x,y,\lambda) \\
s_i(x,y,\lambda) &=& \big(s_{0,i} \ast_\lambda r \big) (x,y,\lambda) \\
b_i(x,y,\lambda) &=& \big(b_{0,i} \ast_\lambda r \big) (x,y,\lambda) \\
\stackrel{(\ref{eqn:d_LSF})}{\Longrightarrow} ~ d_i(x,y,\lambda) &=& s_i(x,y,\lambda) + b_i(x,y,\lambda) \\
D_{xy,i} &=& F_{xy}\cdot\left[ s_i(x,y,\Lambda_{xy,i}) + b_i(x,y,\Lambda_{xy,i}) \right] \\
& \equiv & F_{xy} \cdot S_{xy,i} + B_{xy,i} ~.
\end{eqnarray*}
Note that in contrast to the source component of the data, 
\begin{equation}
S_{xy,i} = s_i(x,y,\Lambda_{xy,i})~,
\end{equation}
we have absorbed the flatfield in the definition of the data's background component,
\begin{equation}
B_{xy,i} = F_{xy} \cdot b_i(x,y,\Lambda_{xy,i})~. \label{B_xyi}
\end{equation}
Due to linearity of the wavelength assignment (\ref{eqn:T_xyli}),
\begin{eqnarray*}
D_{xyl,i} &=& T_{xyl,i}\,F_{xy}\,S_{xy,i}+T_{xyl,i}\,B_{xy,i} \\
& \equiv & F_{xyl,i}\,S_{xy,i} + B_{xyl,i}~.
\end{eqnarray*}
Here we have defined the {\it background subcube}, $B_{xyl,i}=T_{xyl,i} \, B_{xy,i}$ and the {\it flatfield sub-cube} $F_{xyl,i}=T_{xyl,i} F_{xy}$, which are the background component of the data and the flatfield image (the response to a flat spectrum), respectively transformed to a sub-cube using the same slice-assignment (\ref{eqn:T_xyli}) that was used for the data itself. Also, as for the data, background and flatfield sub-cubes are projected onto the same frame and co-added:
\begin{eqnarray*}
F_{xyl,i} & \stackrel{\alpha,\,\delta}{\rightarrow} & F_{x'y'l,i} \\
S_{xy,i} & \stackrel{\alpha,\,\delta}{\rightarrow} & S_{x'y',i} \\
B_{xyl,i} & \stackrel{\alpha,\,\delta}{\rightarrow} & B_{x'y'l,i} \\
D_{x'y'l} &=& \sum_i \left( F_{x'y'l,i}\,S_{x'y',i} + B_{x'y'l,i} \right)
\end{eqnarray*}
$S_{x'y',i}=s_i(x',y',\Lambda_{x'y',i})$ is $s_i(x,y,\Lambda_{xy,i})$ with $x=x_i(x'),\,y=y_i(y')$ given by the astrometric solution. The same transformation applies to $b$. Since the co-added frame's $(x',y')$ corresponds to a unique sky position and we assume the filament emission (but not necessarily the background!) to be constant between exposures, $s_i(x',y',\lambda)=s(x',y',\lambda)$. Then,
\begin{align}
D_{x'y'l} &= \sum_i F_{x'y'l,i}\,s(x',y',\Lambda_{x'y',i}) + F_{x'y'l,i}\,b_i(x',y',\Lambda_{x'y',i}) \notag \\
& \approx \left(\sum_i F_{x'y'l,i}\right) \, s(x',y',\lambda_l) + \sum_i F_{x'y'l,i}\,b_i(x',y',\lambda_l)~ \label{eqn:D_x'y'l} \\
& \equiv F_{x'y'l} \cdot s(x',y',\lambda_l) + B_{x'y'l}~. \notag
\end{align}
In the second line, we have used that $F_{x'y'l,i}$ is non-zero only for $l$ nearest to the sampled wavelength, $\lambda_l \approx \Lambda_{x'y',i}$. Hence, $F_{x'y'l,i} \cdot s(x',y',\Lambda_{x'y',i}) \approx F_{x'y'l,i} \cdot s(x',y',\lambda_l)$ for all $l=1,\dots,48$, and analogous for $b$. However, the background does not factor out of the co-addition, as it may change between exposures. In the last step, we have defined the flatfield and background cubes $F_{x'y'l}$ and $B_{x'y'l}$ as the sum of the flatfield and background sub-cubes.

\subsubsection*{Background and flatfield model}

In order to isolate the source flux component of the data, we still need to model the flatfield and background. 

We begin by investigating the background on the coadded data (\ref{eqn:D_cube},\ref{eqn:D_x'y'l}). Using \sex\, we mask sources (stars, galaxies, and the filament itself) on the wavelength-integrated frame $D_{x'y'}=\sum_{l=1}^{48} D_{x'y'l}$. This frame is the deepest, highest signal-to-noise ratio ($S/N$) data product we have available, hence the mask is as complete as possible. We obtain the set of unmasked pixels, $(x'_u,y'_u)$ which we can use to measure the background without "contamination" by celestial sources:
\begin{equation*}
D_{x'_u,y'_u,l} = \sum_i B_{x_u'y_u'l,i} = B_{x'_u y'_u l}~.
\end{equation*}
Next, in order to ameliorate potential undetected low surface-brightness source flux incursion, to measure variability of $B$, and to average over spectral sensitivity variations in $D_{x'y'l}$, we select 32 boxes of $(25\,\mathrm{pix})^2 \approx$\,(5\arcsec)$^2$ in particularly object-poor locations. For each box $\mathcal{B}_j = \{(x'_u,y'_u)_j\}$, we measure the spatially integrated spectrum:
\begin{eqnarray*}
B_l^{(j)} &=& \sum_{(x'_u,y'_u)\,\in\,\mathcal{B}_j} B_{x'_u y'_u l} \\
& \stackrel{(\ref{eqn:D_x'y'l})}{=} & \sum_{(x'_u,y'_u)\,\in\,\mathcal{B}_j} \sum_i F_{x'_u y'_u l,i}\,b_i(x'_u,y'_u,\lambda_l)~.
\end{eqnarray*}

\begin{figure}
		\includegraphics[width=0.9\linewidth]{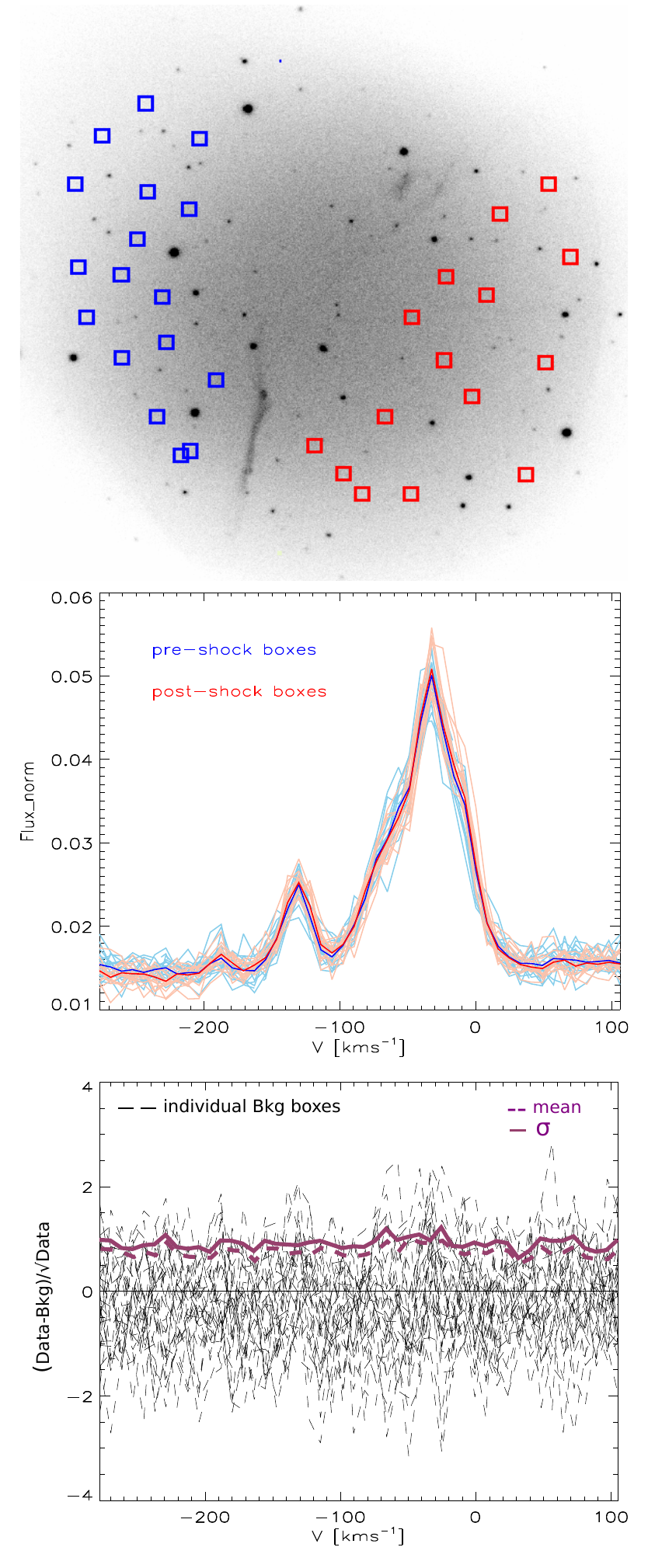}
	\vspace{-0.2cm}	
	\caption{\small The {\it top panel} shows the wavelength-integrated data cube $D_{x'y'}$. The effect of the flatfield is clearly visible as a radial surface-brightness pattern of the background flux in addition to inhomogeneous coverage (empty corners). Overlaid are $(25\,\mathrm{pix})^2 \approx$\,(5\arcsec)$^2$ boxes in the pre-shock/post-shock (blue/red) region, which we used to estimate the incident background spectrum $b(\lambda)$ and, in turn, the flatfield. The {\it middle panel} presents the normalized spectra of the boxes (light blue/red lines), while thick blue and red lines are the mean profiles of the pre- and post-shock boxes. The shapes of the background spectra are similar across the entire FOV.
	The {\it bottom panel} indicates differences between the actual background ("Data") and the background model ("Bkg"), normalized by the Poisson noise. Dashed-black lines represent the differences in the 32 background boxes separately, dashed-purple line is the mean absolute difference, and solid-purple line is the standard deviation between the boxes.} 
	\label{fig:boxes}
\end{figure}

It turns out that, apart from flux normalization, {\it all $B_l^{(j)}$ are nearly the same}, with differences at the percent level (see Figure \ref{fig:boxes}). This allows us to conclude that

\begin{enumerate}[label=\roman{enumi})]
\item the spectral part of the co-added flatfield is spatially invariant, as desired and expected after co-adding $\sim 2500$ exposures with different pointing, orientation and wavelength tuning plus averaging over a $\sim$\,5\arcsec\ box 
\item the spectral shape of the background component is spatially invariant:
\begin{equation}
b_i(x',y',\lambda) = a_i(x',y') \cdot b(\lambda)~,
\end{equation}
where $a_i$ is the amplitude of the background.
\end{enumerate}
Proof comes by contradiction: {\it if} the spectral shape of the 5\arcsec\ box-averaged flatfield or background varied with location, $B_l^{(j)}$ would vary. Notably, we also see no systematic change of the background between the pre- and post-shock regions. Variations of the background flux on smaller ($<$\,5\arcsec) spatial scales are physically unlikely and in any case must be small as the observed variability can entirely be accounted for by Poisson (measurement) noise (middle and bottom panels in Figure~\ref{fig:boxes}). In the bottom panel of Figure~\ref{fig:boxes} we present differences between the actual background and the modeled background. There is no systematic effect, and the spatial variation is within the measurement uncertainty, which in turn is much smaller than the signal. The variation between background boxes, as measured by the 32-element sample standard deviation normalized by the photon noise, is $0.95$ on average across the spectrum (solid-purple line), with maximum value of $1.22$. The average absolute deviation (dashed-purple line) is 0.77 on average, close to the theoretical value of 0.8 expected for the absolute value of a standard-normally distributed variable. Therefore, any differences between the actual background (``Data'') and model (``Bkg'') are fully explained by measurement uncertainties.

We can therefore measure $b(\lambda)$ on the co-added data cube. By combining all 32 background boxes, we additionally minimize noise and residual systematics:
\begin{equation*}
b(\lambda) = \sum_j B_l^{(j)} \equiv \tilde{b}(\lambda)~.
\end{equation*}
The tilde indicates that $\tilde{b}$ is a model of the background spectrum. Normalization of $b(\lambda)$ is absorbed in the $a_i$. We now {\it assume} that $a_i$ is spatially invariant, i.e. that the background amplitude changes only between exposures but not across the field-of-view:
\begin{equation*}
a_i(x',y') = a_i
\end{equation*}
Therefore, also $a_i(x,y)=a_i$ is spatially constant, and we have
\begin{equation*}
D_{x_u y_u,i} = B_{x_u y_u,i} = F_{x_u y_u} \cdot a_i\,\tilde{b}(\Lambda_{x_u y_u,i})~.
\end{equation*}
Here, $(x_u,y_u)$ are the non-masked pixels of the co-added frame, reprojected onto the individual exposures. We do not know the background amplitudes $a_i$ yet, but can already use the knowledge of $b(\lambda)$ to model $F_{xy}$:
\begin{equation}
\left( \sum_i \frac{D_{\tilde{x}_u \tilde{y}_u,i}}{\tilde{b}(\Lambda_{\tilde{x}_u \tilde{y}_u, i})} \right) ~\stackrel{\text{fit+norm.}}{\longrightarrow}~ \tilde{F}_{xy}~.\label{eqn:Fmodel}
\end{equation}
This way, we "divide out" the non-constant background spectrum, which is imprinted on the measurement via $\Lambda_{xy,i}$. The sum on the left side is only carried out for pixels $(\tilde{x}_u,\tilde{y}_u)$ that are not masked in any of the exposures. The result of the sum is a non-normalized flatfield image, which is then modeled by a fourth-order polynomial. The fit eliminates pixel noise and interpolates over masked pixels. It is followed by normalization, such that $\sum_{x,y}\tilde{F}_{xy} = N_\mathrm{pix}$.

Now we justify that the background amplitude is spatially constant, and even that the flatfield is indeed constant as normally expected. We cannot prove this for individual exposures, as the fluxes are impractically small. However, we can restrict the sum in (\ref{eqn:Fmodel}) to different subsets of exposures. This way, we derived $\tilde{F}_{xy}$ for exposures of only one specific "channel" (tuning), and for subsets of observations at different times. The results are invariant within a few percent: random variations of 3.2\% (mean pixel standard deviation) between flatfields of different observing runs, and 1.5\% among channel-specific flatfields. Again, the conclusion is by contradiction: given that exposures vary in pointing, field rotation, and tuning, a non-constant flatfield or a spatially variable background amplitude would lead to systematic variations of $\tilde{F}$, but we do not observe such variations.

With $\tilde{F}_{xy}$ at hand, we measure $a_i$ as the flux scaling required to match the data,
\begin{equation*}
a_i = \frac{\sum_{x_u,y_u} D_{x_u y_u,i}}{\sum_{x_u,y_u} \tilde{F}_{xy} \cdot \tilde{b}(\Lambda_{x_u y_u,i})}~,
\end{equation*}
and reconstruct the background in each exposure:
\begin{equation}
\tilde{B}_{xy,i} = \tilde{F}_{xy} \cdot a_i\,\tilde{b}(\Lambda_{xy,i})~.
\end{equation}
We then transform the flatfield and background models $\tilde{F}_{xy}$ and $\tilde{B}_{xy,i}$ to sub-cubes in the common astrometric frame and co-add them, in the same way the data (\ref{eqn:T_xyli}, \ref{eqn:D_x'y'li},\ref{eqn:D_cube}). As for the "real" flatfield and background, (\ref{eqn:D_x'y'l}) ensures that the resulting coadded cubes correspond to the actual flatfield and background components in the data cube.

\begin{figure}
	\centering
		\includegraphics[width=\linewidth]{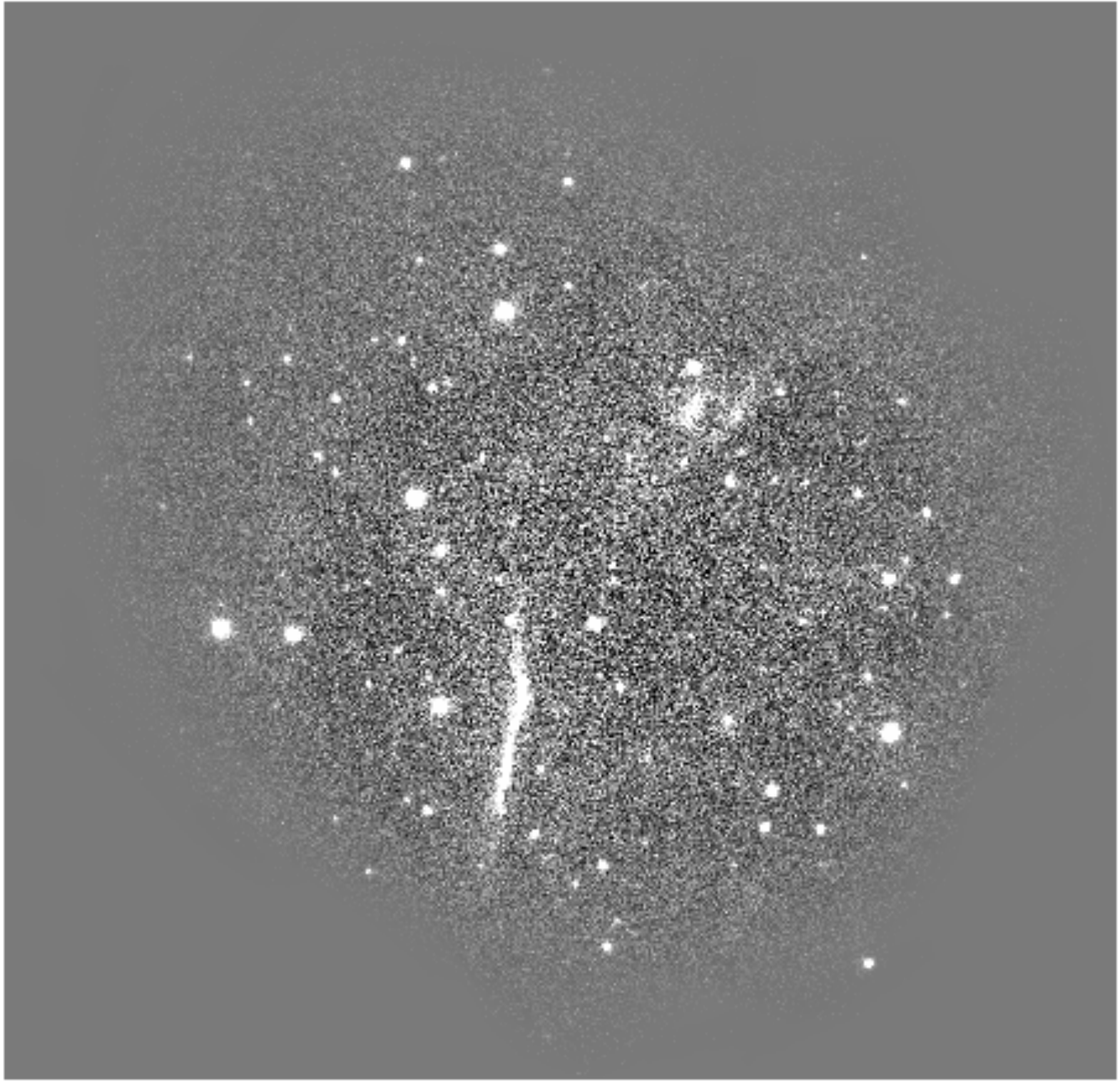}
\caption{\small Wavelength-integrated background residuals $(D-B)_{x'y'}$. The grey scale is linear from -10 to +10 counts/pixel. The smooth gray regions have low signal or are even zero due low effective exposure time and hence lower flux (the flatfield was not divided out). The sources (white) were masked in the flatfield and background construction procedure and additionally avoided by the background-probing boxes. This applies in particular also to the apparent faint but extended brightness around the eastern (left) and northern (top) filaments. }
\label{fig:D-B}
\end{figure}

We use $\tilde{B}_{x'y'l}$ and $\tilde{F}_{x'y'l}$ to check once more whether our assumption of a uniform background and invariant flatfield are fulfilled: the wavelength-integrated $(B/F)_{x'y'}=\sum_l B_{x'y'l}/F_{x'y'l}$ is indeed flat, and background residuals $(D-B)_{x'y'}$ are zero apart from Poisson noise and objects (see Figure~\ref{fig:D-B}). 

%PIP 
\subsection{Photo-ionization precursor (PIP)}
\label{pip}

\begin{figure*}
	\centering
	\vspace{-0.8cm}
		\includegraphics[width=0.7\linewidth,angle=-90]{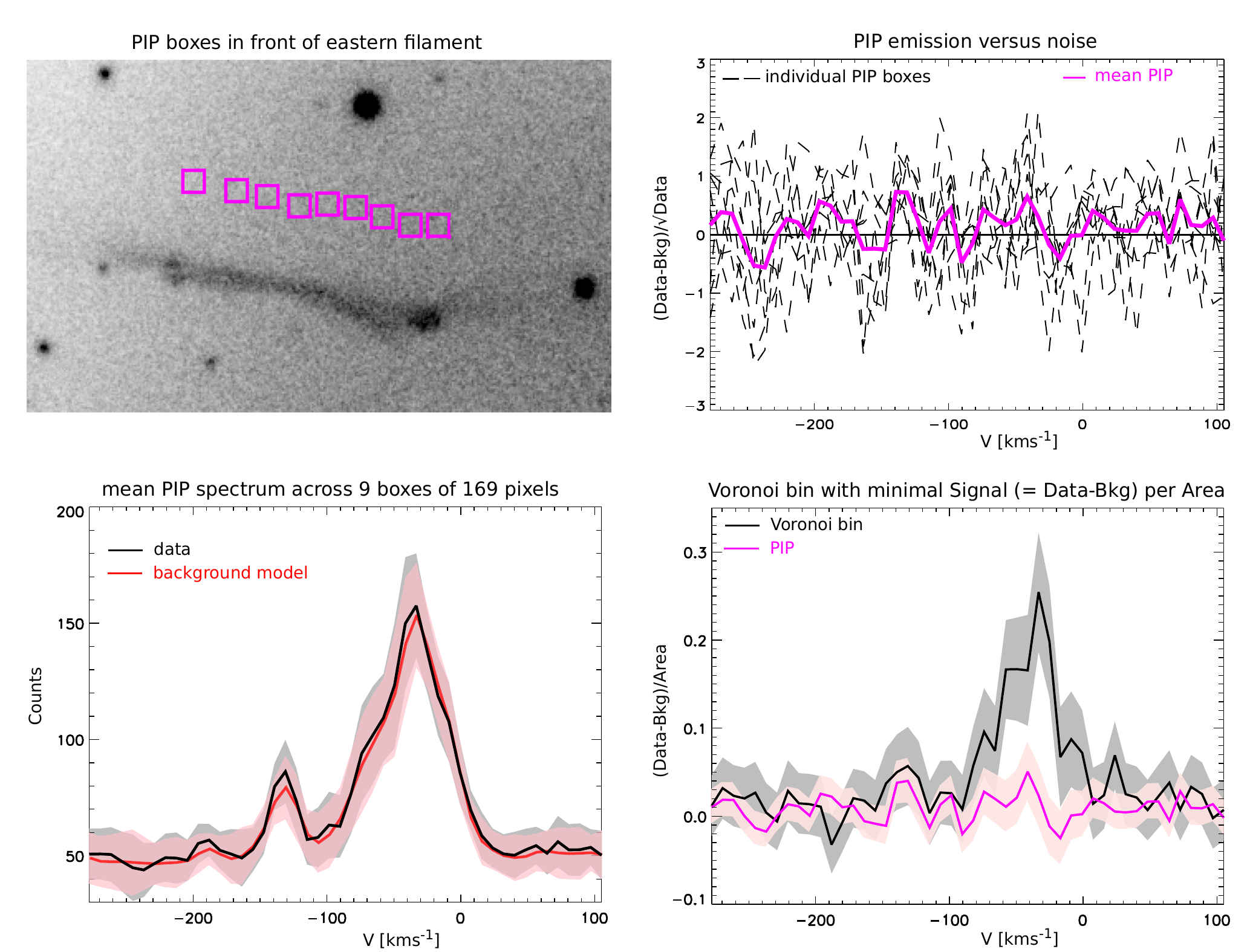}
\caption{\small Magenta boxes in the top-left panel indicate the region where we extracted spectra in search for a potential PIP signal. The difference between data and background model in individual pre-shock boxes (black-dashed) and mean PIP level (solid-magenta), divided by the measurement errors, shows that residual emission is consistent with Poisson noise. The mean flux (solid-black) and the background model (solid-red) in the regions of putative PIP emission are directly compared in bottom-left panel, with their $1-\sigma$ uncertainties are indicated with the corresponding color-shaded regions. Putative PIP flux and the background are entirely consistent with one another. In bottom-right panel, we compare filament emission (black line and grey-shaded region) in the bin with smallest surface brightness to the PIP-region signal (magenta line and pink-shaded region), again illustrating that any possible PIP signal is negligible.    
} 
\label{fig:pip}
\end{figure*}

When gas starts to cool and recombine downstream, the produced photons escape to the pre-shock region and form a PIP \citep{ray79}.
Although non-radiative shocks lack recombination zones, PIP can still be created, where the main sources of the photons produced downstream are HeI$\lambda$584\,\AA\ and HeII$\lambda$304\,\AA. \cite{ghava00} reported on diffuse \Ha, but also [NII] and [SII] emission extending over 1\arcmin\ in front of the Tycho's NE rim, and suggested that it arises in a PIP. They predicted that the pre-shock gas was heated in the PIP to $\sim$\,12\,000\,K. Subsequently, \cite{lee07} measured PIP spectrum in front of 'knot\,g': narrow \Ha\ with \wnl $\approx$\,34\,\kms, and [NII] $\lambda$6583\,\AA\ with the width of 23\,\kms.\\ 
\indent
Similarity of the spectra in the pre- and post-shock background boxes (middle panel in Figure~\ref{fig:boxes}), where pre-shock boxes partially cover the region of the suspected PIP, unambiguously shows that the PIP emission in our data is negligible. Furthermore, since the signal of a PIP increases towards the shock front, we searched for its signature in 9 pre-shock, $13^2$-pixel regions (magenta boxes in Figure~\ref{fig:pip}) that were taken to be closer to the filament than the background boxes ($\approx$\,7\arcsec away from the filament), but still far enough so that we do not pick up on projected filament emission and emission in the CR precursor. The top-right panel shows the comparison of the putative PIP signal and background model, normalized by measurement noise. The mean normalized PIP level is indicated by the solid-magenta line, and is 0.13 on average. Therefore, the possible PIP signal is consistent with zero. We also show the comparison between the mean PIP and background flux, including their 1-sigma uncertainties (bottom-left panel). Finally, in the bottom-right panel we demonstrate that PIP signal is negligible in comparison to the filament flux. We chose here the location (bin) with the smallest Signal per area; the signal is larger in all other bins, and the putative PIP even smaller in relation to the signal. We therefore conclude that there is none or negligible PIP contribution to our filament flux models.

%Voronoi binning
\subsection{Spatial binning of the data}
\label{voronoi}

The Voronoi binning \citep{cc03} was performed setting two criteria: the targeted S/N and the minimum bin size. The minimum bin 
size has to be set in order to account for the seeing of $\simeq$ 1\arcsec, which for the spatial scale of 0.2\arcsec/pixel gives the minimal bin size of 
$\simeq$ 19 pixels. We set the targeted S/N=10. The code of \cite{cc03} in its standard version tends to create elongated-shaped bins in the direction perpendicular to the shock filament. The HST image of Tycho's SNR seen in Figures~1 \& 2 in \cite{lee10} shows complex filamentary structure including several very bright knots as a result of different shock emission projected along the LOS. Small-scale differential gradients are very notable in the direction of the shock normal, the reason why we used the Weighted Voronoi Tessellation adaptation of the Cappellari \& Copin (2003) \citep{ds06} that created a couple of rounder bins in the same direction instead. This way we created 85 spatial Voronoi bins in the eastern and 15 in the northern part of the rim (Figure~\ref{fig:fig_spatial}). Although we see only 2\%-level variations between boxes of 25$^2$ pixel size, these residual background variations become important relative to the S/N for bin sizes of 400 pixels and larger. This yields the upper size limit of 400 pixels for the bins used in our analysis. 73 bins (out of 85) in the eastern and 9 bins (out of 15) in the northern filament fulfill the above criteria and are further considered while the bins with more than 400 pixels are excluded from the analysis. 

%parameters and prior
\subsection{Model parameters and prior PDFs}
\label{param}

\renewcommand{\thetable}{S\arabic{table}}
\setcounter{table}{0}
\begin{table*}
\begin{minipage}{\textwidth}\centering
\caption{\small Model parameters and their prior PDFs. All parameters apart from ln$(F_\mathrm{tot})$ are defined in (0,1) range. }
\begin{tabular}{ l | l |  l} 
 \hline \hline
 parameters & meaning & prior \\ 
 \hline
 ln$(F_\mathrm{tot})$ & natural log-based total flux & flat prior \\\cline{3-3} 
 $f_i, i=\mathrm{[c,NL,NL1,NL2,IL]}$ & flux fractions & Dirichlet prior: $\prod_i f^{\alpha - 1}_i$, $\alpha = 1.5$; $\sum_i f_i = 1, f_i \in (0,1)$  \\\cline{3-3} 
 $\mathrm { \mu^{\prime}_{NL} ,  \langle \mu^{\prime}_{NL} \rangle }$ & NL centroid, NL centroid mean & \multirow{4}{*}{Beta prior: $x^{\alpha - 1}(1-x)^{\beta - 1}$,  $\alpha = \beta = 1.5$; $x$ $\in$ (0, 1)}\\ 
 $\Delta \mu^{\prime}_\mathrm{NL}$ &  separation between the two NLs & \\
 $\Delta \mu^{\prime}_\mathrm{IL}$ &  IL centroid offset from NL centroid (mean)&  \\ 
 $w^{\prime}_\mathrm{NL(IL)}$ &  NL (IL) natural log-width & \\
 \hline \hline
 model & \multicolumn{2}{|l}{model parameters}\\
 \hline 
 NL & \multicolumn{2}{|l}{ln$(F_\mathrm{tot}),~f_\mathrm{NL}, ~f_\mathrm{c}, ~\mu^{\prime}_\mathrm{NL}, ~w^{\prime}_{NL} $} \\
 NLNL & \multicolumn{2}{|l}{ln$(F_\mathrm{tot}),~ f_\mathrm{NL1}, ~f_\mathrm{NL2}, ~f_\mathrm{c}, ~\langle \mu^{\prime}_{NL} \rangle,  ~\Delta \mu^{\prime}_\mathrm{NL},~ w^{\prime}_\mathrm{NL1},  ~w^{\prime}_\mathrm{NL2}$}\\
 NLIL & \multicolumn{2}{|l}{ln$(F_\mathrm{tot}),~ f_\mathrm{NL}, ~f_\mathrm{IL}, ~f_\mathrm{c},  ~\mu^{\prime}_\mathrm{NL} , ~ \Delta \mu^{\prime}_\mathrm{IL}, ~w^{\prime}_\mathrm{NL},  ~w^{\prime}_\mathrm{IL}$}\\
 NLNLIL & \multicolumn{2}{|l}{ln$(F_\mathrm{tot}),~ f_\mathrm{NL1}, ~f_\mathrm{NL2}, ~f_\mathrm{IL}, ~f_\mathrm{c}, ~\langle \mu^{\prime}_{NL} \rangle,  ~\Delta \mu^{\prime}_\mathrm{NL},  ~\Delta \mu^{\prime}_\mathrm{IL}, ~w^{\prime}_\mathrm{NL1},  ~w^{\prime}_\mathrm{NL2}, ~w^{\prime}_\mathrm{IL}$}\\
 \hline
\end{tabular}
\label{tab:param}
\end{minipage}
\end{table*}

As already mentioned in Section~\ref{models}, we define the total flux, continuum and line flux fractions, line centroids and widths as models parameters. In case of the NLNL and NLNLIL models, 
we define the NL centroid mean $\langle \mu_\mathrm{NL} \rangle$ and the separation between the two NLs $\Delta \mu_\mathrm{NL}$ as parameters. IL centroid is introduced with its offset from the NL centroid (mean) $\Delta \mu_\mathrm{IL}$.

The parameters that we actually sample from are slightly different from the parameters that we use in Section~\ref{sec:analysis}; the difference enables direct application of the prior PDFs (Dirichlet or Beta distributions) on the parameters. 
These prior distributions require parameter sets defined in the range (0, 1), except for the naturally based logarithm of the total flux ln$(F_\mathrm{tot})$ for which we use flat unbound prior.
The continuum and line flux fractions are by definition in (0, 1) range, where we set the continuum flux fraction to be dependent on flux fractions in the lines $f_\mathrm{c}=1-\sum f_\mathrm{i}$. 
The fact that flux fractions sum up to 1 and that they are in the range (0, 1) makes the Dirichlet distribution $ \prod_\mathrm{i} f^\mathrm{\alpha_i - 1}_i,~i=[NL,NL1,NL2,IL,c] $ perfect choice for their prior PDF. 
Since we do not favor any of the flux components, we use symmetric Dirichlet distribution with the same index $\alpha$ which we set to $\alpha = 1.5$ to disfavor zero fluxes. 

The line parameters $\mu_\mathrm{NL}$, $\langle \mu_\mathrm{NL} \rangle$, $\Delta \mu_\mathrm{NL}$, $\Delta \mu_\mathrm{IL}$ are all defined with the following functional form: 
$x^{\prime} = (x-x_\mathrm{min})/(x_\mathrm{max}-x_\mathrm{min})$. $\mu_\mathrm{NL}$ or $\langle \mu_\mathrm{NL} \rangle$ are defined in the range [$ V_\mathrm{cen} - V_\mathrm{FSR}/4$, $ V_\mathrm{cen} + V_\mathrm{FSR}/4$], where $V_\mathrm{cen}$ is the center of the free velocity range 
$ V_\mathrm{FSR}$ (FSR in velocity units) and $\Delta \mu_\mathrm{NL}$ in the range [0, $ V_\mathrm{FSR}/2$] so that in the most extreme case the two NL centroids are at the edges of the spectral coverage. $\Delta \mu_\mathrm{IL}$ is within $\pm V_\mathrm{FSR}/4$ having the absolute upper boundary set to the upper (lower) boundary of the NL (IL) width $\approx$100\,\kms. 
Instead of line widths $W_\mathrm{NL}$ and $W_\mathrm{IL}$ defined in the range [15, 100]\,\kms\ and [100, 350]\,\kms\ respectively \citep{mor2i12,mor3i13}, we used their log-widths (natural logarithm) denoted as $w_\mathrm{NL(IL)} = \mathrm{ln}W_\mathrm{NL(IL)}$ in the same functional form as for the centroids and separations.
For all these parameters we define a symmetric Beta distribution prior $x^{\alpha - 1}(1-x)^{\beta - 1}$ with the arguments $\alpha=\beta=1.5$ - slightly favoring the central values of the defined parameters ranges. 
Model parameters and their priors are summarized in the Table~\ref{tab:param}.

\begin{figure*}[t!]
	\centering
		\includegraphics[width=0.5\textwidth,height=\textwidth,angle=-90]{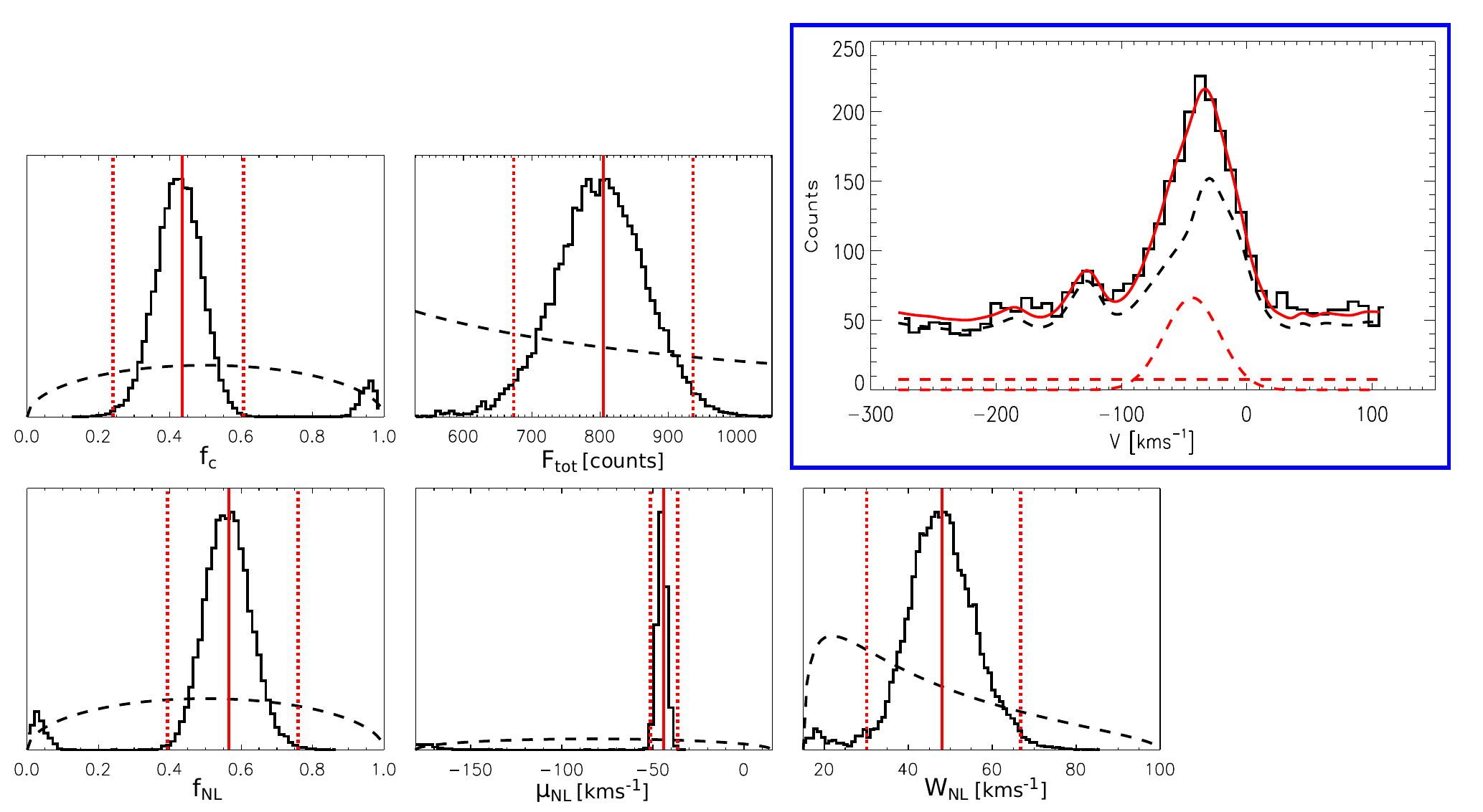}
	\caption{\small Parameter estimation of an NL model via Bayesian inference for the bin in the NE filament of Tycho's SNR for which we presented posterior of NLIL model parameters in Figure~\ref{fig:fig_post} (see its caption for explanation). 
 }
	\label{fig:fig_post_nl}
\end{figure*}

\begin{figure*}[t!]
	\centering
		\includegraphics[width=0.9\textwidth,height=\textwidth,angle=-90]{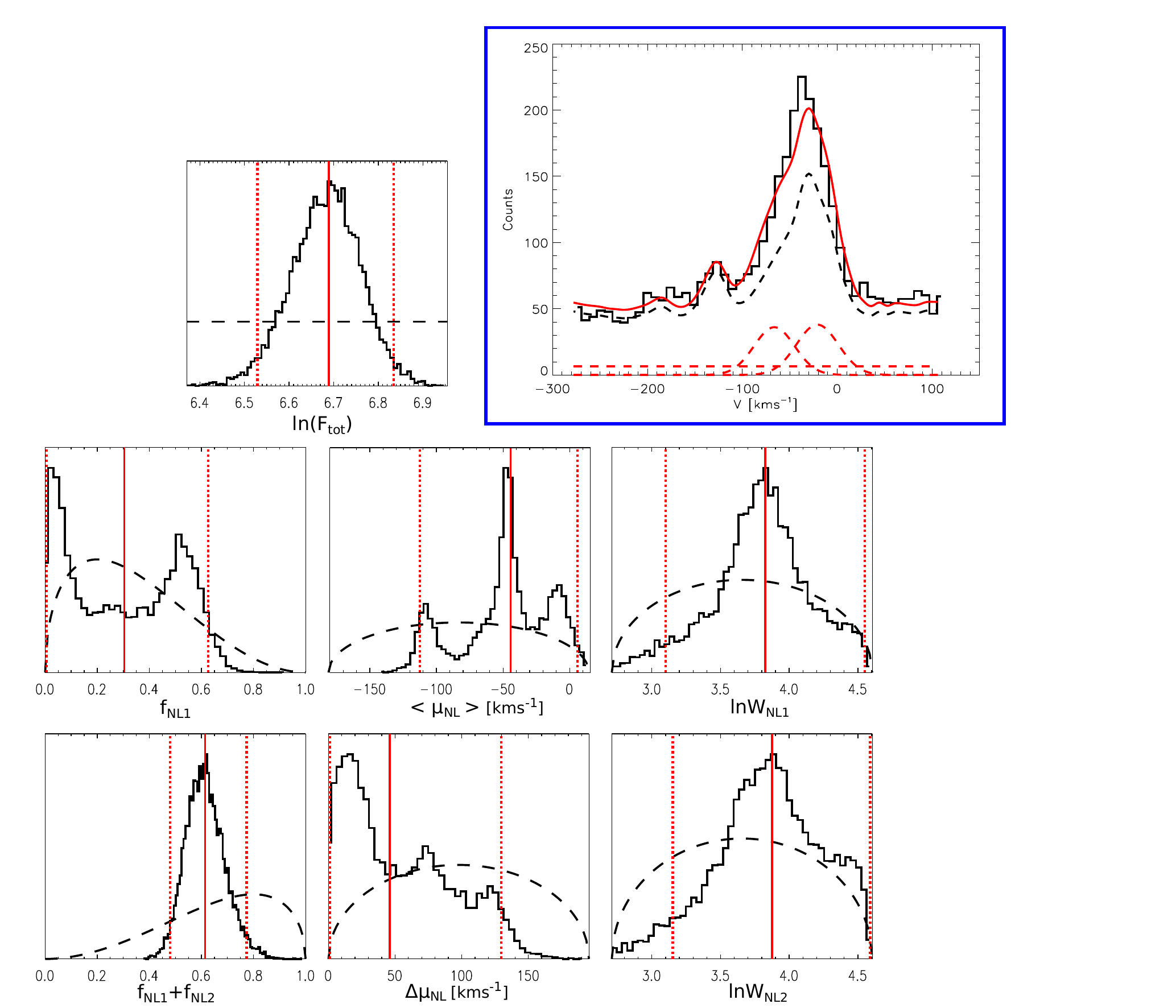}
	\caption{\small Parameter estimation of an NLNL model via Bayesian inference for the bin in the NE filament of Tycho's SNR for which we presented posterior of NLIL model parameters in Figure~\ref{fig:fig_post} (see its caption for explanation). 
                 Instead of $F_\mathrm{tot}$, $W_\mathrm{NL1}$, $W_\mathrm{NL2}$, $f_\mathrm{NL1}$, $f_\mathrm{NL2}$, we present log (natural logarithm) of total intrinsic flux and intrinsic line widths (ln$(F_\mathrm{tot})$, ln$W_\mathrm{NL1}$, ln$W_\mathrm{NL2}$), and cumulative flux fractions ($f_\mathrm{NL1}$, $f_\mathrm{NL1}$+$f_\mathrm{NL2}$). 
 }
	\label{fig:fig_post_nlnl}
\end{figure*}

\begin{figure*}[t!]
	\centering
		\includegraphics[width=\textwidth,height=1.1\textwidth]{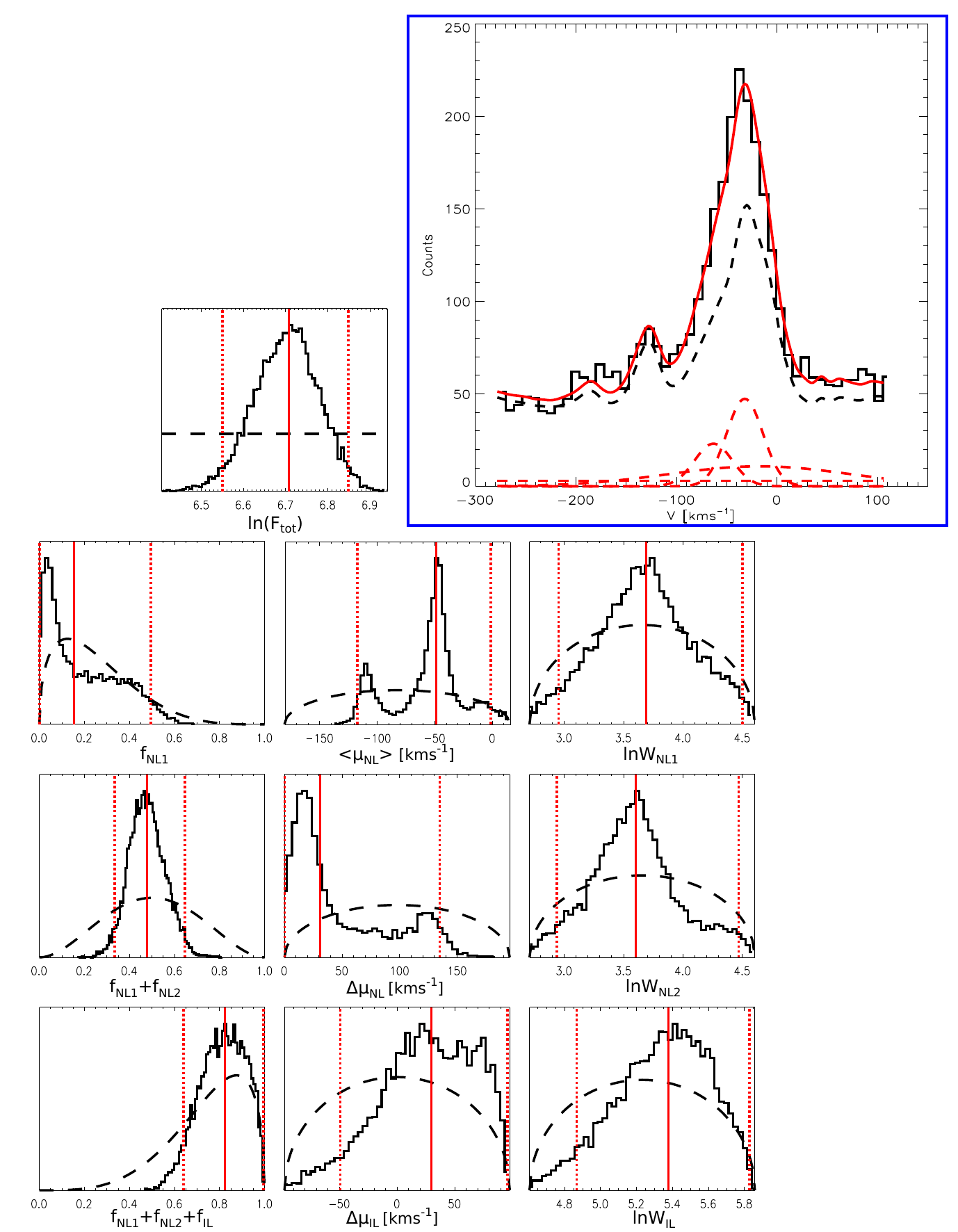}
	\caption{\small Parameter estimation of an NLNLIL model via Bayesian inference for the bin in the NE filament of Tycho's SNR for which we presented posterior of NLIL model parameters in Figure~\ref{fig:fig_post} (see its caption for explanation). 
                 Instead of $F_\mathrm{tot}$, $W_\mathrm{NL1}$, $W_\mathrm{NL2}$, $W_\mathrm{IL}$, $f_\mathrm{NL1}$, $f_\mathrm{NL2}$, $f_\mathrm{IL}$, we present log (natural logarithm) of total intrinsic flux and intrinsic line widths (ln$(F_\mathrm{tot})$, ln$W_\mathrm{NL1}$, ln$W_\mathrm{NL2}$, ln$W_\mathrm{IL}$), and cumulative flux fractions ($f_\mathrm{NL1}$, $f_\mathrm{NL1}$+$f_\mathrm{NL2}$, $f_\mathrm{NL1}$+$f_\mathrm{NL2}$+$f_\mathrm{IL}$). 
 }
	\label{fig:fig_post_nlnlil}
\end{figure*}

In addition to Figure~\ref{fig:fig_post} where we plotted posteriors for the favored NLIL model for one of the bins, in Figures~\ref{fig:fig_post_nl}, \ref{fig:fig_post_nlnl}, \ref{fig:fig_post_nlnlil} we show the posteriors for the NL, NLNL and NLNLIL models, where the latter two figures show the parameters that we actually sample from.

%posterior
\subsection{MCMC sampling}
\label{mcmc}

Posterior samples were drawn ensemble MCMC sampler \citep{gw10}, an implementation of which has been popularized as "emcee" \citep{fmackey13}. 
Among other advantages, this method provides for (near-) optimal tuning at every stage of the sampling, which would otherwise be a substantial challenge and obstacle in 
the way of efficient sampling considering 82 different data sets, four different models for each of them, up to 10 model parameters, and our intent to test the sampler and results based on hundreds of additional simulated data sets.  
We draw the initial parameters for 128 parallel chains (walkers) uniformly between the prior boundaries. The unnormalized 
log-posterior of a model is computed as a sum of unnormalized log-prior and log-likelihood for the proposed walker position. 
Since our data (fluxes in the spectral bins) result from Poisson processes, the likelihood is the product of each datum's probability under a Poisson distribution with expectation value equal to the model prediction. After taking the logarithm,  

\begin{equation}
 \mathrm{ln}L = \sum^\mathrm{N=48}_{i=1} d_\mathrm{i}\mathrm{ln}(m_{i}) - m_{i} - \mathrm{ln}\left(\varGamma(d_{i} + 1)\right) ~,
 \label{eq:poiss}
\end{equation}

\noindent
where $m_{i}$ are the model predictions at each spectral bin $i$, and $d_{i}$ are the corresponding data. The last term in eq.~\ref{eq:poiss} (the factorial term) was left out of the posterior sampling (but not the evidence calculation) because it is 
model independent. We refer to \cite{fmackey13} and \cite{gw10} for details of the sampling algorithm. Before checking the chains' convergence, we disregard the first 25\%, but at least 512, of the samples of each walker ("burn-in"), and further thin the chains until the autocorrelation time of the thinned sample is smaller that 5. 
In order to achieve low noise and to set a first minimum threshold for convergence of the chains, we require at least 2$^{13}$=8096 total samples (all walkers combined) to be kept after thinning. Once this minimum number of samples is reached, we additionally impose the following convergence (stopping) criterion: we split the sample in subsamples and compute the desired estimators 
(maximum-posterior sample, median and 95\%-confidence interval boundaries) for each subsample. The variances of the subsample estimators are then required to be smaller than 5\% of the mean parameter value. 
In order to reduce the probability of coincidentally favorable (small) variances from possible "modes" in the chains, we repeat the process twice, first taking each of the 128 chains as one subsample, and second each of the 128-element walker states as subsamples, and use the arithmetic mean of the resulting eight relative estimator standard deviations towards our 5\%-criterion. \\
\\
\indent
To ensure that the applied procedure gives the correct results, we performed tests of the posterior sampling routine, using simulated data with known model parameters. We tested models with $S/N$ in the range $[5, 50]$ and varying background level (0\%, 50\% and 90\% of the total flux).
First, we checked if the posterior distribution reproduces the prior in the $S/N=0$ limit, but also if the posterior approached a delta function in the infinity limit, i.e. for very large $S/N$. 

%physical uncertainty
Second, we check if the model parameters of the input model are reproduced statistically. We ran the algorithm for 200 different realizations of the same model, i.e. each time drawing the data from the same model prediction with its specified uncertainty included (each a set of 48 Poisson distributions). 
The mean of the distribution of median values is always consistent with the input model parameter and the typical scatter of this mean is 10-20\% for the range of $S/N$ in our real data.  

Finally, we vary model parameters by randomly choosing them 200 times from within the prior boundaries. Again, for each of the resulting simulated data we sample the posterior as described above, and evaluate it in form of the median of the marginalized posteriors. 
We find that the measured median values scatter symmetrically around the 1:1 relation with the input model parameters, with the scatter being roughly equal to the individual posteriors' standard deviation, as desired. 
The distribution of measured values become biased as the input parameters approach the parameter range boundaries, as expected for our priors.

%model comparison
\subsection{Evidence calculation via LOO-CV likelihood}
\label{cv}

We use the cross-validation (CV) likelihood, specifically its "leave-one-out" (LOO) variant, to compute model evidences and to compare models \citep{cbj12}. We prefer it over the standard numeric ("Bayesian") integral, because it draws samples from the posterior instead of the prior. It is hence more efficient, less dependent on the choice of prior, and in some cases numerically more stable.
The idea of CV likelihood is to evaluate the likelihood of part of the data, given the model and the rest of the data. In our case we have 48 data points and measure how well any 47 data points (the complement, $D_{-k}$) under the model $M$ predict the 48th data point (the partition, $D_k$), as quantified by the complement's posterior $P(\theta | D_{-k},M)$ and its prediction for $D_k$. The process is repeated for all possible (48) partitions. Each time we leave out one datum ($D_k$), its {\it partition likelihood} $L_k$ is given by
\begin{equation}
 L_\mathrm{k} = P(D_{k}|D_{-k},M) = \int_\theta P(D_k | \theta, M) P(\theta|D_{-k},M)\,\mathrm{d}\theta  ~.
\end{equation}
The first term in the integrand is the likelihood of $D_k$, while the second term is the posterior PDF after considering the information contained in $D_{-k}$. We can numerically (Monte-Carlo) integrate by drawing a number of samples $N$ from $P(\theta|D_{-k},M)$:
\begin{equation}
 L_k \approx \frac{1}{N} \sum^{N}_{n=1} P(D_k|\theta_n,M) ~.
\end{equation}
Assuming that the data points are independent, the LOO-CV likelihood is the product of all partition likelihoods: $L_\mathrm{LOO-CV} = \prod^{48}_{k=1} L_k$, or
\begin{equation}
 \ln L_\mathrm{LOO-CV} = \sum^{48}_{k=1} \ln L_k ~. 
\end{equation}
It can be shown that $L_\mathrm{LOO-CV}$ for model $M$ is equal to the Bayesian evidence ($E(M)$), and we henceforth use it to compute the Bayes factors, $E(M_1)/E(M_2)$ to compare models.

We tested the ability of the Bayes factors to indicate the correct model by employing simulated data. We find that the typical numerical precision of $\ln L_\mathrm{LOO-CV}$ is better than 0.05\,dex. At the same time, for data generated from parameters and with $S/N$ that are typical for the actual data, the "right" model's $\ln L_\mathrm{LOO-CV}$ is $\gtrsim 0.2\,\mathrm{dex}$ ($\gtrsim 50\%$) better than any of the alternative models. Nevertheless, on heuristic grounds (what probability is considered statistically "significant enough" ?), and in order to bracket the practically limited scope of such tests, we adopt a more conservative $+0.5\,\mathrm{dex}$ threshold ($3:1$ probability) before we consider a model preferable over another.

We have also tested the dependence of the evidence ratios on the choice of prior. We used the same functional form for the priors (Table~\ref{tab:param}), but different $\alpha$, $\beta$ distribution parameters. Specifically, we tested $\alpha_\mathrm{D}$ values for the Dirichlet distribution, and $\alpha_\mathrm{B}=\beta_\mathrm{B}$ for the Beta distribution with values of $(\alpha_\mathrm{D}$, $\alpha_\mathrm{B}) = \{(1.5,\,1.5),\,(1,\,1.5),\,(1.5, 1),\,(1,\,1)\}$, where $(1.5, 1.5)$ was used for the results presented in the main part of the paper. We found that for all bins, the mean standard deviation of the appropriate evidence ratios is $\lesssim$\,0.07\,dex.

\begin{figure*}[t!]
	\centering
		\includegraphics[width=0.9\textwidth,height=\textwidth,angle=-90]{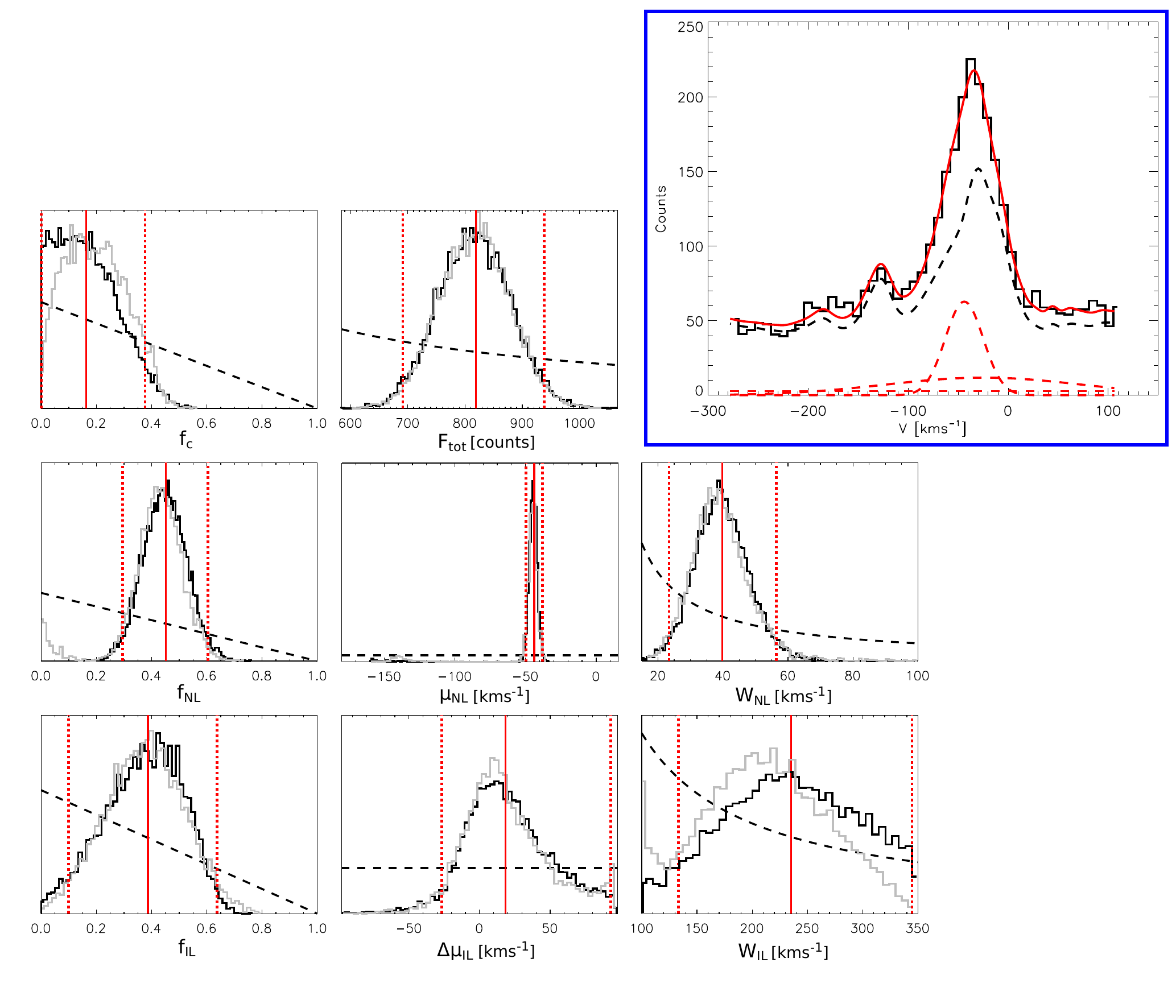}
	\caption{\small Parameter estimation of an NLIL model via Bayesian inference for the bin in the NE filament of Tycho’s SNR for which we presented posterior in Figure~\ref{fig:fig_post} (see its caption for explanation). Posteriors in black are calculated for flat Dirichlet and Beta priors for all parameters. Posteriors from Figure~\ref{fig:fig_post} are overplotted in grey. Among all parameters, the \wil\ posterior is the most sensitive to the prior choice. Even so, the median and 95\%-confidence intervals agree within $\approx$\,10\%, and its basic shape remains robust.
 }
	\label{fig:fig2_flat}
\end{figure*}

\begin{figure*}[t!]
	\centering
		\includegraphics[width=0.9\textwidth,height=\textwidth,angle=-90]{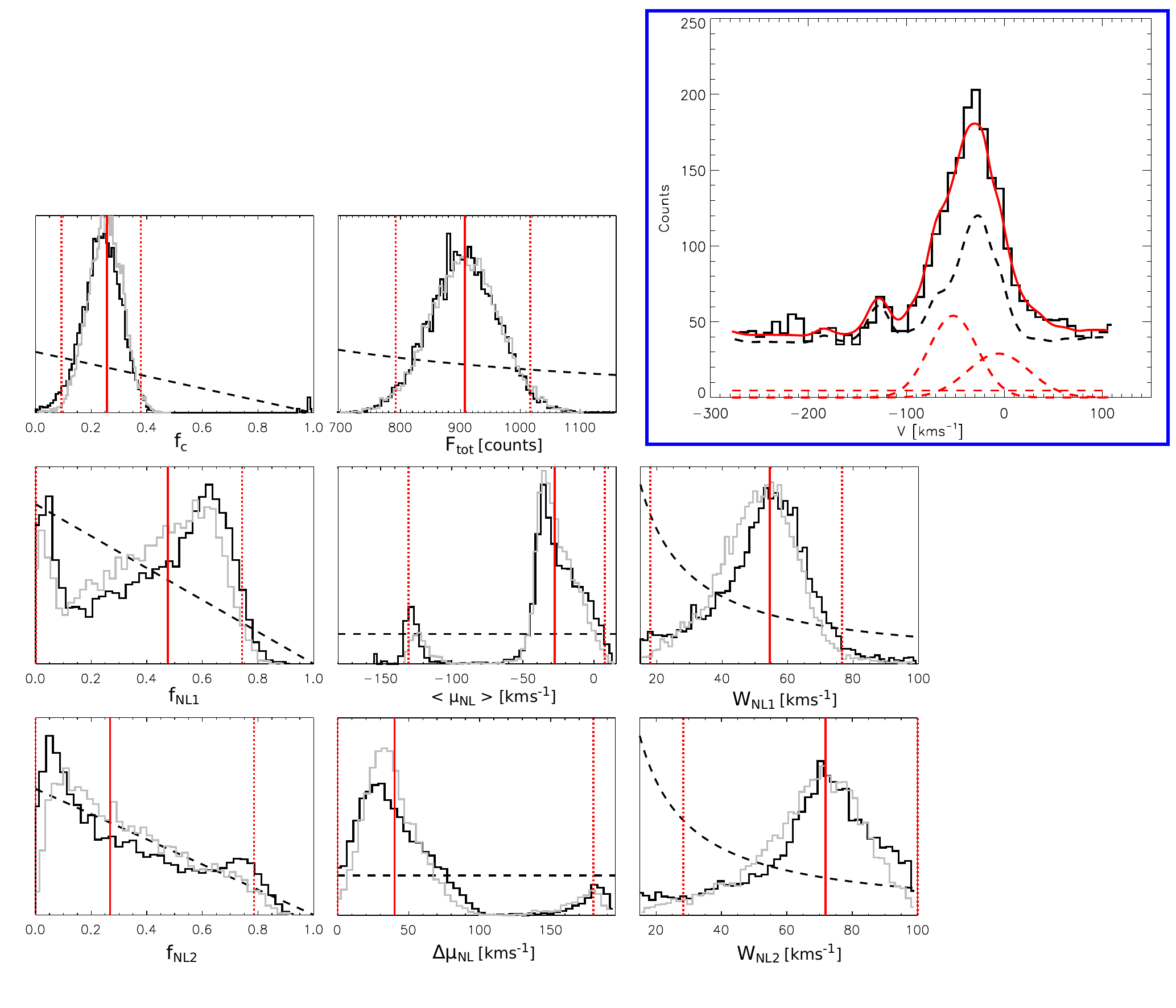}
	\caption{\small Parameter estimation of an NLNL model via Bayesian inference for the bin in the NE filament of Tycho’s SNR for which we presented posterior in Figure~\ref{fig:fig2_add} (see its caption for explanation). Posteriors in black are calculated for flat Dirichlet and Beta priors for all parameters. Posteriors from Figure~\ref{fig:fig2_add} are overplotted in grey. }
	\label{fig:fig5_flat}
\end{figure*}

In addition, in Figures~\ref{fig:fig2_flat} \& \ref{fig:fig5_flat} we show the posteriors, medians and 95\%-confidence intervals when flat priors, i.e. $(\alpha_\mathrm{D}$, $\alpha_\mathrm{B}) = (1,1)$ instead of the fiducial $(1.5, 1.5)$, are applied. We present results for the same two Voronoi bins shown in Figures~\ref{fig:fig_post} \& \ref{fig:fig2_add}. The shape of the posteriors and their medians are very similar to those obtained with the adopted priors, demonstrating that our results do not depend strongly on the choice of prior. \wil\ posteriors for the adopted and flat prior are different, but similar to the degree that median and confidence interval boundaries change only by 30\,\kms\ ($\approx 10\%$). 
Even with the flat prior, 210\,\kms\ is clearly more probable than other \wil\ parameter values, in particular preferred over values close to the \wil\ limits. This shows that the preference of central \wil\ values is not just borne out of the prior shape or range, but genuinely reflects constraints provided by the data, even if they are not as strong as for other parameters.

\clearpage
\newpage
\onecolumngrid

\begin{center}
\begin{minipage}{0.5\textwidth}
\begin{longtable}{cccccccccc}
\caption{Model comparison for the 73 spatial bins in the eastern shock filament of Tycho's SNR. 
Columns 1--5: number of the (Voronoi) bin, $x$ and $y$ coordinates of the bin centroid, number of combined pixels, and signal-to-noise ratio.
Columns 6--10: relative log CV likelihoods of the favored model (denoted with 0) to other models. 
} \label{tab:tycho_cv_E} \\\hline
\hline Bin & x [\arcsec] & y [\arcsec] & Pix & S/N & 0L & NL & NLNL & NLIL & NLNLIL\\
 (1) & (2) & (3) & (4) & (5) & (6) & (7) & (8) & (9) & (10)\\ \hline
\endfirsthead

%\hline %\multicolumn{3}{|r|}{{Continued on next page}} \\ \hline
%\endfoot

%\hline \hline
%\endlastfoot

 1&    22.0&    59.2&   20&   14.40&  27.073&   0.108&   0&   0.077&   0.234\\
 2&    21.4&    58.4&   26&   14.21&  31.122&   0.307&   0.273&   0&   0.098\\
 3&    22.4&    58.2&   21&   14.96&  40.522&   0.245&   0&   0.236&   0.447\\
 4&    22.8&    59.7&   26&   12.52&  13.405&   0&   0.318&   0.220&   0.476\\
 5&    21.6&    57.4&   22&   12.33&  27.723&   0.435&   0.343&   0&   0.110\\
 6&    21.0&    60.7&   46&   13.08&  31.293&   0.459&   0.391&   0&   0.091\\
 7&    19.6&    64.2&  188&   12.04&  39.630&   0.483&   0.325&   0.270&   0\\
 8&    21.1&    59.5&   26&   15.19&  30.855&   0.176&   0.113&   0&   0.160\\
 9&    18.3&    61.1&  378&   12.84&  37.771&   0.376&   0.348&   0&   0.185\\
10&    20.5&    67.3&  234&   11.43&  33.046&   0.410&   0.146&   0.473&   0\\
11&    20.4&    57.1&   51&   12.56&  25.860&   0.236&   0&   0.370&   0.295\\
12&    21.3&    56.3&   32&   11.07&  25.137&   0&   0.217&   0.074&   0.266\\
13&    22.4&    57.2&   22&   11.97&  17.165&   0&   0.161&   0.246&   0.356\\
14&    19.7&    58.5&  104&   13.81&  42.871&   0.594&   0.366&   0.088&   0\\
15&    22.5&    55.9&   48&   10.68&  14.121&   0&   0.182&   0&   0.141\\
16&    21.3&    55.0&   40&   10.47&  19.664&   0.188&   0.178&   0.272&   0\\
17&    19.5&    55.3&  137&   13.05&  30.044&   1.456&   1.307&   0&   0.027\\
18&    22.4&    54.5&   27&    9.69&  15.533&   0&   0.203&   0.013&   0.121\\
19&    21.4&    53.8&   37&   10.44&  12.876&   1.387&   0&   0.475&   0.203\\
20&    20.2&    53.4&   51&   11.33&  17.289&   0.094&   0.360&   0&   0.126\\
21&    23.4&    55.0&   40&    8.49&   9.724&   0.436&   0.084&   0&   0.092\\
22&    21.3&    52.6&   31&   10.51&  20.602&   0&   0.109&   0.050&   0.207\\
23&    22.5&    53.3&   30&   10.98&  19.092&   0.518&   0.212&   0.232&   0\\
24&    18.5&    52.7&  138&   11.15&  23.991&   0.107&   0&   0.333&   0.303\\
25&    22.3&    52.1&   27&    9.17&   4.052&   0&   0.249&   0.132&   0.333\\
26&    20.4&    51.8&   38&   11.57&  10.662&   0&   0.267&   0.281&   0.434\\
27&    21.6&    51.4&   23&    9.86&  18.144&   0.016&   0&   0.081&   0.119\\
28&    19.4&    51.0&   48&   11.10&  21.513&   0.018&   0&   0.456&   0.444\\
29&    20.8&    50.8&   19&   10.63&  20.281&   0.166&   0.223&   0&   0.146\\
30&    23.9&    53.6&   67&    9.73&   9.985&   0.524&   0.488&   0&   0.024\\
31&    19.3&    49.6&   36&   10.09&  23.207&   0&   0.131&   0.041&   0.254\\
32&    20.3&    50.0&   28&   11.16&  18.678&   0.459&   0.628&   0&   0.152\\
33&    20.9&    49.0&   32&   10.58&  16.697&   0.218&   0.105&   0.010&   0\\
34&    19.9&    48.7&   26&   11.08&  21.270&   0.081&   0.098&   0&   0.099\\
35&    18.7&    48.1&   47&   12.01&  23.668&   0&   0.085&   0.084&   0.265\\
36&    20.2&    47.5&   48&   11.38&  21.533&   0&   0.008&   0.052&   0.165\\
37&    17.5&    49.6&  165&   11.17&  19.644&   0.316&   0.352&   0&   0.112\\
38&    19.0&    46.9&   37&   11.92&  25.193&   0.730&   0.622&   0&   0.070\\
39&    18.6&    46.0&   21&   12.10&  23.762&   0.603&   0&   0.834&   0.452\\
40&    20.3&    45.5&  115&   12.72&  21.540&   1.005&   0&   0.258&   0.033\\
41&    17.3&    44.3&   59&   11.28&  23.175&   0.503&   0.713&   0&   0.072\\
42&    18.9&    44.9&   35&   11.55&  23.385&   0.731&   0.432&   0&   0.084\\
43&    18.5&    44.2&   16&    9.10&  11.179&   1.116&   0.790&   0.065&   0\\
\end{longtable}
\end{minipage}
\end{center}

\begin{center}
\begin{minipage}{0.5\columnwidth}
\begin{longtable}{cccccccccc}

\multicolumn{10}{c}%
{Table~\ref{tab:tycho_cv_E} -- continued from previous page} \\\hline
\hline Bin & x [\arcsec] & y [\arcsec] & Pix & S/N & 0L & NL & NLNL & NLIL & NLNLIL\\
 (1) & (2) & (3) & (4) & (5) & (6) & (7) & (8) & (9) & (10)\\ \hline
\endhead

\hline 
\endfoot

44&    16.4&    46.4&  279&   12.59&  27.058&   0.125&   0.066&   0&   0.082\\
45&    18.4&    43.4&   31&   11.30&  21.451&   0.033&   0.072&   0&   0.173\\
46&    16.8&    42.6&   98&   12.74&  22.116&   1.082&   0.673&   0&   0.020\\
47&    18.7&    42.3&   52&   11.63&  20.054&   0&   0.303&   0.172&   0.404\\
48&    16.9&    40.6&   58&   12.11&  26.149&   0.207&   0&   0.051&   0.041\\
49&    17.9&    41.3&   29&   11.62&  18.944&   0.195&   0.282&   0&   0.118\\
50&    18.0&    39.7&   42&   12.24&  20.337&   0.324&   0.289&   0&   0.112\\
51&    15.5&    35.6&   58&   11.36&  28.214&   0.685&   0.432&   0.459&   0\\
52&    17.4&    38.1&   21&   10.47&  17.933&   0.324&   0&   0.113&   0.045\\
53&    16.9&    39.0&   38&   12.27&  22.999&   0.265&   0&   0.404&   0.107\\
54&    18.5&    38.0&   58&   10.56&  12.943&   0&   0.022&   0.292&   0.268\\
55&    16.9&    37.3&   29&   12.04&  18.049&   0.466&   0.152&   0.195&   0\\
56&    17.9&    36.7&   29&   11.74&  18.795&   0.198&   0.316&   0&   0.151\\
57&    16.8&    36.2&   23&   12.53&  16.550&   0.589&   0.619&   0&   0.083\\
58&    19.5&    33.6&  344&   11.89&  16.533&   0.187&   0.395&   0&   0.117\\
59&    17.8&    35.7&   34&   12.40&  30.850&   0.695&   0&   0.361&   0.150\\
60&    16.6&    35.2&   19&   11.81&  16.934&   0.091&   0&   0.096&   0.119\\
61&    17.2&    33.5&   25&   11.15&  17.640&   0.033&   0.203&   0&   0.118\\
62&    17.3&    34.7&   28&   10.86&  19.534&   0&   0.038&   0.172&   0.264\\
63&    16.4&    34.2&   25&   11.63&  26.490&   0.648&   0.649&   0&   0.110\\
64&    17.6&    32.1&   42&   12.86&  26.277&   2.415&   1.670&   0&   0.098\\
65&    16.5&    32.8&   29&   13.09&  22.555&   0&   0.123&   0.128&   0.278\\
66&    16.5&    31.7&   34&   12.12&  27.519&   0.086&   0.126&   0&   0.154\\
67&    16.0&    30.7&   32&   12.02&  17.226&   0&   0.131&   0.088&   0.203\\
68&    14.4&    31.7&  155&   12.61&  32.667&   0&   0.148&   0.165&   0.393\\
69&    13.2&    22.1&   59&    9.33&  18.851&   0.095&   0.241&   0&   0.171\\
70&    12.3&    18.4&  285&   10.04&  15.999&   0.047&   0.152&   0&   0.112\\
71&    13.8&    26.8&   47&    8.78&   8.551&   0&   0.239&   0.030&   0.194\\
72&    15.3&    33.6&   61&   11.76&  19.728&   0.090&   0.377&   0&   0.185\\
73&    13.3&    20.6&   59&    9.43&  11.914&   0&   0.208&   0.138&   0.300\\
\end{longtable}
\end{minipage}
\end{center}

\addtocounter{table}{-1}
\begin{center}
\begin{minipage}{0.5\columnwidth}
\begin{longtable}{cccccccccc}
\caption{Same as Table~\ref{tab:tycho_cv_E} just for the 9 bins in the northern filament.} \label{tab:tycho_cv_N} \\\hline
\hline Bin & x [\arcsec] & y [\arcsec] & Pix & S/N & 0L & NL & NLNL & NLIL & NLNLIL\\
 (1) & (2) & (3) & (4) & (5) & (6) & (7) & (8) & (9) & (10)\\ \hline
\endfirsthead

\hline %\multicolumn{3}{|r|}{{Continued on next page}} \\ \hline
\endfoot

 1&    13.0&    23.7&  351&   14.30&  33.932&   0.846&   0&   0.667&   0.357\\
 2&    10.4&    20.6&   62&   15.31&  43.600&   0.483&   0.479&   0&   0.264\\
 3&     8.9&    22.1&  169&   15.36&  53.524&   1.645&   0.423&   0&   0.009\\
 4&    12.3&    20.8&  169&   14.77&  41.332&   0.003&   0&   0.068&   0.205\\
 5&     8.6&    19.7&  129&   15.50&  49.913&   1.194&   0&   0.644&   0.517\\
 6&    11.7&    17.3&  355&   16.72&  34.586&   0.470&   0&   0.186&   0.077\\
 7&     9.0&    17.8&   75&   16.31&  42.528&   0.959&   0&   0.741&   0.269\\
 8&     9.1&    15.0&  190&   14.65&  40.034&   1.090&   0.479&   0.329&   0\\
 9&     7.4&    16.5&  104&   14.73&  37.961&   0.086&   0.246&   0.025&   0\\
\end{longtable}
\end{minipage}
\end{center}

\begin{center}
\begin{minipage}{\columnwidth}
\setlength\LTcapwidth{0.75\columnwidth}
{\footnotesize
\begin{longtable}{ c c c c c c c}
 \caption{Median and highest density 95\%-confidence interval of parameter posteriors in the favored model of Voronoi bins. Shown here are the spatial bins in the eastern filament of Tycho's SNR. 
              If none of multi-line models is at least $\approx$ 3 times (or 0.5\,dex) more likely, the single-line (NL) model is taken as the favored one.}
\label{tab:tycho_param_E} \\ \hline
\hline Bin & Model & $W_\mathrm{NL1}$ & $W_\mathrm{NL2}$ & $W_\mathrm{IL}$ & 
$f_\mathrm{IL}/f_\mathrm{<NL>}$ & $\Delta \mu_\mathrm{NL}$\\
&  &  [kms$^{-1}$] &  [kms$^{-1}$] &  [kms$^{-1}$] &  &  [kms$^{-1}$] \\ \hline

\endfirsthead

%\hline %\multicolumn{3}{|r|}{{Continued on next page}} \\ \hline
%\endfoot

\multirow{2}{*}{ 1 } & \multirow{2}{*}{ NL } &
 60.53&\multirow{2}{*}{-}&\multirow{2}{*}{-}&\multirow{2}{*}{-}&\multirow{2}{*}{-}\\
& &[43.84, 80.68]& & & & \\%\hline
\multirow{2}{*}{ 2 } & \multirow{2}{*}{ NL } &
 65.12&\multirow{2}{*}{-}&\multirow{2}{*}{-}&\multirow{2}{*}{-}&\multirow{2}{*}{-}\\
& &[48.10, 84.66]& & & & \\%\hline
\multirow{2}{*}{ 3 } & \multirow{2}{*}{ NL } &
 59.23&\multirow{2}{*}{-}&\multirow{2}{*}{-}&\multirow{2}{*}{-}&\multirow{2}{*}{-}\\
& &[44.61, 76.77]& & & & \\%\hline
\multirow{2}{*}{ 4 } & \multirow{2}{*}{ NL } &
 72.07&\multirow{2}{*}{-}&\multirow{2}{*}{-}&\multirow{2}{*}{-}&\multirow{2}{*}{-}\\
& &[49.95, 94.80]& & & & \\%\hline
\multirow{2}{*}{ 5 } & \multirow{2}{*}{ NL } &
 60.90&\multirow{2}{*}{-}&\multirow{2}{*}{-}&\multirow{2}{*}{-}&\multirow{2}{*}{-}\\
& &[44.14, 81.94]& & & & \\%\hline
\multirow{2}{*}{ 6 } & \multirow{2}{*}{ NLIL } &
 40.63&\multirow{2}{*}{-}&191.02&  0.51&\multirow{2}{*}{-}\\
& &[24.78, 56.31]& &[100.15, 308.96]&[0.01, 1.55]& \\%\hline
\multirow{2}{*}{ 7 } & \multirow{2}{*}{ NLNLIL } &
 35.23& 31.68&195.54&  0.26& 21.65\\
& &[15.02, 63.35]&[15.11, 79.41]&[102.05, 310.43]&[0, 0.64]&[0.04, 108.74]\\%\hline
\multirow{2}{*}{ 8 } & \multirow{2}{*}{ NL } &
 60.78&\multirow{2}{*}{-}&\multirow{2}{*}{-}&\multirow{2}{*}{-}&\multirow{2}{*}{-}\\
& &[45.34, 78.82]& & & & \\%\hline
\multirow{2}{*}{ 9 } & \multirow{2}{*}{ NL } &
 46.85&\multirow{2}{*}{-}&\multirow{2}{*}{-}&\multirow{2}{*}{-}&\multirow{2}{*}{-}\\
& &[32.28, 61.53]& & & & \\%\hline
\multirow{2}{*}{ 10 } & \multirow{2}{*}{ NL } &
 40.87&\multirow{2}{*}{-}&\multirow{2}{*}{-}&\multirow{2}{*}{-}&\multirow{2}{*}{-}\\
& &[28.68, 54.37]& & & & \\%\hline
\multirow{2}{*}{ 11 } & \multirow{2}{*}{ NL } &
 54.04&\multirow{2}{*}{-}&\multirow{2}{*}{-}&\multirow{2}{*}{-}&\multirow{2}{*}{-}\\
& &[35.29, 73.60]& & & & \\%\hline
\multirow{2}{*}{ 12 } & \multirow{2}{*}{ NL } &
 78.70&\multirow{2}{*}{-}&\multirow{2}{*}{-}&\multirow{2}{*}{-}&\multirow{2}{*}{-}\\
& &[62.54, 95.65]& & & & \\%\hline
\multirow{2}{*}{ 13 } & \multirow{2}{*}{ NL } &
 70.71&\multirow{2}{*}{-}&\multirow{2}{*}{-}&\multirow{2}{*}{-}&\multirow{2}{*}{-}\\
& &[49.50, 92.40]& & & & \\%\hline
\multirow{2}{*}{ 14 } & \multirow{2}{*}{ NLNLIL } &
 40.34& 35.18&191.06&  0.33& 22.74\\
& &[15.76, 81.33]&[15.50, 72.99]&[102.59, 311.14]&[0, 0.80]&[0.16, 101.29]\\%\hline
\multirow{2}{*}{ 15 } & \multirow{2}{*}{ NL } &
 71.88&\multirow{2}{*}{-}&\multirow{2}{*}{-}&\multirow{2}{*}{-}&\multirow{2}{*}{-}\\
& &[43.78, 99.09]& & & & \\%\hline
\multirow{2}{*}{ 16 } & \multirow{2}{*}{ NL } &
 75.86&\multirow{2}{*}{-}&\multirow{2}{*}{-}&\multirow{2}{*}{-}&\multirow{2}{*}{-}\\
& &[56.59, 96.57]& & & & \\%\hline
\multirow{2}{*}{ 17 } & \multirow{2}{*}{ NLIL } &
 39.13&\multirow{2}{*}{-}&212.36&  0.88&\multirow{2}{*}{-}\\
& &[22.30, 56.71]& &[100.18, 307.88]&[0.01, 6.15]& \\%\hline
\multirow{2}{*}{ 18 } & \multirow{2}{*}{ NL } &
 82.43&\multirow{2}{*}{-}&\multirow{2}{*}{-}&\multirow{2}{*}{-}&\multirow{2}{*}{-}\\
& &[64.85, 98.87]& & & & \\%\hline
\multirow{2}{*}{ 19 } & \multirow{2}{*}{ NLNL } &
 45.04& 65.68&\multirow{2}{*}{-}&\multirow{2}{*}{-}& 91.34\\
& &[15.39, 90.16]&[39.58, 90.64]& & &[41.02, 115.24]\\%\hline
\multirow{2}{*}{ 20 } & \multirow{2}{*}{ NL } &
 74.45&\multirow{2}{*}{-}&\multirow{2}{*}{-}&\multirow{2}{*}{-}&\multirow{2}{*}{-}\\
& &[54.71, 96.58]& & & & \\%\hline
\multirow{2}{*}{ 21 } & \multirow{2}{*}{ NL } &
 73.00&\multirow{2}{*}{-}&\multirow{2}{*}{-}&\multirow{2}{*}{-}&\multirow{2}{*}{-}\\
& &[47.82, 98.64]& & & & \\%\hline

\end{longtable}}
\end{minipage}
\end{center}

\begin{center}
\begin{minipage}{\columnwidth}
{\footnotesize
\begin{longtable}{ccccccc}

\multicolumn{7}{c}%
{Table~\ref{tab:tycho_param_E} -- continued from previous page} \\ \hline
\hline Bin & Model & $W_\mathrm{NL1}$ & $W_\mathrm{NL2}$ & $W_\mathrm{IL}$ & 
$f_\mathrm{IL}/f_\mathrm{<NL>}$ & $\Delta \mu_\mathrm{NL}$\\
&  &  [kms$^{-1}$] &  [kms$^{-1}$] &  [kms$^{-1}$] &  &  [kms$^{-1}$] \\ \hline
\endhead

\multirow{2}{*}{ 22 } & \multirow{2}{*}{ NL } &
 68.76&\multirow{2}{*}{-}&\multirow{2}{*}{-}&\multirow{2}{*}{-}&\multirow{2}{*}{-}\\
& &[50.01, 88.42]& & & & \\%\hline
\multirow{2}{*}{ 23 } & \multirow{2}{*}{ NLNLIL } &
 48.04& 49.68&168.44&  0.42& 51.99\\
& &[15.86, 88.08]&[19.25, 84.92]&[100.05, 295.69]&[0, 1.78]&[11.42, 153.20]\\%\hline
\multirow{2}{*}{ 24 } & \multirow{2}{*}{ NL } &
 59.97&\multirow{2}{*}{-}&\multirow{2}{*}{-}&\multirow{2}{*}{-}&\multirow{2}{*}{-}\\
& &[43.21, 77.48]& & & & \\%\hline
\multirow{2}{*}{ 25 } & \multirow{2}{*}{ NL } &
 66.63&\multirow{2}{*}{-}&\multirow{2}{*}{-}&\multirow{2}{*}{-}&\multirow{2}{*}{-}\\
& &[36.35, 96.93]& & & & \\%\hline
\multirow{2}{*}{ 26 } & \multirow{2}{*}{ NL } &
 70.97&\multirow{2}{*}{-}&\multirow{2}{*}{-}&\multirow{2}{*}{-}&\multirow{2}{*}{-}\\
& &[47.59, 95.35]& & & & \\%\hline
\multirow{2}{*}{ 27 } & \multirow{2}{*}{ NL } &
 58.66&\multirow{2}{*}{-}&\multirow{2}{*}{-}&\multirow{2}{*}{-}&\multirow{2}{*}{-}\\
& &[39.30, 81.32]& & & & \\%\hline
\multirow{2}{*}{ 28 } & \multirow{2}{*}{ NL } &
 56.87&\multirow{2}{*}{-}&\multirow{2}{*}{-}&\multirow{2}{*}{-}&\multirow{2}{*}{-}\\
& &[41.50, 74.12]& & & & \\%\hline
\multirow{2}{*}{ 29 } & \multirow{2}{*}{ NL } &
 56.07&\multirow{2}{*}{-}&\multirow{2}{*}{-}&\multirow{2}{*}{-}&\multirow{2}{*}{-}\\
& &[36.44, 79.31]& & & & \\%\hline
\multirow{2}{*}{ 30 } & \multirow{2}{*}{ NLIL } &
 59.03&\multirow{2}{*}{-}&181.53&  1.44&\multirow{2}{*}{-}\\
& &[21.68, 93.56]& &[103.90, 288.14]&[0, 7.08]& \\%\hline
\multirow{2}{*}{ 31 } & \multirow{2}{*}{ NL } &
 56.38&\multirow{2}{*}{-}&\multirow{2}{*}{-}&\multirow{2}{*}{-}&\multirow{2}{*}{-}\\
& &[37.83, 74.36]& & & & \\%\hline
\multirow{2}{*}{ 32 } & \multirow{2}{*}{ NLIL } &
 41.53&\multirow{2}{*}{-}&150.76&  1.02&\multirow{2}{*}{-}\\
& &[15.31, 76.64]& &[100.17, 278.80]&[0, 3.29]& \\%\hline
\multirow{2}{*}{ 33 } & \multirow{2}{*}{ NL } &
 38.17&\multirow{2}{*}{-}&\multirow{2}{*}{-}&\multirow{2}{*}{-}&\multirow{2}{*}{-}\\
& &[22.16, 56.19]& & & & \\%\hline
\multirow{2}{*}{ 34 } & \multirow{2}{*}{ NL } &
 47.04&\multirow{2}{*}{-}&\multirow{2}{*}{-}&\multirow{2}{*}{-}&\multirow{2}{*}{-}\\
& &[30.20, 65.05]& & & & \\%\hline
\multirow{2}{*}{ 35 } & \multirow{2}{*}{ NL } &
 72.60&\multirow{2}{*}{-}&\multirow{2}{*}{-}&\multirow{2}{*}{-}&\multirow{2}{*}{-}\\
& &[56.45, 89.57]& & & & \\%\hline
\multirow{2}{*}{ 36 } & \multirow{2}{*}{ NL } &
 59.26&\multirow{2}{*}{-}&\multirow{2}{*}{-}&\multirow{2}{*}{-}&\multirow{2}{*}{-}\\
& &[40.97, 80.03]& & & & \\%\hline
\multirow{2}{*}{ 37 } & \multirow{2}{*}{ NL } &
 42.92&\multirow{2}{*}{-}&\multirow{2}{*}{-}&\multirow{2}{*}{-}&\multirow{2}{*}{-}\\
& &[23.91, 62.21]& & & & \\%\hline
\multirow{2}{*}{ 38 } & \multirow{2}{*}{ NLIL } &
 41.82&\multirow{2}{*}{-}&170.06&  0.74&\multirow{2}{*}{-}\\
& &[21.28, 66.79]& &[100.12, 294.73]&[0, 2.26]& \\%\hline
\multirow{2}{*}{ 39 } & \multirow{2}{*}{ NLNL } &
 51.41& 56.03&\multirow{2}{*}{-}&\multirow{2}{*}{-}& 60.15\\
& &[16.46, 90.57]&[29.12, 80.18]& & &[9.50, 84.17]\\%\hline
\multirow{2}{*}{ 40 } & \multirow{2}{*}{ NLNL } &
 38.06& 63.21&\multirow{2}{*}{-}&\multirow{2}{*}{-}& 39.90\\
& &[15.61, 61.48]&[26.59, 96.32]& & &[1.32, 144.16]\\%\hline
\multirow{2}{*}{ 41 } & \multirow{2}{*}{ NLIL } &
 48.34&\multirow{2}{*}{-}&201.48&  0.62&\multirow{2}{*}{-}\\
& &[24.99, 70.31]& &[103.46, 310.50]&[0.03, 1.67]& \\%\hline
\multirow{2}{*}{ 42 } & \multirow{2}{*}{ NLIL } &
 48.09&\multirow{2}{*}{-}&162.60&  0.61&\multirow{2}{*}{-}\\
& &[27.07, 72.44]& &[100.39, 291.90]&[0.01, 1.82]& \\%\hline
\multirow{2}{*}{ 43 } & \multirow{2}{*}{ NLNLIL } &
 44.57& 33.42&171.73&  0.75& 55.59\\
& &[15.43, 87.20]&[15.03, 77.23]&[100.31, 295.94]&[0.01, 2.23]&[1.57, 97.03]\\%\hline

\end{longtable}}
\end{minipage}
\end{center}

\begin{center}
\begin{minipage}{\columnwidth}
{\footnotesize
\begin{longtable}{ccccccc}

\multicolumn{7}{c}%
{Table~\ref{tab:tycho_param_E} -- continued from previous page} \\ \hline
\hline Bin & Model & $W_\mathrm{NL1}$ & $W_\mathrm{NL2}$ & $W_\mathrm{IL}$ & 
$f_\mathrm{IL}/f_\mathrm{<NL>}$ & $\Delta \mu_\mathrm{NL}$\\
&  &  [kms$^{-1}$] &  [kms$^{-1}$] &  [kms$^{-1}$] &  &  [kms$^{-1}$] \\ \hline
\endhead

\multirow{2}{*}{ 44 } & \multirow{2}{*}{ NL } &
 65.49&\multirow{2}{*}{-}&\multirow{2}{*}{-}&\multirow{2}{*}{-}&\multirow{2}{*}{-}\\
& &[48.42, 85.13]& & & & \\%\hline
\multirow{2}{*}{ 45 } & \multirow{2}{*}{ NL } &
 55.89&\multirow{2}{*}{-}&\multirow{2}{*}{-}&\multirow{2}{*}{-}&\multirow{2}{*}{-}\\
& &[38.00, 76.15]& & & & \\%\hline
\multirow{2}{*}{ 46 } & \multirow{2}{*}{ NLIL } &
 35.29&\multirow{2}{*}{-}&221.61&  0.89&\multirow{2}{*}{-}\\
& &[17.55, 53.39]& &[100.39, 317.73]&[0, 10.90]& \\%\hline
\multirow{2}{*}{ 47 } & \multirow{2}{*}{ NL } &
 71.18&\multirow{2}{*}{-}&\multirow{2}{*}{-}&\multirow{2}{*}{-}&\multirow{2}{*}{-}\\
& &[51.66, 91.45]& & & & \\%\hline
\multirow{2}{*}{ 48 } & \multirow{2}{*}{ NL } &
 49.41&\multirow{2}{*}{-}&\multirow{2}{*}{-}&\multirow{2}{*}{-}&\multirow{2}{*}{-}\\
& &[32.16, 66.58]& & & & \\%\hline
\multirow{2}{*}{ 49 } & \multirow{2}{*}{ NL } &
 43.21&\multirow{2}{*}{-}&\multirow{2}{*}{-}&\multirow{2}{*}{-}&\multirow{2}{*}{-}\\
& &[27.87, 63.95]& & & & \\%\hline
\multirow{2}{*}{ 50 } & \multirow{2}{*}{ NL } &
 61.30&\multirow{2}{*}{-}&\multirow{2}{*}{-}&\multirow{2}{*}{-}&\multirow{2}{*}{-}\\
& &[42.56, 83.35]& & & & \\%\hline
\multirow{2}{*}{ 51 } & \multirow{2}{*}{ NLNLIL } &
 50.84& 37.90&201.87&  0.31& 37.67\\
& &[16.21, 81.76]&[16.10, 66.80]&[104.16, 313.03]&[0, 0.89]&[1.03, 90.84]\\%\hline
\multirow{2}{*}{ 52 } & \multirow{2}{*}{ NL } &
 61.42&\multirow{2}{*}{-}&\multirow{2}{*}{-}&\multirow{2}{*}{-}&\multirow{2}{*}{-}\\
& &[39.35, 85.21]& & & & \\%\hline
\multirow{2}{*}{ 53 } & \multirow{2}{*}{ NL } &
 54.00&\multirow{2}{*}{-}&\multirow{2}{*}{-}&\multirow{2}{*}{-}&\multirow{2}{*}{-}\\
& &[38.21, 71.73]& & & & \\%\hline
\multirow{2}{*}{ 54 } & \multirow{2}{*}{ NL } &
 47.73&\multirow{2}{*}{-}&\multirow{2}{*}{-}&\multirow{2}{*}{-}&\multirow{2}{*}{-}\\
& &[23.04, 68.53]& & & & \\%\hline
\multirow{2}{*}{ 55 } & \multirow{2}{*}{ NLNLIL } &
 33.67& 50.83&187.38&  0.50& 34.31\\
& &[15.24, 72.78]&[17.70, 88.41]&[101.07, 312.58]&[0, 1.49]&[1.00, 82.17]\\%\hline
\multirow{2}{*}{ 56 } & \multirow{2}{*}{ NL } &
 58.08&\multirow{2}{*}{-}&\multirow{2}{*}{-}&\multirow{2}{*}{-}&\multirow{2}{*}{-}\\
& &[38.88, 80.69]& & & & \\%\hline
\multirow{2}{*}{ 57 } & \multirow{2}{*}{ NLIL } &
 45.30&\multirow{2}{*}{-}&199.89&  1.12&\multirow{2}{*}{-}\\
& &[17.24, 72.11]& &[107.28, 312.43]&[0.02, 3.11]& \\%\hline
\multirow{2}{*}{ 58 } & \multirow{2}{*}{ NL } &
 74.99&\multirow{2}{*}{-}&\multirow{2}{*}{-}&\multirow{2}{*}{-}&\multirow{2}{*}{-}\\
& &[54.48, 97.77]& & & & \\%\hline
\multirow{2}{*}{ 59 } & \multirow{2}{*}{ NLNL } &
 57.49& 58.12&\multirow{2}{*}{-}&\multirow{2}{*}{-}& 63.14\\
& &[19.54, 93.95]&[36.03, 82.85]& & &[11.53, 93.23]\\%\hline
\multirow{2}{*}{ 60 } & \multirow{2}{*}{ NL } &
 66.49&\multirow{2}{*}{-}&\multirow{2}{*}{-}&\multirow{2}{*}{-}&\multirow{2}{*}{-}\\
& &[46.53, 89.49]& & & & \\%\hline
\multirow{2}{*}{ 61 } & \multirow{2}{*}{ NL } &
 73.42&\multirow{2}{*}{-}&\multirow{2}{*}{-}&\multirow{2}{*}{-}&\multirow{2}{*}{-}\\
& &[52.04, 96.65]& & & & \\%\hline
\multirow{2}{*}{ 62 } & \multirow{2}{*}{ NL } &
 55.28&\multirow{2}{*}{-}&\multirow{2}{*}{-}&\multirow{2}{*}{-}&\multirow{2}{*}{-}\\
& &[37.46, 75.62]& & & & \\%\hline
\multirow{2}{*}{ 63 } & \multirow{2}{*}{ NLIL } &
 49.93&\multirow{2}{*}{-}&202.95&  0.49&\multirow{2}{*}{-}\\
& &[32.62, 69.52]& &[106.47, 316.80]&[0, 1.34]& \\%\hline
\multirow{2}{*}{ 64 } & \multirow{2}{*}{ NLIL } &
 38.86&\multirow{2}{*}{-}&191.52&  1.12&\multirow{2}{*}{-}\\
& &[19.78, 60.12]& &[107.16, 298.62]&[0.12, 2.53]& \\%\hline
\multirow{2}{*}{ 65 } & \multirow{2}{*}{ NL } &
 68.07&\multirow{2}{*}{-}&\multirow{2}{*}{-}&\multirow{2}{*}{-}&\multirow{2}{*}{-}\\
& &[47.10, 92.23]& & & & \\%\hline

\end{longtable}}
\end{minipage}
\end{center}

\begin{center}
\begin{minipage}{\columnwidth}
{\footnotesize
\begin{longtable}{ccccccc}

\multicolumn{7}{c}%
{Table~\ref{tab:tycho_param_E} -- continued from previous page} \\ \hline 
\hline bin & model & $W_\mathrm{NL1}$ & $W_\mathrm{NL2}$ & $W_\mathrm{IL}$ & 
$f_\mathrm{IL}/f_\mathrm{<NL>}$ & $\Delta \mu_\mathrm{NL}$\\
&  &  [kms$^{-1}$] &  [kms$^{-1}$] &  [kms$^{-1}$] &  &  [kms$^{-1}$] \\ \hline
\endhead

\hline 
\endfoot

\multirow{2}{*}{ 66 } & \multirow{2}{*}{ NL } &
 57.62&\multirow{2}{*}{-}&\multirow{2}{*}{-}&\multirow{2}{*}{-}&\multirow{2}{*}{-}\\
& &[42.10, 77.46]& & & & \\%\hline
\multirow{2}{*}{ 67 } & \multirow{2}{*}{ NL } &
 65.35&\multirow{2}{*}{-}&\multirow{2}{*}{-}&\multirow{2}{*}{-}&\multirow{2}{*}{-}\\
& &[46.28, 87.53]& & & & \\%\hline
\multirow{2}{*}{ 68 } & \multirow{2}{*}{ NL } &
 72.89&\multirow{2}{*}{-}&\multirow{2}{*}{-}&\multirow{2}{*}{-}&\multirow{2}{*}{-}\\
& &[57.12, 89.85]& & & & \\%\hline
\multirow{2}{*}{ 69 } & \multirow{2}{*}{ NL } &
 60.22&\multirow{2}{*}{-}&\multirow{2}{*}{-}&\multirow{2}{*}{-}&\multirow{2}{*}{-}\\
& &[41.25, 82.71]& & & & \\%\hline
\multirow{2}{*}{ 70 } & \multirow{2}{*}{ NL } &
 46.09&\multirow{2}{*}{-}&\multirow{2}{*}{-}&\multirow{2}{*}{-}&\multirow{2}{*}{-}\\
& &[27.58, 67.10]& & & & \\%\hline
\multirow{2}{*}{ 71 } & \multirow{2}{*}{ NL } &
 50.96&\multirow{2}{*}{-}&\multirow{2}{*}{-}&\multirow{2}{*}{-}&\multirow{2}{*}{-}\\
& &[29.85, 76.82]& & & & \\%\hline
\multirow{2}{*}{ 72 } & \multirow{2}{*}{ NL } &
 78.28&\multirow{2}{*}{-}&\multirow{2}{*}{-}&\multirow{2}{*}{-}&\multirow{2}{*}{-}\\
& &[61.10, 96.81]& & & & \\%\hline
\multirow{2}{*}{ 73 } & \multirow{2}{*}{ NL } &
 64.24&\multirow{2}{*}{-}&\multirow{2}{*}{-}&\multirow{2}{*}{-}&\multirow{2}{*}{-}\\
& &[42.79, 89.21]& & & & \\%\hline
\end{longtable}}
\end{minipage}
\end{center}

\addtocounter{table}{-3}
\begin{center}
\begin{minipage}{\columnwidth}
\setlength\LTcapwidth{0.75\columnwidth}
{\footnotesize
\begin{longtable}{ c c c c c c c}
 \caption{Same as Table~\ref{tab:tycho_param_E} just for Voronoi bins in the Tycho's northern filament.}
\label{tab:tycho_param_N} \\ \hline
\hline Bin & Model & $W_\mathrm{NL1}$ & $W_\mathrm{NL2}$ & $W_\mathrm{IL}$ & 
$f_\mathrm{IL}/f_\mathrm{<NL>}$ & $\Delta \mu_\mathrm{NL}$\\
&  &  [kms$^{-1}$] &  [kms$^{-1}$] &  [kms$^{-1}$] &  &  [kms$^{-1}$] \\ \hline
\endhead

\hline 
\endfoot

\multirow{2}{*}{ 1 } & \multirow{2}{*}{ NLNL } &
 33.85& 53.16&\multirow{2}{*}{-}&\multirow{2}{*}{-}& 28.04\\
& &[15.04, 61.64]&[21.03, 87.55]& & &[0.29, 156.41]\\%\hline
\multirow{2}{*}{ 2 } & \multirow{2}{*}{ NLIL } &
 58.50&\multirow{2}{*}{-}&195.28&  0.39&\multirow{2}{*}{-}\\
& &[41.77, 75.74]& &[102.92, 316.25]&[0, 1.13]& \\%\hline
\multirow{2}{*}{ 3 } & \multirow{2}{*}{ NLIL } &
 54.32&\multirow{2}{*}{-}&156.07&  0.41&\multirow{2}{*}{-}\\
& &[37.83, 69.52]& &[100.12, 274.97]&[0.02, 1.08]& \\%\hline
\multirow{2}{*}{ 4 } & \multirow{2}{*}{ NL } &
 59.63&\multirow{2}{*}{-}&\multirow{2}{*}{-}&\multirow{2}{*}{-}&\multirow{2}{*}{-}\\
& &[37.09, 78.26]& & & & \\%\hline
\multirow{2}{*}{ 5 } & \multirow{2}{*}{ NLNL } &
 51.95& 69.86&\multirow{2}{*}{-}&\multirow{2}{*}{-}& 39.81\\
& &[25.10, 73.85]&[34.12, 97.51]& & &[2.97, 173.42]\\%\hline
\multirow{2}{*}{ 6 } & \multirow{2}{*}{ NLNL } &
 42.30& 56.58&\multirow{2}{*}{-}&\multirow{2}{*}{-}& 20.37\\
& &[15.65, 83.53]&[20.46, 91.93]& & &[0.05, 67.38]\\%\hline
\multirow{2}{*}{ 7 } & \multirow{2}{*}{ NLNL } &
 37.89& 60.23&\multirow{2}{*}{-}&\multirow{2}{*}{-}& 42.00\\
& &[15.09, 66.94]&[26.97, 93.47]& & &[6.73, 184.78]\\%\hline
\multirow{2}{*}{ 8 } & \multirow{2}{*}{ NLNLIL } &
 44.66& 34.55&149.45&  0.36& 29.63\\
& &[15.68, 70.78]&[15.22, 80.12]&[100.36, 291.76]&[0, 0.89]&[0.23, 86.09]\\%\hline
\multirow{2}{*}{ 9 } & \multirow{2}{*}{ NL } &
 56.38&\multirow{2}{*}{-}&\multirow{2}{*}{-}&\multirow{2}{*}{-}&\multirow{2}{*}{-}\\
& &[37.58, 74.56]& & & & \\%\hline

\end{longtable}}
\end{minipage}
\end{center}

\clearpage
\newpage
%\twocolumn

\end{document}